%% file: neurips_2026.tex
\documentclass{article}

\usepackage[preprint]{neurips_2026}

\usepackage[utf8]{inputenc} 
\usepackage[T1]{fontenc}    
\usepackage{hyperref}       
\usepackage{url}            
\usepackage{booktabs}       
\usepackage{amsfonts}       
\usepackage{nicefrac}       
\usepackage{microtype}      
\usepackage{xcolor}         

\usepackage{graphicx}
\usepackage{multirow}
\usepackage{amsmath}
\usepackage{amsthm}
\usepackage{mathrsfs}
\usepackage[title]{appendix}
\usepackage{xcolor}
\usepackage{textcomp}
\usepackage{manyfoot}
\usepackage{booktabs}
\usepackage{algorithm}
\usepackage{algorithmicx}
\usepackage{algpseudocode}
\usepackage{listings}
\usepackage{subcaption}
\usepackage[ruled,noend,algo2e]{algorithm2e}
\usepackage{xspace}
\usepackage{etoc}
\usepackage{minitoc}

\theoremstyle{plain}

\theoremstyle{definition}

\theoremstyle{remark}

\begin{document}

\title{Tackling the Data-Parallel Load Balancing Bottleneck in LLM Serving: \\ Practical Online Routing at Scale}

\author{%
  Tianci Bu$^{1}$\thanks{Equal contribution.} \quad
  Yuan Lyu$^{2}$\footnotemark[1] \quad
  Zixi Chen$^{2}$\thanks{Equal contribution.} \quad
  Chendong Song$^{1}$\footnotemark[2] \quad
  Hong Liang$^{2}$ \\
  \textbf{Tsepten Gurung}$^{2}$ \quad
  \textbf{Yuwei Fan}$^{2}$\thanks{Corresponding authors.} \quad
  \textbf{Yinyu Ye}$^{1}$\footnotemark[3] \quad
  \textbf{Zijie Zhou}$^{1}$\footnotemark[3] \\[6pt]
  $^{1}$Department of Industrial Engineering and Decision Analytics, HKUST \\
  $^{2}$Theory Laboratory, Central Research Institute, 2012 Laboratories, Huawei\\
  \texttt{\{tbuaa, csongap\}@connect.ust.hk}\\
  \texttt{\{yyye, jerryzhou\}@ust.hk} \\
  \texttt{\{lyu.yuan, chenzixi, liang.hong3, gurung.tsepten, fanyuwei2\}@huawei.com} \\
}

\maketitle

\begin{abstract}
Data-parallel (DP) load balancing has emerged as a first-order
bottleneck in large-scale LLM serving. When a model is sharded across
devices via tensor parallelism (TP) or expert parallelism (EP) and
replicated across many DP workers, every decode step ends in a
synchronization barrier whose latency is set by the most heavily
loaded worker; even modest persistent imbalance across DP workers
compounds, step after step, into a substantial fraction of wasted
compute. The problem is hard for reasons specific to LLM decoding:
assignments are sticky (migrating KV caches has a high cost), per-request
loads grow over time, arrivals are non-stationary, and the router
must decide within a sub-100\,ms decode budget over hundreds of
waiting requests and tens of workers. We present \textbf{BalanceRoute},
a family of practical online routing algorithms that target this
bottleneck. The first, \textbf{BR-0}, requires no prediction
infrastructure and uses a piecewise-linear F-score that captures the
sharp asymmetry between admissions that fill safe margin and those
that overflow into the envelope; a two-stage decomposition keeps
per-step cost compatible with millisecond-scale scheduling. The
second, \textbf{BR-H}, generalizes BR-0 with a short, constant
lookahead $H$ and a lightweight termination-classifier interface,
extending the F-score to a horizon-discounted form. 
We deploy BalanceRoute on a 144-NPU cluster and evaluate
against vLLM baselines on both a proprietary production trace and the
public Azure-2024 trace. Across both workloads, BalanceRoute
substantially reduces average DP imbalance and improves end-to-end
serving throughput.
\end{abstract}

\input{Intro}

\input{Background}

\input{BF0}

\input{system}

\input{Experiment}

\section{Conclusion, Limitations, and Future Work}
\label{sec:conclusion}

We presented \textbf{BalanceRoute}, a family of practical online routers for DP-decode load balancing. Evaluated on a 144-NPU cluster against four standard load-balancing baselines on both a proprietary production trace and the public Azure-2024 trace, BalanceRoute substantially reduces DP imbalance and improves end-to-end throughput, with the advantage over the strongest baseline rising from $+13.4\%$ at $G{=}4$ to $+34.5\%$ at $G{=}16$. 

\noindent \textbf{Limitations and future works. } Each cell is a single replay; consistent 144-NPU availability precludes systematic replicate sweeps, though qualitative orderings are stable across two datasets and three cluster sizes. A potential future direction is to overlap cross-tier KV transfer with decode under $H{>}0$ to characterise how predictor noise interacts with $G$ across the full range of cluster sizes, and to explore decentralised or hardware-offloaded variants of the control plane.

\clearpage
\bibliographystyle{plainnat}
\bibliography{reference}

\clearpage
\appendix


\input{AppendRelated}

\input{AppendAlg}

\input{Appendix.tex}

\end{document}

%% file: Intro.tex
\section{Introduction} \label{sec:intro}

Modern large language models (LLMs) are too large to fit on a single accelerator, so production serving systems shard each model across the devices within a worker group via tensor parallelism or expert parallelism (TP/EP)~\citep{shoeybi2019megatron,fedus2022switch,liu2024deepseek}, and replicate the resulting group across many parallel workers via data parallelism (DP) to scale throughput~\citep{kwon2023efficient}.
Within-group sharding creates collective communication (TP all-reduce or EP all-to-all) at every decode step, imposing a synchronization barrier across all DP replicas \cite{vllm_data_parallel_deployment}; each decode step ends only when the slowest replica is done. At large scale, this barrier turns cross-replica load imbalance into a dominant source of inefficiency.
By a standard order-statistics argument, the gap between the maximum and the mean per-worker load grows with the number of workers, so the idle fraction of every decode step worsens as the system scales up;~\citet{chen2026universal} reports that more than 40\% of accelerator time is lost to barrier idle in production LLM deployments. The cost compounds beyond throughput: idle accelerators continue to draw substantial power~\citep{ozcan2025quantifying}, converting wasted compute into wasted energy. DP load balancing has therefore become a first-order bottleneck on industrial LLM serving efficiency.
 
Despite this importance, DP load balancing in LLM serving is poorly served by the heuristics deployed in practice. Production routers in open-source serving stacks~\citep{vllm-router} rely on generic load-balancing heuristics: round-robin, join-shortest-queue, random, power-of-two-choices~\citep{mitzenmacher2001power}. These heuristics are agnostic to the features distinguishing LLM decoding from classical parallel computing: sticky assignments tied to per-request KV caches, per-step load that grows with each generated token, and routing decisions made at sub-100\,ms decode latency over hundreds of waiting requests and tens to hundreds of replicas (more backgrounds and details are provided in Section~\ref{sec:bg}). \citet{zhou2026positionllmservingneeds}
identifies this DP-decode routing problem as a representative case where the gap between generic load-balancing heuristics and
LLM-specific structure calls for principled algorithm design. In addition, \citet{chen2026universal} recently formalized this problem by an analytical model along with theoretical insights based on a short-horizon lookahead integer optimization formulation. Building on their insights, we focus on a complementary algorithmic question: \emph{how to design routers that
achieve strong DP load balance at real serving scale within the
millisecond per-step budget faced by production deployments. }Other related work is deferred to Appendix~\ref{append:related}.
 
In this work, we present \emph{BalanceRoute}, a family of practical online DP load-balancing routers, and validate them at production scale. Our contributions are:

\textbf{(i) A prediction-free online router (BR-0,
        Section~\ref{sec:br0}).} We derive a piecewise-linear F-score from the marginal effect of an admission on single-step imbalance and use it to drive a two-stage algorithm. 

\textbf{(ii) A lookahead-aware online router (BR-H,
        Section~\ref{sec:brh}).} When a lightweight termination
        classifier is available, BR-H extends the F-score to a
        constant-horizon discounted form within the same two-stage structure. BR-0 emerges as the $H{=}0$, so the two algorithms form a single family.
    
\textbf{(iii) Production-scale empirical validation      (Section~\ref{sec:experiments}).} We deploy BalanceRoute on a 144-NPU Ascend 910C cluster and evaluate it against vLLM baselines on a proprietary production trace and the public Azure-2024 trace. We observe substantial reductions in average DP imbalance and corresponding throughput gains, with the gains growing in the regimes our analysis predicts.

%% file: Background.tex
\section{Background and Problem Formulation} \label{sec:bg}

This section describes the data-parallel (DP) decode setting in which our
routers operate, identifies the structural reasons why DP load balancing
is hard, and introduces the formal model. 

\subsection{DP-decode load imbalance}
\label{subsec:dp_decode}

\noindent \textbf{Why DP is the relevant unit of imbalance.}
Modern LLMs are too large for a single worker to host a full model
replica. Production serving systems shard parameters across devices using
tensor parallelism (TP) or, for mixture-of-experts models, expert
parallelism (EP), and replicate the resulting model across
multiple groups using data parallelism. Each DP group serves an independent
batch of decoding requests. Load imbalance \emph{across} DP workers is determined
entirely by how the router assigns requests, and is the bottleneck we
target.

\noindent \textbf{Sticky assignments and growing KV caches.}
We focus on the router which sends completed prefill requests to decode workers. Under prefill–decode (PD) disaggregation \cite{zhong2024distserve,patel2023splitwise}, prompts are
processed in a separate prefill cluster; once a request finishes prefill,
it is routed to a decode worker which then generates output tokens
autoregressively. Each decoded token appends one entry to the request's
key–value (KV) cache, so the per-request memory and attention cost grow
monotonically over the lifetime of the request. Migrating an in-flight
request would require transferring its full KV cache across devices at
every step; in practice, therefore, assignments are \emph{sticky}: once a
request lands on a worker, it remains there until completion. 

\begin{figure}[t]
    \centering
    \includegraphics[width=0.95\linewidth]{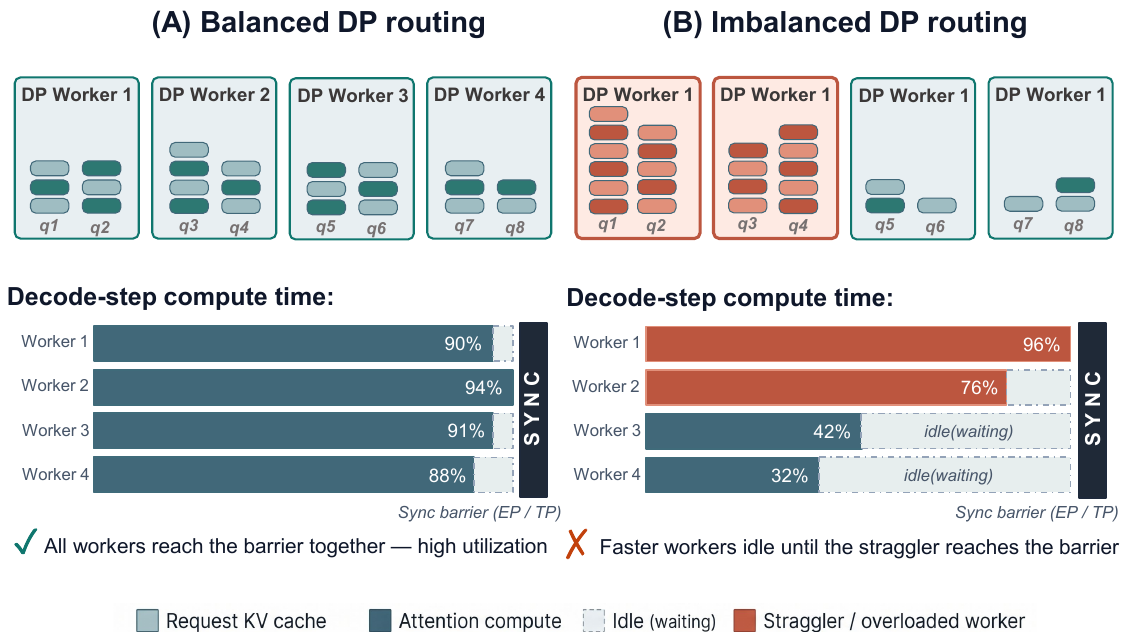}
    \caption{%
        Data-parallel decode under barrier synchronization.
    }
    \label{fig:dp_routing}
\end{figure}

\noindent \textbf{Barrier synchronization.}
The bulk of per-step compute on a decode worker is to read each active request's KV cache; the per-step load on a worker is therefore
approximately linear in the total KV-cache footprint of its active
batch. Within a single decode step, the workers in a DP fleet do not run
independently. Each step contains collective communication—EP all-to-all
when the model uses MoE layers \cite{liu2024deepseek}, TP all-reduce when attention or MLP
weights are sharded—that imposes a step-wise synchronization barrier
across all DP workers \cite{vllm_data_parallel_deployment}. Concretely, every worker must finish its per-step
compute before any worker can proceed, and step latency is set by the
slowest participant. Figure~\ref{fig:dp_routing} contrasts a balanced and
an imbalanced assignment under this constraint: in the balanced case, all
workers reach the barrier together, whereas in the imbalanced case the
lighter workers spend a substantial fraction of every step waiting for the
straggler. Because the per-worker idle accumulates at \emph{every} decode
step, even a modest persistent imbalance compounds into a large fraction
of wasted compute over the lifetime of a request.

\noindent \textbf{Why is DP load-balancing problem hard?}
Four properties together place the DP-decode setting outside the regime in
which classical online makespan or queueing-based load balancing applies: 
\textit{(i) Sticky assignment.} Decisions cannot be revised: migrating a request has a huge cost since we need to carry all its KVs. \textit{(ii) Unknown, dynamic per-request workload.} The remaining
        output length of a request is not known at routing time, and the
        per-step attention cost grows as the KV cache grows. The router
        must therefore reason about not just the current load but the
        evolving load implied by its assignments.
\textit{(iii) Non-stationary arrivals.} Requests arrive from prefill in bursts whose composition is workload-dependent and shifts over time.
\textit{(iv) Millisecond decision budget.} The router runs at every
        decode step (sub-100\,ms in production). This budget rules out per-step
        solutions to decision rules with heavy computations.

\noindent \textbf{When does DP load imbalance matter most?}
Two regimes amplify the impact of DP-decode imbalance and explain why this
problem has become a first-order concern only at modern serving scale: \textbf{(i) Large scale LLM inference.} As the number of DP workers $G$ grows, by a standard order-statistics argument, the gap between the maximum load and the mean load grows with $G$. Barrier idle scales with this
        gap, so the load-balancing penalty is super-linear in $G$. This is the
        regime in which industrial serving operates and in which our
        large-scale experiments (Section~\ref{sec:experiments}) reside.
\textbf{(ii) Bandwidth-driven per-step cost.}
        The per-step time on a worker is approximately $a x + b$, where
        $x$ is the total KV-cache footprint of the worker's active
        batch, $a$ is the bandwidth, and $b$ is a load-independent
        overhead. DP load balancing is
        therefore most consequential when the bandwidth-driven term
        $a x$ is large relative to the fixed overhead $b$—the regime
        of long contexts, high effective batch sizes, or hardware with
        relatively low bandwidth-to-overhead ratios.

\subsection{Mathematical model}
\label{subsec:model}

We now formalize the prefill-to-decode routing problem. Let $G$ be the number of DP workers, indexed by $g\in[G]$, each with a
maximum concurrency of $B\in\mathbb{N}$ active requests. Decode step is discrete
and indexed by $k\in\{0,1,2,\ldots\}$: at step $k$, every active request
on every worker advances by one decode iteration, after which the workers
synchronize at the barrier described above.

\noindent \textbf{Requests and workload profiles.}
Let $\mathcal{I}$ be the set of all requests in a given trace. Each
request $i\in\mathcal{I}$ becomes available to the router at some step
$k_i$ (its prefill-completion time) and is assigned to a worker at some
step $x_i\geq k_i$. Once assigned, $i$ is processed continuously on that
worker until completion; assignments are sticky and we never preempt.
The processing requirement of $i$ is summarized by a \emph{workload profile} $W_i \;=\; \bigl(w_i^{(1)},\, w_i^{(2)},\, \ldots,\, w_i^{(o_i)}\bigr),$ where $o_i\in\mathbb{N}$ is the total number of decode steps performed for
$i$, and $w_i^{(j)}\geq 0$ is the KV workload contributed by $i$ during its
$j$-th decode step on the assigned worker. Specifically, a request $i$
whose prefill produced $s_i$ tokens has $w_i^{(j)} = s_i+j-1$. The profile $W_i$ is partially unobserved by the router at assignment time:
the router knows $i$ has arrived and the value of $w_i^{(1)}=s_i$ (prefill size) but does not know how long it will run ($o_i$ is unknown).

\noindent \textbf{Per-step state.}
Let $g(i)$ denote the worker to which request $i$ is assigned. For each
worker $g$ and step $k$, define the active set $\mathcal{A}_g(k) \;=\; \bigl\{\, i\in\mathcal{I}\;:\;
g(i) = g,\ x_i \leq k < x_i + o_i \,\bigr\},$ i.e., the requests currently being processed on worker $g$ at step $k$.
The instantaneous load on worker $g$ at step $k$ is $L_g(k) \;=\; \sum_{i\in\mathcal{A}_g(k)} w_i^{(k - x_i + 1)},$ 
the sum of step-$(k-x_i+1)$ workloads of all requests currently active
on $g$. Let $R_{\text{wait}}(k)$ denote the
\emph{waiting set} of requests that have completed prefill by step $k$
but have not yet been assigned to a decode worker.

\noindent \textbf{Routing policy.}
A non-anticipative routing policy $\pi$ specifies, at each step $k$, an
assignment $\{S_g(k)\}_{g\in[G]}$ where $S_g(k)\subseteq R_{\text{wait}}(k)$
is the (possibly empty) set of waiting requests dispatched to worker $g$
at step $k$, subject to the capacity constraint $|\mathcal{A}_g(k)| + |S_g(k)| \;\leq\; B$, $\text{for all } g\in[G],$ and to the disjointness constraint $S_g(k)\cap S_{g'}(k) = \emptyset$
for $g\neq g'$. The decision at step $k$ may depend only on observable
state up to step $k$; in particular, it cannot depend on the latent
profiles $\{W_i\}$.

%% file: BF0.tex
\section{BR-0: Prediction-Free Balance Routing}
\label{sec:br0}

We begin with the first router, \emph{BalanceRoute-0} (BR-0), where no decode length prediction information is needed. At step $k$, the scheduler observes only the
current state $\{\mathcal{A}_g(k), L_g(k)\}_{g\in[G]}$ and the waiting
set $R_{\text{wait}}(k)$. For request $i$, write $s_i := w_i^{(1)}$ for its load at admission (the
prefill length). When the router
admits a subset $Q \subseteq R_{\text{wait}}(k)$ to worker $g$ at step
$k$, the immediate increment to $L_g(k)$ is
\(
\Delta_s(Q) := \sum_{i \in Q} s_i.
\)

\subsection{Marginal current-step imbalance}
\label{subsec:br0-marginal}

We define the total imbalance at decode step $k$ as $I(k) \;=\; G \cdot M(k) \;-\; \sum_{g \in [G]} L_g(k)$, where $M(k) := \max_{g \in [G]} L_g(k)$, i.e., the work that lighter workers would have done if loaded as heavily
as the heaviest. Routing decisions at step $k$ change $\{L_g(k)\}$ for
admitted requests but not the active sets of already-running ones; the
policy's lever on $I(k)$ is therefore the choice of admissions
$\{S_g(k)\}_{g \in [G]}$.

For a candidate subset $Q \subseteq R_{\text{wait}}(k)$ and worker $g$,
let $m_g \;:=\; M(k) - L_g(k) \;\geq\; 0$
denote the \emph{safe margin}: the additional load $g$ can absorb before
overtaking the current heaviest worker. Admitting $Q$ to worker $g$ changes
$L_g(k)$ to $L_g(k) + \Delta_s(Q)$ and $M(k)$ to
$\max\bigl(M(k),\, L_g(k) + \Delta_s(Q)\bigr)$, so the change in $I(k)$ is $\Delta I_g(Q) \;=\; G \cdot \bigl(\Delta_s(Q) - m_g\bigr)_+ \;-\; \Delta_s(Q).$ 
We define a per-worker metric, \emph{F-score}, which measures the
imbalance reduction induced by admitting $Q$ to $g$:
\begin{equation}
\label{eq:br0-Fscore}
F_g(Q) \;:=\; -\Delta I_g(Q)
\;=\; \Delta_s(Q) \;-\; G \cdot \bigl(\Delta_s(Q) - m_g\bigr)_+.
\end{equation}
Equation~\eqref{eq:br0-Fscore} is piecewise linear in $\Delta_s(Q)$ with
two regimes:
\begin{itemize}
\item \textbf{Safe} ($\Delta_s(Q) \leq m_g$). The score is purely linear,
    $F_g(Q) = \Delta_s(Q)$; placing more load on $g$ strictly
    reduces $I(k)$ as long as $g$ does not overtake the current
    heaviest worker.
\item \textbf{Overflow} ($\Delta_s(Q) > m_g$). The score becomes
    $F_g(Q) = G\, m_g - (G-1)\,\Delta_s(Q)$; once $g$ overtakes
    the previous heaviest worker, every additional unit of load on $g$
    raises $M(k)$ and \emph{increases} imbalance by a factor of $G - 1$.
\end{itemize}
The crossover at $\Delta_s(Q) = m_g$ is sharp: pushing slightly above
$m_g$ flips a $1$-per-unit benefit into a $(G{-}1)$-per-unit cost. At
large scale, this asymmetry is severe, and a useful one-line summary of BR-0 is: \emph{fill safe margins
preferentially, and when overflow is unavoidable, route to the worker
where it costs least.}

\subsection{Two-stage algorithm}
\label{subsec:br0-algo}
BR-0 splits the per-step decision
into two stages, controlled by a threshold $S_{\mathrm{greedy}}$ on the
total free capacity $S_{\mathrm{tot}} := \sum_g \texttt{cap}[g]$, where
$\texttt{cap}[g] := B - |\mathcal{A}_g(k)|$.

\noindent \textbf{Stage 1: greedy fill (abundant capacity).}
While $S_{\mathrm{tot}} > S_{\mathrm{greedy}}$ and $R_{\text{wait}}(k)$
is nonempty, BR-0 selects the worker $g$ with the largest free capacity and admits the single request
$i \in R_{\text{wait}}(k)$ that maximizes $F_g(\{i\})$. Because
$g$ has the largest spare capacity, it is in the safe
regime: $F_g(\{i\}) = s_i$ and we send the largest waiting request to the lightest-loaded worker with
the most spare slots.  

\noindent \textbf{Stage 2: refined allocation (scarce capacity).}
Once $S_{\mathrm{tot}} \leq S_{\mathrm{greedy}}$, the few remaining
slots have outsized leverage on $I(k)$, and crossing the safe-regime
boundary becomes likely. BR-0 places workers with $\texttt{cap}[g] > 0$
into a priority queue keyed by $(\texttt{cap}[g], m_g)$, favoring
workers with both larger free capacity and larger safe margin. For each
worker $g$ popped from the queue, BR-0 considers the head $R_{\max}$
candidates of $R_{\text{wait}}(k)$ (sorted by $s_i$) and selects the
subset $Q^\star$ with $|Q^\star| \leq \min(\texttt{cap}[g], R_{\max})$
maximizing $F_g(Q)$. If the best
subset has nonpositive score, BR-0 admits the single best request
anyway to avoid starvation; the worker is reinserted into the queue if
capacity remains. The full procedure is given in Algorithm~\ref{alg:br0} in Appendix \ref{append:alg}.

\section{BR-H: Lookahead-Aware Balance Routing}
\label{sec:brh}
 
In practice, predictions of each active request's remaining decode length $r_i(k) := o_i - a_i(k)$ are often available at routing time~\citep{zheng2023response,qiu2024efficient,fu2024efficient,shahout2024don,chen2025adaptively}, where $a_i(k) := k - x_i$. A full prediction is, however, neither necessary nor accurate: \citet{chen2026universal} show that a constant short lookahead $H$ already controls $\sum_{k} I(k)$, and beyond the binary threshold ``terminates within $H$ steps,'' the precise heavy-tailed value of $r_i(k)$ has only second-order effect on routing---so a binary classifier captures the routing-relevant signal more reliably than a full-length regressor. Even given exact $r_i(k)$, longer horizons $H' \gg H$ depend on unobservable future arrivals, whose noise grows faster than the signal. We therefore extend BR-0 with a constant $H$-step lookahead, yielding BR-H. 

For a fix horizon $H$, we develop a \emph{termination classifier} and \emph{conditional-mean regressor} for predicting the workload in the future $H$ decode steps. The detail of the short-horizon prediction interface can be found in Appendix \ref{subsec:brh-prediction}.

\subsection{Horizon F-score}
\label{subsec:brh-fscore}
 
The F-score in BR-0~\eqref{eq:br0-Fscore} measures the marginal effect of admitting $Q$ to $g$ on a single step. We extend it to the horizon by summing per-step marginal effects over $h = 0, \dots, H$, with two modifications: (i) a discount vector $\mathbf{d} := (1, \gamma, \dots, \gamma^H)^\top$ with $\gamma \in (0, 1]$ down-weights far-future imbalance, reflecting that near-term predictions are more reliable; and (ii) the binary penalty coefficient $G$ in~\eqref{eq:br0-Fscore} is replaced by a tunable coefficient $\beta$, allowing empirical calibration of the overflow penalty under prediction noise. A scaling constant $\alpha > 0$ on the reward term plays the symmetric role for the safe regime.
 
Approximating each request in $Q$ as contributing a constant $s_i$ of additional load over the horizon, the per-step margin consumed by $Q$ at step $k+h$ is $(\Delta_s(Q) - m_{g,h})_+$, while the per-step reduction in total load deficit is $\Delta_s(Q)$ whenever $g$ remains under the envelope. Aggregating with $\mathbf{d}$ gives the horizon F-score
\begin{equation}
\label{eq:brh-Fscore}
F_g(Q) \;=\; \alpha \, (\mathbf{1}^\top \mathbf{d}) \, \Delta_s(Q)
\;-\; \beta \, \bigl(\Delta_s(Q)\, \mathbf{1} - \mathbf{m}_g \bigr)_+^\top \mathbf{d},
\end{equation}
with tunable parameters $(\alpha, \beta, \gamma)$. As in BR-0, $F_g(Q)$ is piecewise linear in $\Delta_s(Q)$ with two regimes: \emph{horizon-safe} ($\Delta_s(Q) \leq \min_h m_{g,h}$), where the penalty term vanishes and $F_g(Q)$ is linear in $\Delta_s(Q)$; and \emph{overflow} ($\Delta_s(Q) > \min_h m_{g,h}$), where the penalty activates step by step as the safe envelope is exceeded. A smaller $\gamma$ trusts predictions less and pulls BR-H closer to BR-0's myopic behavior.
 
\noindent \textbf{Reduction to BR-0.}
Substituting $H = 0$ into~\eqref{eq:brh-Fscore} yields $\mathbf{d} = (1)$, $\mathbf{m}_g = (M(k) - L_g(k))$, 
which coincides with the BR-0 score~\eqref{eq:br0-Fscore} when $(\alpha, \beta) = (1, G)$.
 
\subsection{Two-stage algorithm}
\label{subsec:brh-algorithm}
 
BR-H follows the same two-stage structure as BR-0, with three modifications to use horizon information: (i) scoring uses the horizon F-score~\eqref{eq:brh-Fscore} instead of~\eqref{eq:br0-Fscore}, (ii) the priority-queue key in Stage 2 uses the horizon-min margin $\min_h m_{g,h}$ rather than the single-step margin, and (iii) once per scheduling round, BR-H computes the horizon projection $\{L_g(k+h)\}_{h=0}^{H}$, $\mathbf{M}$, and $\{\mathbf{m}_g\}$ from the predicted profiles, with subsequent admissions updating these incrementally. The full procedure is given in Algorithm~\ref{alg:brh} in Appendix \ref{append:alg}.

\begin{algorithm2e}[htbp]
\caption{\textbf{BalanceRoute-H (BR-H) (Intuition).} Full algorithm with predictor refresh, tie-breaking, starvation guard, and state-update bookkeeping in Algorithm~\ref{alg:brh} in Appendix \ref{append:alg}. }
\label{alg:brh-sketch}
\KwIn{Waiting pool $R_{\text{wait}}$, active sets $\{\mathcal{A}_g\}$, horizon $H$, F-score $F_g$~\eqref{eq:brh-Fscore}}
\KwOut{Per-worker admissions at step $k$}
Compute horizon-projected margins $\{\mathbf{m}_g\}$ from cached predictions $\{\widehat{c}_i\}$\;
\textbf{Stage 1 --- abundant capacity:} while total free slots $> S_{\mathrm{greedy}}$, pick the worker $g$ with the most free slots and admit the single request $i^\star = \arg\max_{i \in R_{\text{wait}}} F_g(\{i\})$\;
\textbf{Stage 2 --- scarce capacity:} for each remaining worker, in priority order $(\texttt{cap}_g, \min_h m_{g,h})$, admit subset $Q^\star \subseteq R_{\text{wait}}$ with $|Q^\star| \le R_{\max}$ maximizing $F_g(Q)$\;
\end{algorithm2e}

%% file: system.tex
\section{System Implementation}
\label{sec:system}

We deploy BR-H on \texttt{vllm-ascend}~\citep{vllm-ascend}, the Ascend-NPU port of vLLM~\citep{kwon2023efficient}, in a prefill--decode (PD) disaggregated configuration, validated on Qwen3-30B-A3B (MoE) on Atlas~910B and DeepSeek-V3 W8A8 on Atlas~910C (full topologies in Section~\ref{sec:experiments}). The BR-H rule of Section~\ref{sec:brh} runs in milliseconds once the cluster snapshot
\begin{equation}
\label{eq:cluster-snapshot}
\Bigl( \{\mathcal{A}_g(k), L_g(k)\}_{g\in[G]},\;\; R_{\mathrm{wait}}(k),\;\; \{\widehat{c}_i(k)\}_{i \in \cup_g \mathcal{A}_g(k)} \Bigr)
\end{equation}
is available at every decode step; the engineering task is to maintain it under non-preemptive cross-tier KV transfer, no engine introspection, and a sub-100\,ms decision window.

\begin{figure}[htbp]\centering
\includegraphics[width=0.95\linewidth]{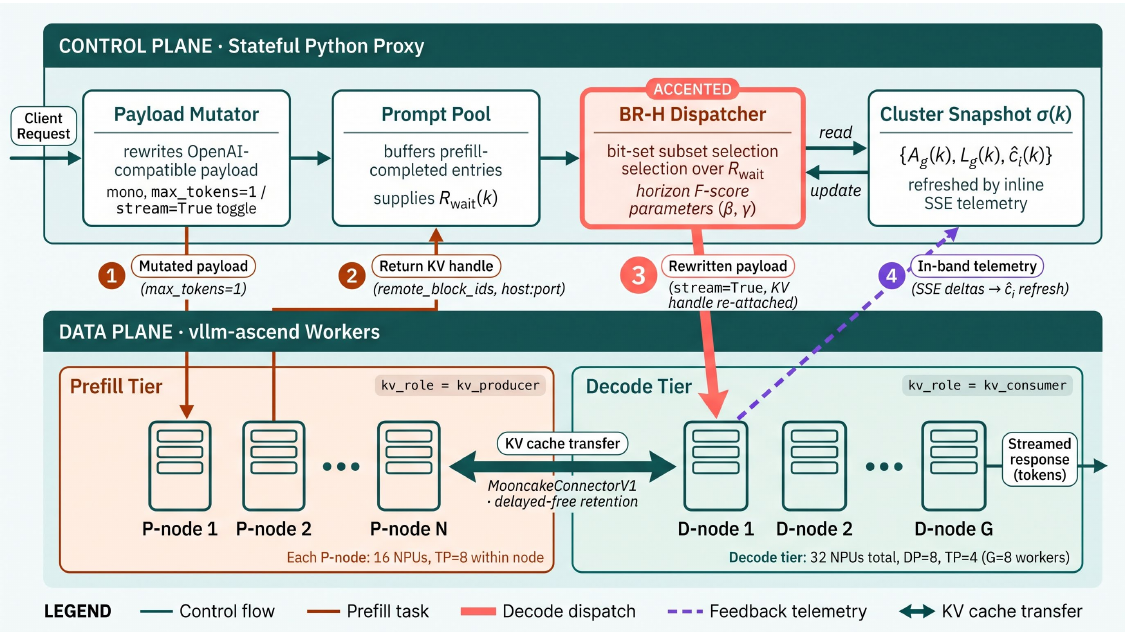}
\caption{\textbf{System architecture.} A stateful proxy maintains the cluster snapshot and runs the BR-H rule. It drives the prefill tier (\texttt{kv\_role}{=}\texttt{kv\_producer}) via payload mutation, buffers completed requests in a \emph{Prompt Pool}, and then dispatches them in batches to the decode tier (\texttt{kv\_role}{=}\texttt{kv\_consumer}). Load projections are rectified by monitoring per-token SSE telemetry from the decoders.}
\label{fig:system_arch}
\end{figure}

\subsection{Architecture and design principles}
\label{subsec:system-arch}

Three principles shape the design. \emph{Centralized state}: the horizon F-score~\eqref{eq:brh-Fscore} requires a cross-worker view of projected loads and Stage~2 selects subsets from a pool, so a stateless router cannot serve~\eqref{eq:cluster-snapshot}; we therefore place a single stateful Python proxy in front of unmodified \texttt{vllm-ascend} workers. \emph{Above-the-engine integration}: we drive the prefill--decode lifecycle through payload mutation on the standard OpenAI-compatible API and recover per-token decode progress by inline parsing of Server-Sent Events (SSE) streams, requiring no engine-binary modification. \emph{Pool decoupling}: prefill-completed requests buffer in a \emph{prompt pool} until the decode dispatcher wakes with a global view, enabling Stage~2 subset selection at the cost of late-binding KV transfer---bounded empirically, since deployed BR-H still beats every baseline on TPOT-P95 (Table~\ref{tab:main_combined}).

The system has three tiers (Figure~\ref{fig:system_arch}): a prefill tier and a decode tier of \texttt{vllm-ascend} instances, with the stateful proxy (Python/\texttt{uvloop}) between client and engines; cross-tier KV transfer uses \texttt{MooncakeConnector}~\citep{mooncake}. The proxy maintains snapshot~\eqref{eq:cluster-snapshot} in process memory through per-worker \texttt{DecodeInstanceState}$[g]$ mirroring the executor's active batch (with $L_g(k)$ derived in $O(1)$), a \texttt{PromptPool} supplying $R_{\mathrm{wait}}(k)$, and a \texttt{PrefillerBudget}$[i]$ caching KV-utilization samples from each prefiller's Prometheus endpoint as back-pressure. Event-driven coroutines update the structures of prefill dispatch, decode dispatch, telemetry polling, and pool reaping; the payload-mutation contract, retry/cancellation logic, a latency-optimized BR-0 pool-bypass path, and how to solve the BR-H rule in milliseconds are detailed in Appendix~\ref{app:engineering}.

%% file: Experiment.tex
\section{Experiments}
\label{sec:experiments}
\subsection{Experimental setup}
\label{subsec:exp_setup}

\noindent \textbf{Serving stack.}
All experiments use the deployment of Section~\ref{sec:system}: \texttt{vllm-ascend}~\citep{vllm-ascend} with V1 scheduler, continuous batching, PagedAttention~\citep{kwon2023efficient}, and \texttt{cudagraph\_mode}{=}\texttt{FULL\_DECODE\_ONLY} on the decode tier; \texttt{MooncakeConnector}~\citep{mooncake} handles cross-tier KV transfer. Our 144-NPU cluster runs DeepSeek-V3~671B (W8A8) on Atlas~910C in a 4P1D topology by default: four prefill nodes ($16$ NPUs each, TP$=8$) and one decode node ($32$ NPUs, DP$=8$, TP$=4$), giving $G{=}8$ decode workers; batch limit $B = \texttt{max\_num\_seqs}$.

\noindent \textbf{Datasets.}
We evaluate on two prefill--decode traces with markedly different output-length distributions, both run at heavy load (cluster near saturation). \textbf{Proprietary Data} is a proprietary production trace with heavy-tailed outputs ($8{,}000$ requests, mean prompt $3{,}197$, mean output $1{,}185$). \textbf{Azure-2024}~\citep{azure2024llmtrace,stojkovic2025dynamollm} is the conversation split of the public Azure LLM Inference Trace, filtered to output length $> 1000$ tokens---the regime in which decode-stage imbalance dominates and outputs are cap-bounded ($10{,}000$ requests, mean prompt $4{,}652$, mean output $1{,}052$).

\noindent \textbf{Baselines.}
We compare against four load-balancing policies from the vLLM router~\citep{vllm-router}: \textbf{Random}, \textbf{Round-Robin (RR)}, \textbf{Power-of-Two-Choices (P2C)}~\citep{mitzenmacher2001power}, and \textbf{Join-Shortest-Queue (JSQ)}, the upstream \texttt{vllm-ascend} default. Workload-aware endpoint routers~\citep{jain2024intelligent,jain2025performance} target a structurally different setting (request-to-instance assignment without cross-replica barrier synchronization) and are not directly comparable; see Appendix~\ref{append:related}.

\noindent \textbf{Our methods.}
We deploy two routers: \textbf{BR-0}, the prediction-free router of Section~\ref{sec:br0}, and \textbf{BR-H} ($H{=}80$) of Section~\ref{sec:brh}, in three configurations. \emph{BR-H oracle} replaces $\widehat{c}_i(k)$ with the trace ground-truth $\min(r_i(k), H)$ as a reference for perfect lookahead. \emph{BR-H deployed} uses the two-stage estimator of Section~\ref{subsec:brh-prediction} with one of two realizations: \textbf{Survival} reads both stages off the marginal training-output CDF, and \textbf{ExactMatch} maintains a prompt-hash-keyed CDF and falls back to Survival on key miss. Offline predictor accuracy on both workloads is in Appendix~\ref{subsec:exp_predictor}.

\noindent \textbf{Metrics.}
\textbf{Average imbalance} is the trace mean of per-step max\,$-$\,min KV workload across the $G$ decode workers, the step-level proxy for barrier-induced idle. \textbf{TPOT P95} is the 95th-percentile time-per-output-token (ms), capturing decode-stage tail latency. \textbf{Throughput} is output tokens per second. We report decode-stage metrics only, since prefill is unaffected by decode-tier routing.

\subsection{Main results on Proprietary Data and Azure-2024}
\label{subsec:exp_main}

\begin{figure}[htbp]
\centering
\begin{subfigure}[t]{0.48\linewidth}
    \centering
    \includegraphics[width=\linewidth]{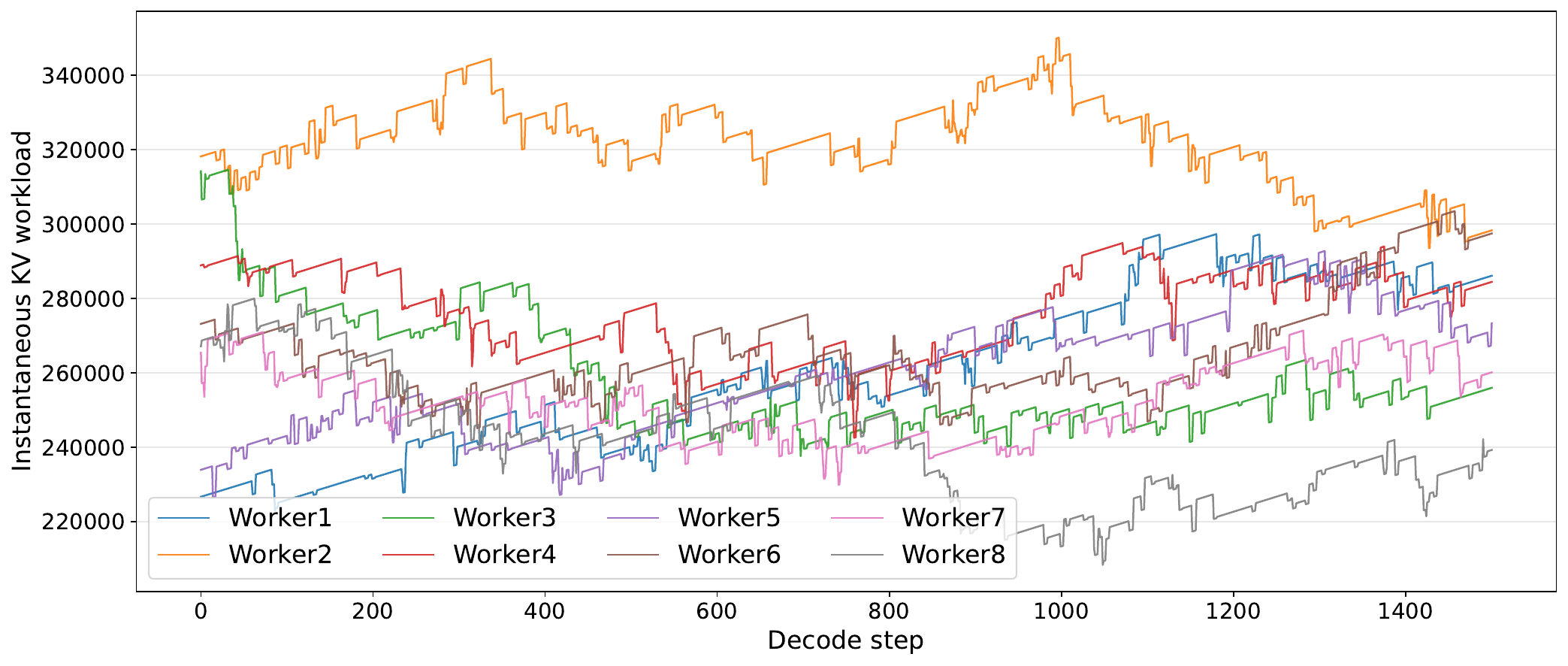}
    \caption*{\centering\footnotesize Random}
\end{subfigure}\hfill
\begin{subfigure}[t]{0.48\linewidth}
    \centering
    \includegraphics[width=\linewidth]{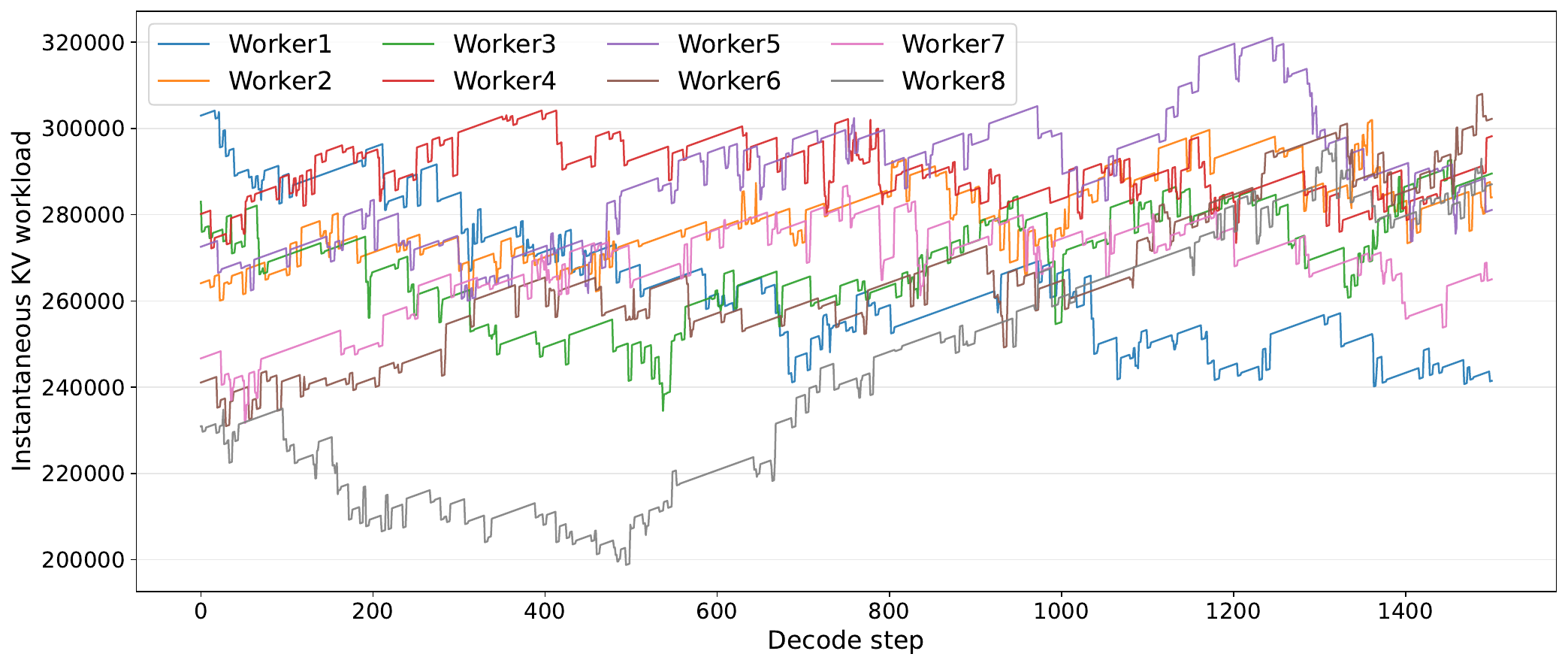}
    \caption*{\centering\footnotesize JSQ }
\end{subfigure}

\vspace{0.7em}
\begin{subfigure}[t]{0.48\linewidth}
    \centering
    \includegraphics[width=\linewidth]{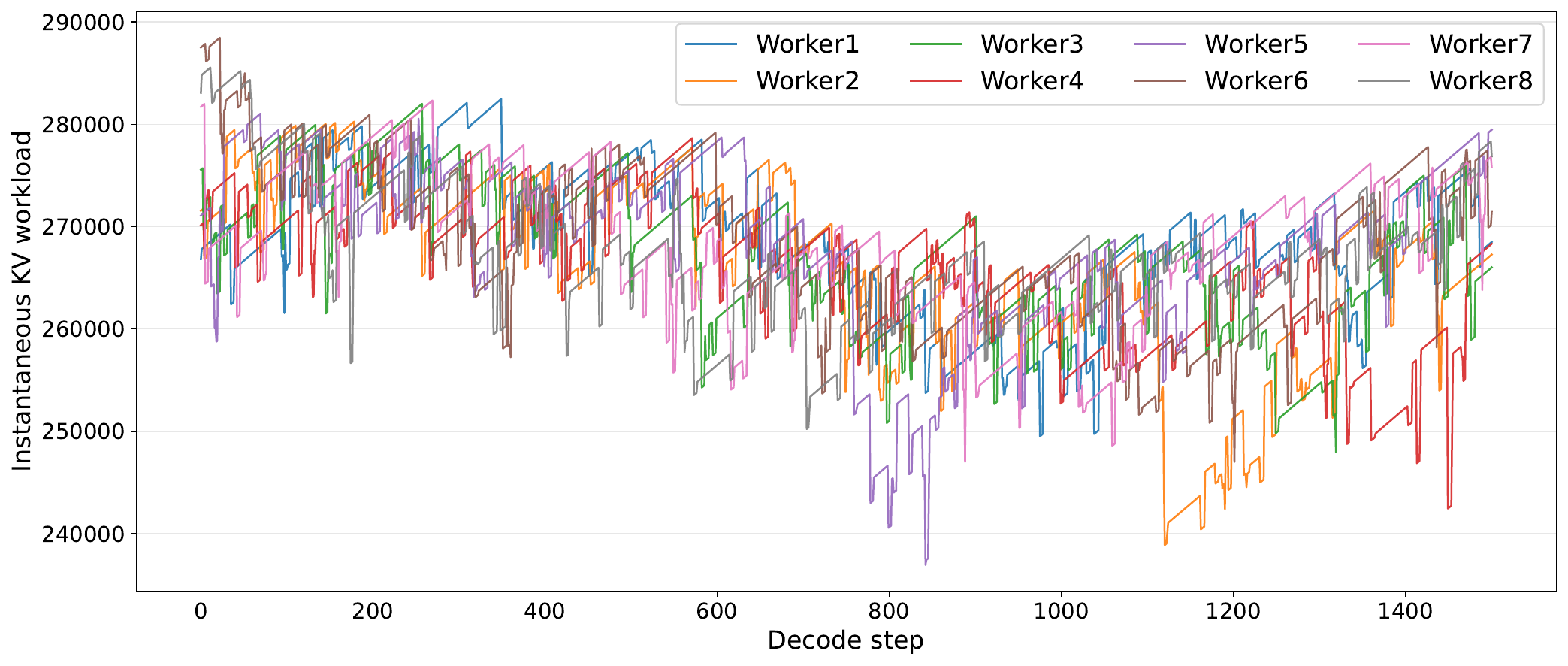}
    \caption*{\centering\footnotesize \textbf{BR-0}}
\end{subfigure}\hfill
\begin{subfigure}[t]{0.48\linewidth}
    \centering
    \includegraphics[width=\linewidth]{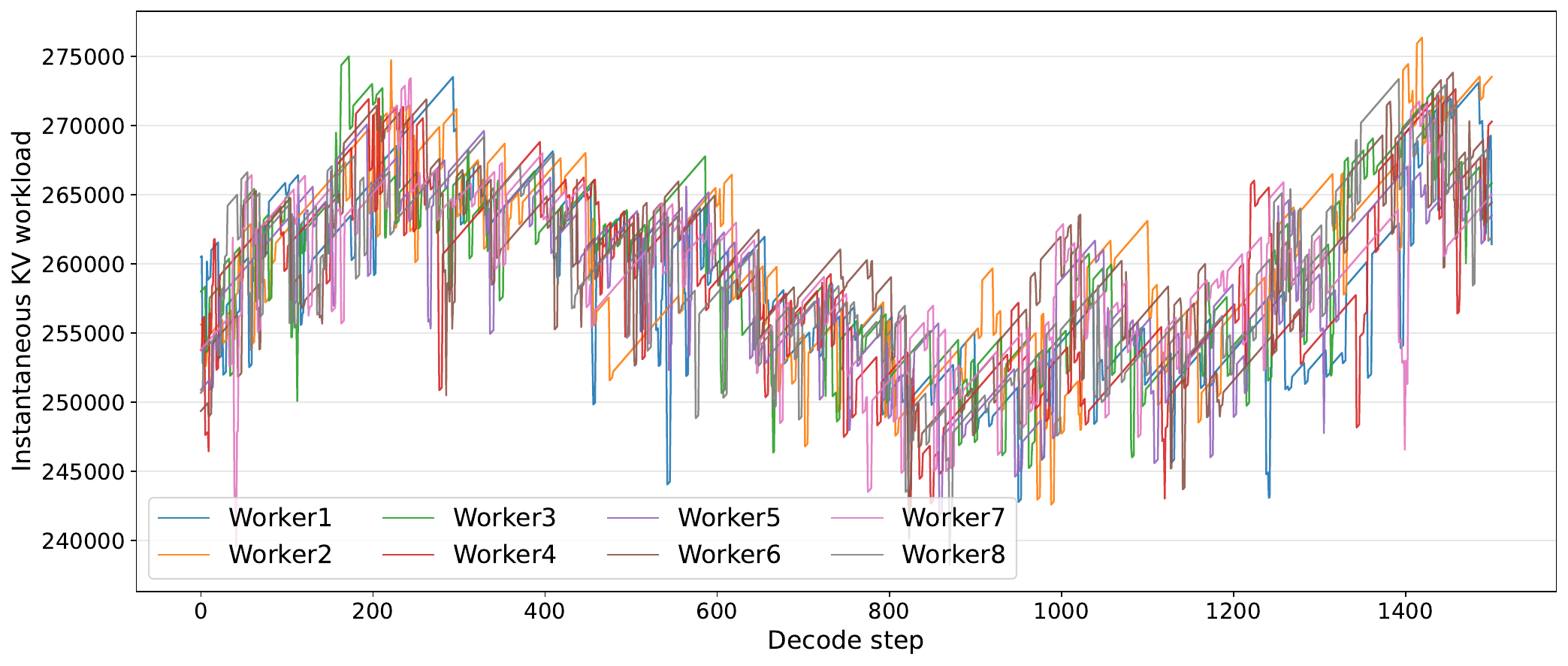}
    \caption*{\centering\footnotesize \textbf{BR-H} predicted, ExactMatch}
\end{subfigure}

\caption{\textbf{Per-worker KV-cache workload on Proprietary Data, $G{=}8$.} Each panel plots all eight workers over the same $1{,}500$-step decode segment; vertical spread is the instantaneous imbalance. The remaining panels (RR, P2C, BR-H oracle) are in Appendix~\ref{app:full-kv-grids}.}
\label{fig:kv-workload-grid}
\end{figure}

\begin{table}[htbp]
\centering
\caption{\textbf{Main results on Proprietary Data and Azure-2024.} $G{=}8$, heavy load (Azure-2024 filtered to output length $>1000$). Imbalance in tokens, TPOT P95 in ms, throughput (Tput) in tok/s; $\downarrow$ lower is better, $\uparrow$ higher is better. Primary deployment in \textbf{bold}.}
\label{tab:main_combined}
\begin{tabular}{lcccccc}
\toprule
& \multicolumn{3}{c}{\textbf{Proprietary Data}} & \multicolumn{3}{c}{\textbf{Azure-2024}} \\
\cmidrule(lr){2-4} \cmidrule(lr){5-7}
Method & Imbal $\downarrow$ & TPOT $\downarrow$ & Tput $\uparrow$ & Imbal $\downarrow$ & TPOT $\downarrow$ & Tput $\uparrow$ \\
\midrule
Random                & 438{,}225 & 85.2 & 811     & 146{,}564 & 90.3 & 706 \\
Round-Robin           & 254{,}798 & 83.2 & 847     & 110{,}049 & 89.1 & 771 \\
Power-of-Two-Choices  & 262{,}940 & 84.0 & 836     & 109{,}033 & 89.3 & 794 \\
Join-Shortest-Queue   & 215{,}110 & 83.2 & 843     & 104{,}737 & 89.3 & 785 \\
\midrule
BR-0 \emph{prediction-free}                & 51{,}927          & 79.3          & 943             & 54{,}051          & 83.3          & 854 \\
BR-H \emph{oracle}, $(43, 0.86)$           & 23{,}975          & 78.0          & 1{,}022         & 37{,}831          & 78.1          & 951 \\
BR-H \emph{oracle}, $(14.67, 0.64)$        & 23{,}577          & 78.0          & 1{,}029         & 35{,}301          & 77.3          & 957 \\
BR-H \emph{deployed}, Survival             & 40{,}757          & 78.7          & 979             & 44{,}010          & 82.1          & 893 \\
\textbf{BR-H} \emph{deployed}, ExactMatch  & \textbf{36{,}277} & \textbf{78.8} & \textbf{977}    & \textbf{38{,}496} & \textbf{82.3} & \textbf{909} \\
\bottomrule
\end{tabular}
\end{table}

Table~\ref{tab:main_combined} reports the heavy-load comparison at $G{=}8$. Figure~\ref{fig:kv-workload-grid} visualizes the load imbalance between different algorithms for the Proprietary Data trace, and Figure~\ref{fig:kv-workload-azure-grid-paper} in Appendix \ref{app:full-kv-grids} visualizes the one for the Azure-2024 trace. The qualitative ordering is identical across datasets: every BR row dominates every baseline row on every metric. BR-0 alone reduces imbalance by $4.1\times$ on Proprietary Data and $1.94\times$ on Azure-2024 over the strongest baseline (JSQ), lifting throughput by $11.8\%$ and $8.8\%$ respectively. Oracle BR-H tightens imbalance by a further $\sim\!2\times$ on Proprietary Data and $\sim\!1.5\times$ on Azure, reaching $\sim\!1{,}029$ and $957$\,tok/s; the gap between BR-0 and oracle BR-H quantifies the structural reward available to lookahead under perfect prediction (Section~\ref{subsec:brh-fscore}). The deployed ExactMatch predictor recovers most of this lookahead gain end-to-end ($977$ and $909$\,tok/s, $+15.4\%$ and $+14.5\%$ over the strongest baseline), and Survival delivers comparable throughput on both traces. ExactMatch is the primary deployment for workloads with prompt-level recurrence (Appendix~\ref{subsec:exp_predictor} characterizes it as aligned with this regime) while Survival is the principled fallback when recurrence is absent. Deployed BR-H also achieves the lowest TPOT P95 of any method on both datasets, confirming that throughput gains do not come at the cost of tail latency. Per-worker KV-workload visualizations and the trace-level Survival analysis are in Appendices~\ref{app:full-kv-grids} and~\ref{app:survival-30002}.

\subsection{Scaling with system size}
\label{subsec:exp_scaling}

We measure how each router's performance evolves with $G$ on three configurations: $G{=}4$ on 2P1D (48 NPUs), $G{=}8$ on 4P1D (96 NPUs, the main configuration), and $G{=}16$ on 5P1D (144 NPUs). Per-worker offered load is held constant by scaling request rate proportionally with $G$. To isolate router behaviour from predictor-quality effects, we run BR-H with oracle prediction at the operating point $(\beta, \gamma){=}(14.67, 0.64)$ used in the $G{=}8$ main results.

\begin{table}[htbp]
\centering
\caption{\textbf{Scaling summary on Proprietary Data.} $G{=}4$ (2P1D, 48 NPUs), $G{=}8$ (4P1D, 96 NPUs), $G{=}16$ (5P1D, 144 NPUs). BR-H runs with oracle prediction at $H{=}80$, $(\beta, \gamma){=}(14.67, 0.64)$. Each cell reports avg.\ imbalance / TPOT P95 / throughput. Best in each column in \textbf{bold}.}
\label{tab:scaling_summary}
\begin{tabular}{lccc}
\toprule
Method  & $G{=}4$ (2P1D) & $G{=}8$ (4P1D) & $G{=}16$ (5P1D) \\
\midrule
Random       &  80{,}118 / 66.30 /  364.7 & 438{,}225 / 85.15 /  810.5 & 638{,}062 / 77.80 /  923.6 \\
Round-Robin  &  68{,}108 / 65.27 /  365.7 & 254{,}798 / 83.16 /  846.7 & 616{,}089 / 77.91 /  925.3 \\
P2C          &  64{,}012 / 65.40 /  366.4 & 262{,}940 / 84.01 /  836.3 & 688{,}283 / 77.46 /  927.3 \\
JSQ          &  75{,}326 / 66.45 /  365.4 & 215{,}110 / 83.20 /  842.8 & 675{,}607 / 77.40 /  926.6 \\
\midrule
BR-0         &   9{,}711 / 63.19 /  407.1 &  51{,}927 / 79.32 /  942.5 & 215{,}170 / 74.92 / 1{,}117.8 \\
BR-H \emph{oracle}
             & \textbf{6{,}576} / \textbf{62.52} / \textbf{415.3}
             & \textbf{23{,}577} / \textbf{78.04} / \textbf{1{,}028.9}
             & \textbf{117{,}067} / \textbf{72.04} / \textbf{1{,}247.4} \\
\bottomrule
\end{tabular}
\end{table}

\begin{figure}[htbp]
\centering
\includegraphics[width=0.95\linewidth]{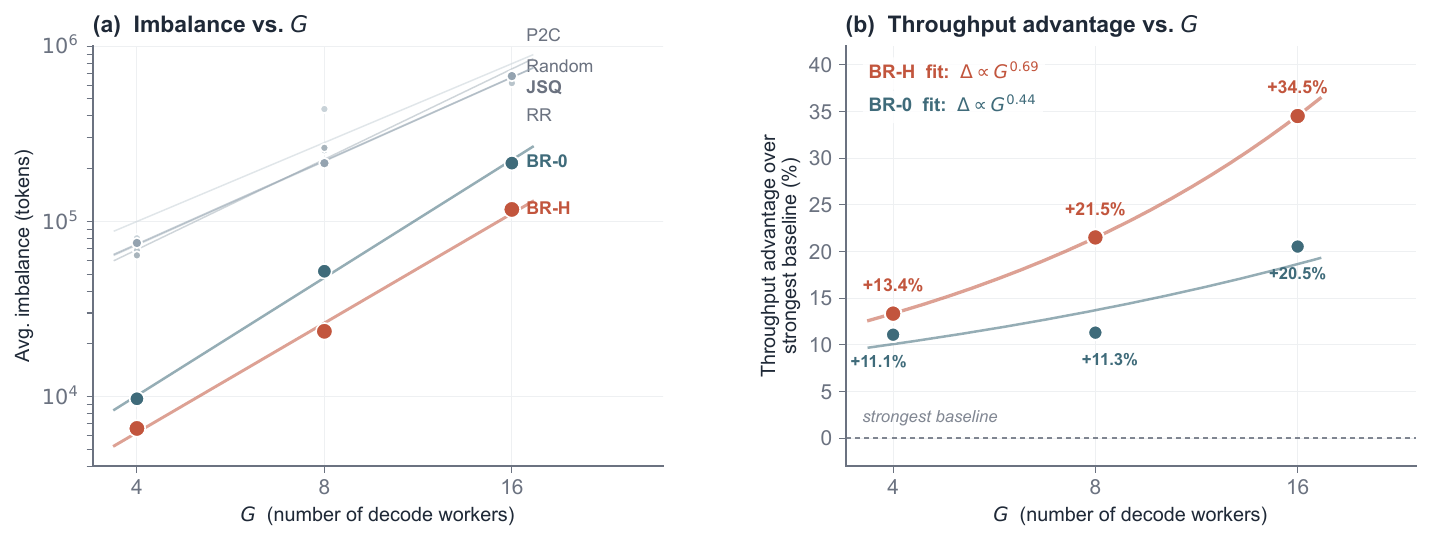}
\caption{\textbf{Scaling on Proprietary Data across $G \in \{4, 8, 16\}$.} }
\label{fig:scaling}
\end{figure}

Table~\ref{tab:scaling_summary} and Figure~\ref{fig:scaling} report the headline numbers. BR-H stays well below every baseline at every tested $G$: $6.6$k tokens at $G{=}4$ vs.\ $64$--$80$k for baselines, $24$k at $G{=}8$ vs.\ $215$--$438$k, and $117$k at $G{=}16$ vs.\ $616$--$688$k. All bands grow with $G$, consistent with the order-statistics argument of Section~\ref{subsec:dp_decode}. The throughput advantage over the strongest baseline rises with $G$: $+13.4\%$ at $G{=}4$, $+21.5\%$ at $G{=}8$, $+34.5\%$ at $G{=}16$. BR-0's advantage follows a similar but flatter trajectory ($+11\%, +11\%, +20.5\%$); the BR-H-over-BR-0 gap also widens with $G$ ($+2\%, +9\%, +12\%$). Absolute throughput at $G{=}16$ is sublinear because 5P1D adds only one prefill node for twice the decode workers, dropping the prefill-to-decode ratio from $4{:}1$ to $5{:}2$; per-worker advantages at fixed $G$ are unaffected. Per-worker traces underlying every cell are in Appendix~\ref{app:scaling}.


\subsection{Robustness and dispatch overhead}
\label{subsec:exp_robust_overhead}

\noindent \textbf{Sensitivity to $(\beta, \gamma)$.}
BR-H is robust across the explored parameter region. We sweep a cross-shape around $(\beta{=}48, \gamma{=}0.9)$ on Proprietary Data with $\beta \in \{1, 24, 48, 96\}$ at $\gamma{=}0.9$ and $\gamma \in \{0.5, 0.7, 0.9, 1.0\}$ at $\beta{=}48$ (seven configurations, oracle, $H{=}80$). Throughput across the sweep ranges $945$--$1{,}042$\,tok/s and TPOT P95 ranges $77.3$--$79.5$\,ms; every swept configuration improves on the strongest baseline by a wide margin. Full sweep tables (Table \ref{tab:sensitivity_sweep_g8_app},\ref{tab:sensitivity_sweep_g16_app}), Figure \ref{fig:betagamma} and per-worker traces are in Appendix~\ref{app:sensitivity-traces}.

\noindent \textbf{Dispatch overhead.}
Figure~\ref{fig:overhead} in Appendix \ref{subsec:exp_overhead} measures the per-tick cost of the BR-H dispatcher itself. We instrument the proxy on two deployments---DeepSeek-V3 671B and Qwen3-30B-A3B, both at $G{=}8$, $R_{\max}{=}4$. Total per-loop dispatch cost is $P_{50}{=}1.20$\,ms (mean $1.34$\,ms, $P_{99}{=}2.77$\,ms) on DeepSeek-V3 and $P_{50}{=}1.03$\,ms ($P_{99}{=}2.52$\,ms) on Qwen3, the two distributions agreeing within $17\%$ at $P_{50}$. Against the $\sim\!60$\,ms per-step engine budget, this places the dispatcher $\sim\!50\times$ below the engine step at the median and $\sim\!22\times$ below at $P_{99}$. The per-stage decomposition (Stage~2 dominates, $\sim\!69\%$ of total) is in Appendix~\ref{app:engineering}.

%% file: AppendRelated.tex
\section{Other Related Works} \label{append:related}

This appendix expands on related work that lies outside the immediate
scope of our routing problem but situates our contribution in the
broader LLM-serving literature. 
 
\subsection{LLM serving systems and architecture}
\label{app:related-systems}
 
Recent serving systems have rearchitected the inference stack to expose
structural levers for throughput and latency. vLLM introduces
PagedAttention to virtualize and compact the KV
cache~\citep{kwon2023efficient}, eliminating fragmentation and enabling
flexible large batches with near-zero KV waste; SGLang co-designs a
program/runtime interface with batching and
caching~\citep{zheng2024sglang} for multi-call LLM applications. At the
cluster level, prefill–decode (PD) disaggregation separates prompt
encoding from autoregressive decoding across worker
pools~\citep{zhong2024distserve, patel2023splitwise} so that TTFT and
TPOT can be optimized independently. These structural choices interact
with parallelism: decode commonly uses data parallelism (DP) over
requests, while model parallelism mixes tensor parallelism
(TP)~\citep{shoeybi2019megatron} for dense layers, expert parallelism
(EP)~\citep{fedus2022switch, liu2024deepseek} for MoE layers, and
pipeline parallelism for very large models. Together, these form the
substrate on which scheduling and load-balancing policies operate; our
work targets the policy layer that sits on top of this substrate.
 
\subsection{Single-worker scheduling and batching}
\label{app:related-scheduling}
 
Production engines adopt iteration-level scheduling with continuous
batching as the default control primitive. Orca introduces
iteration-level scheduling and selective batching to keep accelerators
busy under heterogeneous request lengths~\citep{yu2022orca}, and vLLM
co-designs PagedAttention with preemptive request scheduling at
iteration granularity~\citep{kwon2023efficient}. Building on these
primitives, several systems explore complementary schedulers:
Sarathi-Serve mitigates prefill–decode interference via chunked
prefill~\citep{agrawal2024taming}; FastServe adds token-granular
preemption via a multi-level feedback queue to reduce job completion
time~\citep{wu2023fast}. A parallel line uses learned or proxy signals
to approximate shortest-job-first scheduling, including proxy-model
predictors~\citep{qiu2024efficient} and embedding-based
estimators~\citep{shahout2024don}. These works optimize \emph{within} a
single worker—how to batch and order requests already routed to it—and
are complementary to our cross-worker routing problem.
 
\subsection{Cross-worker routing and load balancing}
\label{app:related-routing}
 
A growing body of work targets routing requests across multiple LLM
workers, with several lines of attack. \emph{Generic load-balancing
heuristics} dominate production deployments: the vLLM
router~\citep{vllm-router} and SGLang router~\citep{zheng2024sglang}
implement round-robin, random, power-of-two-choices, consistent hashing,
and prefix-cache-aware routing—policies that originate from classical
distributed-systems load balancing~\citep{mitzenmacher2001power,
karger1996better} and are agnostic to the LLM-specific structure
that our paper exploits. \emph{Workload-aware request routing} systems
treat LLM instances as independent endpoints and optimize end-to-end
latency by routing each arriving request to the most
suitable instance. The Intelligent Router of
\citet{jain2024intelligent} and the follow-up Performance-Aware LLM
Load Balancer~\citep{jain2025performance} exemplify this thread:
both train an output-length classifier (DistillBERT) to bucket request
decode lengths, derive an analytical workload-mixing penalty that
estimates within-instance prefill–decode interference, and use a
heuristic-guided reinforcement-learning agent to choose the target
instance. They report up to 11\% end-to-end latency reduction over
round-robin on small clusters (4–8 LLM instances). \emph{MoE expert
balancing} addresses imbalance \emph{within} a worker that uses expert
parallelism, redistributing tokens across redundant
experts~\citep{eplb2025, lplb2025} or via dynamic
prefetching; these
operate at a different granularity from cross-replica DP balancing.
 
The DP-decode setting we address differs from each of these threads
along an axis that none of them captures: \emph{barrier
synchronization}. Workload-aware request routers
\citep{jain2024intelligent, jain2025performance} treat each LLM
instance as an independent endpoint whose latency depends only on its
own active batch; the optimization target is end-to-end response
latency, and a slow instance harms only the requests it serves. In our
setting, each DP replica is parametric-sharded with TP/EP and tied to
all other replicas through a step-wise collective barrier, so a slow
replica delays \emph{every} replica at \emph{every} decode step. The
routing decision is a fleet-level imbalance-control problem rather
than an instance-selection problem, and the relevant time scale is the
sub-100\,ms decode step rather than the request lifetime. RL-based
routers with non-trivial inference overhead per decision are not
operationally viable at this granularity, and instance-selection
objectives such as ``best target for this request'' do not capture
imbalance-induced barrier idle. \citet{chen2026universal} are the first
to formalize the DP-decode load-balancing problem with explicit barrier
synchronization, and \citet{zhou2026positionllmservingneeds} highlight
it as a representative LLM-serving problem requiring principled
algorithm design beyond generic heuristics. Our work develops practical
schedulers in this setting with deployable per-step cost.
 
\subsection{Theoretical foundations for LLM scheduling}
\label{app:related-theory}
 
A recent theory stream formalizes LLM serving as online scheduling with
KV-cache constraints. \citet{jaillet2025online} develop a
hindsight-optimal integer-program benchmark, prove impossibility in the
fully online adversarial setting, and design near-optimal algorithms
under a semi-online model; \citet{chen2025adaptively}
and~\citet{wang2025llm} relax assumptions in this model, addressing
heterogeneous prefill, unknown decode lengths with interval
predictions, and variable prefill–decode lengths.
\citet{ao2025optimizing} introduce a fluid-guided framework with memory
constraints and derive Nested-WAIT policies that are near-optimal
against a fluid benchmark in heavy traffic. These works provide
principled scheduling \emph{within a single worker}; the multi-worker
setting under barrier synchronization—where our paper
operates—is the focus of~\citet{chen2026universal} and the present
work.

%% file: AppendAlg.tex
\section{Algorithm Pseudocode} \label{append:alg}

In this section, we provide algorithm pseudocode for BR-0 and BR-H respectively:

\begin{algorithm2e}[htbp]
\caption{BalanceRoute-0 (BR-0)}\label{alg:br0}
\KwIn{Waiting set $R_{\text{wait}}(k)$, active sets
      $\{\mathcal{A}_g(k)\}$, batch limit $B$, thresholds
      $S_{\mathrm{greedy}}, R_{\max}$}
\KwOut{Assignments $\{S_g(k)\}_{g \in [G]}$ at step $k$}
\BlankLine
$\texttt{cap}[g] \leftarrow B - |\mathcal{A}_g(k)|$ for all $g$;\quad
$S_{\mathrm{tot}} \leftarrow \sum_g \texttt{cap}[g]$\;
\lIf{$S_{\mathrm{tot}} = 0$ \textbf{or} $R_{\text{wait}}(k) = \emptyset$}{\textbf{return}}
$M(k) \leftarrow \max_g L_g(k)$;\quad
$m_g \leftarrow M(k) - L_g(k)$ for all $g$\;
\BlankLine
\textbf{Stage 1: Greedy fill}\;
\While{$S_{\mathrm{tot}} > S_{\mathrm{greedy}}$ \textbf{and} $R_{\text{wait}}(k) \neq \emptyset$}{
    $g \leftarrow \arg\max_{g'} \texttt{cap}[g']$ \tcp*{tie-break: smallest $L_{g'}(k)$}
    $i^\star \leftarrow \arg\max_{i \in R_{\text{wait}}(k)} F_g(\{i\})$\;
    $\mathcal{A}_g(k) \leftarrow \mathcal{A}_g(k) \cup \{i^\star\}$;\quad
    $R_{\text{wait}}(k) \leftarrow R_{\text{wait}}(k) \setminus \{i^\star\}$\;
    update $\texttt{cap}[g], S_{\mathrm{tot}}, L_g(k), M(k), \{m_{g'}\}$\;
}
\BlankLine
\textbf{Stage 2: Refined allocation}\;
Initialize priority queue $\mathcal{Q}$ over $\{g : \texttt{cap}[g] > 0\}$,
keyed by $(\texttt{cap}[g], m_g)$\;
\While{$\mathcal{Q} \neq \emptyset$ \textbf{and} $R_{\text{wait}}(k) \neq \emptyset$}{
    Pop $g$ from $\mathcal{Q}$;\quad $c_g \leftarrow \texttt{cap}[g]$\;
    Take up to $R_{\max}$ candidates from head of $R_{\text{wait}}(k)$
    (sorted by $s_i$)\;
    $Q^\star \leftarrow \arg\max_{|Q| \leq \min(c_g, R_{\max})} F_g(Q)$
    over candidate subsets\;
    \lIf{$F_g(Q^\star) \leq 0$}{
        $Q^\star \leftarrow \{\arg\max_i F_g(\{i\})\}$
        \tcp*{starvation guard}}
    $\mathcal{A}_g(k) \leftarrow \mathcal{A}_g(k) \cup Q^\star$;\quad
    $R_{\text{wait}}(k) \leftarrow R_{\text{wait}}(k) \setminus Q^\star$\;
    update $\texttt{cap}[g], L_g(k), M(k), \{m_{g'}\}$;\quad
    \lIf{$\texttt{cap}[g] > 0$}{reinsert $g$ into $\mathcal{Q}$}
}
\end{algorithm2e}

\begin{algorithm2e}[htbp]
\caption{BalanceRoute-H (BR-H)}\label{alg:brh}
\KwIn{Waiting set $R_{\text{wait}}(k)$, active sets
      $\{\mathcal{A}_g(k)\}$, batch limit $B$, horizon $H$, prediction period $\Delta T$ (default as $H/2$)
      thresholds $S_{\mathrm{greedy}}, R_{\max}$, score parameters
      $(\alpha, \beta, \gamma)$}
\KwOut{Assignments $\{S_g(k)\}_{g \in [G]}$ at step $k$}
\BlankLine
$\texttt{cap}[g] \leftarrow B - |\mathcal{A}_g(k)|$ for all $g$;\quad
$S_{\mathrm{tot}} \leftarrow \sum_g \texttt{cap}[g]$\;
\lIf{$S_{\mathrm{tot}} = 0$ \textbf{or} $R_{\text{wait}}(k) = \emptyset$}{\textbf{return}}
For each $i \in \bigcup_g \mathcal{A}_g(k)$, update the cached $\widehat{c}_i$ every $\Delta T$ decode steps:
recompute via~\eqref{eq:composite}; 
derive $\{L_g(k+h)\}_{h=0}^{H}$, envelope $\mathbf{M}$, margins $\{\mathbf{m}_g\}$\;
\BlankLine
\textbf{Stage 1: Greedy fill}\;
\While{$S_{\mathrm{tot}} > S_{\mathrm{greedy}}$ \textbf{and} $R_{\text{wait}}(k) \neq \emptyset$}{
    $g \leftarrow \arg\max_{g'} \texttt{cap}[g']$
    \tcp*{tie-break: smallest $L_{g'}(k)$}
    $i^\star \leftarrow \arg\max_{i \in R_{\text{wait}}(k)} F_g(\{i\})$\;
    $\mathcal{A}_g(k) \leftarrow \mathcal{A}_g(k) \cup \{i^\star\}$;\quad
    $R_{\text{wait}}(k) \leftarrow R_{\text{wait}}(k) \setminus \{i^\star\}$\;
    update $\texttt{cap}[g], S_{\mathrm{tot}}$, projected loads, $\{\mathbf{m}_{g'}\}$\;
}
\BlankLine
\textbf{Stage 2: Refined allocation}\;
Initialize priority queue $\mathcal{Q}$ over $\{g : \texttt{cap}[g] > 0\}$,
keyed by $(\texttt{cap}[g], \min_h m_{g,h})$\;
\While{$\mathcal{Q} \neq \emptyset$ \textbf{and} $R_{\text{wait}}(k) \neq \emptyset$}{
    Pop $g$ from $\mathcal{Q}$;\quad $c_g \leftarrow \texttt{cap}[g]$\;
    Take up to $R_{\max}$ candidates from head of $R_{\text{wait}}(k)$\;
    $Q^\star \leftarrow \arg\max_{|Q| \leq \min(c_g, R_{\max})} F_g(Q)$
    over candidate subsets\;
    \lIf{$F_g(Q^\star) \leq 0$}{
        $Q^\star \leftarrow \{\arg\max_i F_g(\{i\})\}$
        \tcp*{starvation guard}}
    $\mathcal{A}_g(k) \leftarrow \mathcal{A}_g(k) \cup Q^\star$;\quad
    $R_{\text{wait}}(k) \leftarrow R_{\text{wait}}(k) \setminus Q^\star$\;
    update $\texttt{cap}[g]$, projected loads, $\{\mathbf{m}_{g'}\}$;\quad
    \lIf{$\texttt{cap}[g] > 0$}{reinsert $g$ into $\mathcal{Q}$}
}
\end{algorithm2e}

%% file: Appendix.tex


\section{Prediction interface}
\subsection{Short-horizon prediction interface}
\label{subsec:brh-prediction}
 
Fix a horizon $H$. For each active request $i \in \bigcup_g \mathcal{A}_g(k)$ at step $k$, BR-H requires a single bounded scalar
\begin{equation}
\label{eq:horizon-contribution}
\widehat{c}_i(k) \;\approx\; \mathbb{E}\bigl[\,\min\bigl(r_i(k),\, H\bigr) \,\big|\, x_i, a_i(k)\bigr] \;\in\; [0, H],
\end{equation}
the \emph{expected in-window contribution} of $i$ to the next $H$ decode steps. The choice of estimand is dictated by what BR-H actually consumes: the horizon F-score (Section~\ref{subsec:brh-fscore}) reads only how much in-window work each request contributes, truncated at $H$ both because the discount $\gamma^h$ already attenuates beyond it and because longer horizons are dominated by unobservable future arrivals. Anything finer than $\widehat{c}_i(k)$ is not used; anything coarser loses signal at the binary threshold ``$r_i(k) \leq H$''.
 
The estimand admits a clean two-stage decomposition. Conditioning on the event $A_i := \{r_i(k) \leq H\}$ that request $i$ finishes inside the window, and using $\min(r_i(k), H) = r_i(k)$ on $A_i$ and $H$ on $A_i^c$,
\begin{equation}
\label{eq:two-stage}
\mathbb{E}\bigl[\min(r_i(k), H) \,\big|\, x_i, a_i(k)\bigr]
\;=\;
\bigl(1 - p_{\mathrm{fin}}^{(i)}\bigr)\cdot H
\;+\; p_{\mathrm{fin}}^{(i)} \cdot \mu_{\mathrm{rem}}^{(i)},
\end{equation}
where $p_{\mathrm{fin}}^{(i)} := \Pr(r_i(k) \leq H \mid x_i, a_i(k))$ and $\mu_{\mathrm{rem}}^{(i)} := \mathbb{E}[r_i(k) \mid x_i, a_i(k),\, r_i(k) \leq H] \in (0, H]$.
The decomposition gives BR-H two estimators with naturally aligned losses:
\begin{enumerate}
\item \emph{Termination classifier.} A calibrated estimate $\widehat{p}_{\mathrm{fin}}^{(i)}$ of $p_{\mathrm{fin}}^{(i)}$, trained with cross-entropy. Calibration matters because the composite reads $\widehat{p}_{\mathrm{fin}}^{(i)}$ as a probability rather than a hard label.
\item \emph{Conditional-mean regressor.} An estimate $\widehat{\mu}_{\mathrm{rem}}^{(i)}$ of $\mu_{\mathrm{rem}}^{(i)}$, trained with squared error on the finish-positive subsample. Squared error is the proper loss because the target is a conditional mean.
\end{enumerate}
The composite estimate is
\begin{equation}
\label{eq:composite}
\widehat{c}_i(k)
\;=\; \bigl(1 - \widehat{p}_{\mathrm{fin}}^{(i)}\bigr)\cdot H \;+\; \widehat{p}_{\mathrm{fin}}^{(i)}\cdot \widehat{\mu}_{\mathrm{rem}}^{(i)},
\end{equation}
clipped to $[0, H]$. When $\widehat{p}_{\mathrm{fin}}^{(i)}$ is near zero the composite collapses to the conservative anchor $H$; when near one it tracks $\widehat{\mu}_{\mathrm{rem}}^{(i)}$.
 
The interface is deliberately agnostic to how $\widehat{p}_{\mathrm{fin}}^{(i)}$ and $\widehat{\mu}_{\mathrm{rem}}^{(i)}$ are realized. Anything implementing the two-stage contract above plugs in: simple non-parametric estimators based on the empirical training-output CDF, per-prompt memorization, or learned classifier-and-regressor pairs are all valid instantiations, and the BR-H rule does not distinguish among them. We describe specific estimator families used in our deployment, together with their accuracy characterizations, in Section~\ref{sec:experiments}; full implementation details are deferred to Appendix~\ref{app:predictor}.
 
From $\widehat{c}_i(k)$ we obtain the projected per-step workload that the score consumes by aging the contribution down to zero across the window:
\begin{equation}
\label{eq:wprofile-new}
\widehat{w}_i^{(h)}(k) \;\approx\;
\begin{cases}
s_i + a_i(k) + h - 1, & h \leq \widehat{c}_i(k),\\
0, & h > \widehat{c}_i(k),
\end{cases}
\qquad h = 1, \dots, H,
\end{equation}
which preserves the convention that an active request contributes
$s_i + a_i(k) + h - 1$ at step $k+h$ and contributes nothing once
it has departed within the window. Predictions are refreshed every $\Delta T$ decode steps and decremented by 1 per step in between; we defer the choice of $\Delta T$ and the remaining details (confidence gating, floor handling) to Appendix~\ref{app:predictor}.
 
Aggregating per-request profiles over each worker's active set yields a horizon-projected load trajectory $\{L_g(k+h)\}_{h=0}^{H}$, an envelope $M_h := \max_{g'} L_{g'}(k+h)$, and a margin vector
\[
\mathbf{m}_g \;:=\; \bigl( (M_h - L_g(k+h))_+ \bigr)_{h=0}^{H} \;\in\; \mathbb{R}_{\geq 0}^{H+1}.
\]
All three are computed once at the start of each scheduling round and updated incrementally as admissions are made.

\subsection{Implementation of the two-stage prediction interface}
\label{app:predictor}
 
The two-stage interface specified in Section~\ref{subsec:brh-prediction} is a contract between Stage~1 and Stage~2: a calibrated probability $\widehat{p}_{\mathrm{fin}}^{(i)}$ paired with a conditional-mean estimate $\widehat{\mu}_{\mathrm{rem}}^{(i)} \in (0, H]$, combined into the composite $\widehat{c}_i(k)$ via~\eqref{eq:composite}. The contract is agnostic to how the two estimates are realized; multiple concrete realizations satisfy it, and the BR-H rule does not distinguish among them. This appendix describes the realizations evaluated in Section~\ref{sec:experiments}, the training protocol, the refresh logic that maintains $\widehat{c}_i(k)$ as a request progresses, and the latency budget at deployment.
 
\subsubsection{Estimator realizations}
\label{app:predictor-realizations}
 
We evaluate three realizations spanning the spectrum of training cost and feature engineering. All three implement the same Stage-1 / Stage-2 contract.
 
\paragraph{Empirical-survival.}
The empirical-survival estimator reads both stages directly off the empirical training-output CDF $\widehat{F}$. Stage~1 is
\[
\widehat{p}_{\mathrm{fin}}^{(i)} \;=\; \frac{\widehat{F}(a_i(k) + H) - \widehat{F}(a_i(k))}{1 - \widehat{F}(a_i(k))},
\]
the conditional probability that an output sampled from the training distribution would terminate within the next $H$ steps given that it has already produced $a_i(k)$ tokens. Stage~2 is the conditional mean
\[
\widehat{\mu}_{\mathrm{rem}}^{(i)} \;=\; \frac{1}{|\{j : a_i(k) < o_j \leq a_i(k) + H\}|} \sum_{a_i(k) < o_j \leq a_i(k) + H} (o_j - a_i(k))
\]
on the same training corpus. Both stages have $O(\log n)$ inference cost on a sorted training output history and require no per-request features. This is the production default when no auxiliary signal is available.
 
\paragraph{Per-prompt-class memorization.}
When the workload exhibits prompt repetition (multi-turn dialogue, cached prompts, structured templates), a per-prompt empirical CDF tightens both stages substantially over the marginal CDF. We maintain a key-indexed CDF---key derived from a hash of the prompt, or a quantile bucket of $s_i$---and apply the empirical-survival formulas above within the matching bucket; on key miss, the estimator falls back to the marginal empirical-survival baseline. Inference cost is $O(\log n)$ per call with no feature pipeline.
 
\paragraph{Learned classifier and regressor.}
For workloads with content-aware signal beyond the marginal output distribution, we train a learned classifier-and-regressor pair. Stage~1 is a binary classifier minimizing cross-entropy on the label $\mathbb{1}[r_i(k) \leq H]$. Stage~2 is a regressor minimizing squared error on the finish-positive subsample with target $r_i(k) \in (0, H]$. Both estimators consume a feature vector built from $s_i$, $a_i(k)$, exponentially-weighted moving averages of recent observed output lengths, and rolling-window statistics on past outputs and inputs; features are causal by construction. Concrete model classes are pluggable; in Section~\ref{sec:experiments} we use gradient-boosted trees with $200$ estimators of depth $3$, but the interface admits MLP or other replacements without affecting the rest of the pipeline.
 
\subsubsection{Training protocol}
\label{app:predictor-training}
 
For the empirical-survival and memorization variants, ``training'' reduces to sorting an output-length history and is $O(n \log n)$. For the learned variant, training samples are synthesized by walking each historical $(s_j, o_j)$ at age points $T = 0, \Delta T, 2\Delta T, \ldots$ up to $T < o_j$, emitting one $(\text{features},\, \mathbb{1}[o_j \leq T+H],\, o_j - T)$ tuple per age point; the classifier trains on all tuples and the regressor on the finish-positive subset. We use $\Delta T = H/2$, which controls training-set density without inflating cost.
 
For each evaluated dataset we hold out a time-disjoint segment for training and a separate segment for evaluation. Stage-1 metrics (ROC-AUC, Brier score) are computed on the binary task; Stage-2 metrics (conditional MAE) are computed on the finish-positive subsample. Distribution drift is handled by periodic re-fit; the cost of re-fit is dominated by the sort for non-parametric variants and by gradient boosting for the learned variant.
 
\subsubsection{Refresh logic}
\label{app:predictor-refresh}
 
The proxy maintains $\widehat{c}_i(k)$ along three rules.
 
\emph{Periodic refresh.} The proxy re-queries the predictor every $\Delta T$ generated tokens (default $\Delta T = H/2$), recomputing $\widehat{p}_{\mathrm{fin}}^{(i)}$ and $\widehat{\mu}_{\mathrm{rem}}^{(i)}$ at the current age and updating $\widehat{c}_i(k)$ via~\eqref{eq:composite}. Between refreshes the proxy decrements $\widehat{c}_i(k)$ by one per generated token, since each elapsed step is one less step of in-window contribution.
 
\emph{Stage-1 confidence gate.} On each refresh, the predictor's update is accepted only when $\widehat{p}_{\mathrm{fin}}^{(i)} \geq 0.5$; otherwise $\widehat{c}_i(k)$ is reset to the conservative anchor $H$. The gate is necessary because the regressor's output is well-defined only on the finish-positive subsample; on long-tail requests where $p_{\mathrm{fin}}^{(i)}$ is small, $\widehat{\mu}_{\mathrm{rem}}^{(i)}$ is effectively unconstrained and a refresh that uses it would inject a phantom near-departure event into the projection. Gating on $\widehat{p}_{\mathrm{fin}}^{(i)}$ rather than on the magnitude of $\widehat{c}_i(k)$ avoids this failure mode at the natural decision boundary of the two-stage decomposition.
 
\emph{Floor.} While the request remains active, $\widehat{c}_i(k)$ is bounded below by $1$. A floor of zero would let an over-confident initial prediction silently mark the request as already departed between scheduled refreshes, dropping it from the projection while it is still consuming work; the floor closes this gap. Crossing the floor between scheduled refreshes triggers an immediate refresh, so a request that was incorrectly predicted to finish soon is corrected at the moment its prediction would expire.

\subsubsection{Latency budget}
\label{app:predictor-latency}

Per-call inference cost is $\sim 1\,\mu\mathrm{s}$ for the empirical-survival and memorization variants (a binary search on a sorted array) and $\sim 1\,\mathrm{ms}$ for the learned variant (a gradient-boosted tree forward pass; feature extraction dominates). For the learned variant, refresh latency is amortized by batching all active-request refreshes that fall on a common $\Delta T$ boundary into a single forward pass; under the deployed dispatcher this places predictor cost an order of magnitude below the per-tick scheduling budget of Section~\ref{app:engineering-bitset}.

\subsection{Predictor evaluation}
\label{subsec:exp_predictor}

The two-stage interface of Section~\ref{subsec:brh-prediction} requires a calibrated probability $\widehat{p}_{\mathrm{fin}}^{(i)}$ of in-window finish and a conditional-mean estimate $\widehat{\mu}_{\mathrm{rem}}^{(i)} \in (0, H]$ on the finish-positive subsample, combined into $\widehat{c}_i(k)$ via~\eqref{eq:composite}. We deploy two concrete realizations of this interface that share the same admission features, refresh contract, and composite formula, and differ only in how each stage is fit. \textbf{Empirical-survival} reads both stages off the marginal training-output CDF in $O(\log n)$ per call with no per-request features. \textbf{Per-prompt memorization (ExactMatch)} maintains a prompt-hash-keyed empirical CDF and applies the same formulas within the matching bucket, falling back to the marginal CDF on key miss. ExactMatch is therefore a generalization of Empirical-survival: equivalent to it on every request whose prompt has not been seen in training and tighter only when prompt-level recurrence is present. Both realizations share the same proxy binary and are toggled by a configuration flag.

Table~\ref{tab:predictor_accuracy} reports the offline accuracy of both realizations on both workloads. The two metrics read together: Stage-1 AUC is inflated by the positive-class base rate when most active requests sit within $H$, while Stage-2 conditional MAE captures regressor quality on the finish-positive subsample. Outputs concentrating near a cap (Azure) make both stages tight regardless of realization. On Proprietary Data, Empirical-survival sits at the floor reachable from the marginal output distribution alone, while ExactMatch tightens both stages by exploiting per-prompt structure where it exists. The two rows on Proprietary Data bracket what is achievable: Empirical-survival is the lower bound that does not depend on prompt structure; ExactMatch exploits this information effectively.

Both realizations have sub-10ms per-call inference cost. Refreshes occur at the $\Delta T{=}H/2$ boundary on the SSE-parsing path inside the proxy (Section~\ref{subsec:system-telemetry}), so predictor work amortizes over chunks the proxy is already inspecting and stays well below the per-tick scheduling budget of Section~\ref{app:engineering-bitset}.

\begin{table}[h]
\centering
\caption{\textbf{Offline predictor accuracy.} Stage~1: ROC-AUC on $\{r_i(k) \le H\}$. Stage~2: conditional MAE in tokens on the finish-positive subsample.}
\label{tab:predictor_accuracy}
\begin{tabular}{llcc}
\toprule
Dataset & Realization & Stage-1 AUC $\uparrow$ & Stage-2 MAE $\downarrow$ \\
\midrule
Azure-2024 & Empirical-survival       & 0.993 &  5.4 \\
Azure-2024 & Per-prompt memorization  & 0.995 &  5.4 \\
\midrule
Proprietary Data      & Empirical-survival       & 0.700 & 20.1 \\
Proprietary Data      & Per-prompt memorization  & 0.974 &  2.9 \\
\bottomrule
\end{tabular}
\end{table}

\section{Engineering details of the system implementation}
\label{app:engineering}
 
This appendix supplies the implementation details deferred from Section~\ref{sec:system}. Section~\ref{app:engineering-data} expands the proxy's data structures and the coroutines that maintain them. Section~\ref{app:engineering-payload} gives the precise payload-mutation contract used to drive the prefill--decode hand-off, including retry, cancellation, and recomputation handling. Section \ref{subsec:system-telemetry} explains the in-band telemetry. Section~\ref{app:engineering-bitset} describes the bit-set construction behind Stage-2 subset selection and analyses its complexity. Section~\ref{app:engineering-instrumentation} describes the optional engine-side instrumentation we use for closed-loop validation. Section~\ref{app:engineering-h0} describes the pool-bypass deployment path for BR-0. Section \ref{subsec:exp_overhead} displays the dispatch overhead in our experiment.
 
\subsection{Proxy data structures and coroutines}
\label{app:engineering-data}
 
The structures introduced in Section~\ref{sec:system} have the following implementation notes.
 
\paragraph{\texttt{DecodeInstanceState}$[g]$.}
A dictionary of \texttt{RequestTracker} entries keyed by request id. Each tracker also stores the producer-side hand-off metadata returned by the prefill probe (Section~\ref{app:engineering-payload}), which is needed to re-attach the request at decode admission. The aggregate $\mathrm{cap}[g] = B - |\mathcal{A}_g(k)|$ is maintained alongside $L_g(k)$ for $O(1)$ lookup at dispatch time.
 
\paragraph{\texttt{PromptPool}.}
An arrival-ordered registry of prefill-completed entries awaiting decode admission. Each entry holds a tracker plus the producer-side hand-off metadata, so admission to a decode worker requires no further round-trip to the prefill tier. Entries also carry a TTL that bounds the time a request may sit in the pool waiting for a slot; expired entries are aborted by a pool-reaper coroutine, which also issues a release on the corresponding prefiller-side delayed-free record.
 
\paragraph{\texttt{PrefillerBudget}$[i]$.}
The most recent KV-cache utilization sample of prefiller $i$. A parse failure or scrape timeout is treated conservatively as utilization $1.0$, so a transiently unreachable prefiller is automatically excluded from rotation rather than mistakenly preferred for its stale low reading.
 
\paragraph{Cooperating coroutines.}
Four coroutines drive the pipeline. (i) A \emph{prefill dispatcher} dequeues client arrivals, picks an unsaturated prefiller, and forks an \texttt{asyncio} task to issue the prefill probe and insert the completed entry into the pool. (ii) A \emph{decode dispatcher} blocks on a pair of \texttt{asyncio.Event}s (\texttt{pool\_not\_empty} and \texttt{decoder\_slot\_freed}) and, on wake, runs Algorithm~\ref{alg:brh} against the current pool and decode-worker grid, emitting a batch of admissions in one tick. (iii) A \emph{metrics poller} reads each prefiller's Prometheus endpoint at a fixed cadence and updates \texttt{PrefillerBudget}. (iv) A \emph{pool reaper} ages out entries past their TTL.
 
\paragraph{Connection pooling.}
Each \texttt{vllm-ascend} worker is fronted by a dedicated \texttt{httpx.AsyncClient} with a generously sized keep-alive pool over persistent HTTP/1.1 connections, so TCP setup is amortized across many requests rather than paid per dispatch. SSE streams from decoders are forwarded chunk-by-chunk to clients while being parsed inline for state mirroring.
 
\subsection{Payload-mutation protocol}
\label{app:engineering-payload}
 
\paragraph{Prefill probe.}
The proxy clones the incoming client request and applies three overrides:
\begin{itemize}
\item \texttt{stream}{=}\texttt{False},
\item \texttt{max\_tokens}{=}\texttt{min\_tokens}{=}$1$,
\item \texttt{kv\_transfer\_params}{=}\{\texttt{do\_remote\_decode}: \texttt{true}, \texttt{do\_remote\_prefill}: \texttt{false}, $\ldots$\}.
\end{itemize}
The first two overrides force the prefiller to compute the full KV cache, return one token (so the response carries usage statistics, including the realized prompt length), and stop short of autoregressive decode. The third instructs \texttt{MooncakeConnectorV1} to retain the KV blocks on the producer until the consumer signals receipt, and to populate the response with the producer-side handles \texttt{remote\_engine\_id}, \texttt{remote\_block\_ids}, and \texttt{remote\_host:remote\_port}. The proxy reads the realized \texttt{prompt\_tokens} from the response \texttt{usage} field and records it on the request's tracker, then parks the entry in the pool together with the returned \texttt{kv\_transfer\_params}.
 
\paragraph{Decode admission.}
When BR-H selects an entry for decode worker $g$, the proxy reconstitutes the original request---\texttt{stream}{=}\texttt{True}, original \texttt{max\_tokens}, original sampling parameters---and re-attaches the producer-side \texttt{kv\_transfer\_params}. Worker $g$ enters \texttt{WAITING\_FOR\_REMOTE\_KVS}; \texttt{MooncakeConnectorV1} pulls the KV blocks from the producer and, on completion, emits a \texttt{DONE\_RECVING\_MSG} that releases the prefiller's delayed-free queue. The first decoded chunk arriving at the proxy is the unambiguous signal that the cross-tier transfer has completed; the proxy uses it to release the prefiller's KV-cache budget for the next request.
 
\paragraph{Recomputation handling.}
A by-product of the payload-mutation protocol is a clean handler for vLLM's \texttt{stop\_reason}{=}\texttt{recomputed} event, which can occur under preemption: the proxy detects it inline, unregisters the request from worker $g$, augments the original prompt with the already-emitted tokens, decrements \texttt{max\_tokens} accordingly, and re-enters the request as a fresh pool entry, preserving end-to-end completion semantics without engine modification.
 
\paragraph{Cancellation.}
A \texttt{listen\_for\_disconnect} side-task ensures that a client cancellation---whether while the entry is in the pool or after dispatch but before the first decoded token---deterministically releases the corresponding pool slot or decode-worker slot rather than leaking a sticky assignment. Retries follow exponential backoff with capped attempts.

\subsection{In-band telemetry}
\label{subsec:system-telemetry}

BR-H requires two streams of feedback that are not exposed by the standard OpenAI-compatible API.

\noindent\textbf{Per-token decode progress.}
As the proxy forwards SSE chunks from the decoder to the client, it parses each one inline. A content delta triggers a simple increment of the $a_i(k)$ counter for that request; a finish signal triggers cleanup. The heavier predictor recomputation (Section~\ref{subsec:brh-prediction}) is gated to $\Delta T$-step boundaries rather than being run per token, so the per-chunk work remains an integer update plus a marker check; parsing happens before the chunk leaves the proxy, so no token-level latency is added on the critical path.

\noindent\textbf{Prefiller KV budget.}
The delayed-free mechanism in \texttt{MooncakeConnectorV1} holds KV caches on the prefiller until a consumer commits to downloading them. To prevent a prefiller from blocking on full cache, a metrics poller reads the standard Prometheus endpoint of each prefiller at a fixed cadence. The prefill dispatcher uses this data as back-pressure, refusing to send new probes to any prefiller that exceeds a configurable utilization threshold.

\subsection{Bit-set subset selection}
\label{app:engineering-bitset}
 
The Stage-2 subproblem is to choose $Q \subseteq R_{\mathrm{wait}}(k)$ with $|Q| \leq \min(\mathrm{cap}[g], R_{\max})$ that maximizes $F_g(Q)$ in~\eqref{eq:brh-Fscore}. The reduction sketched in Section~\ref{app:engineering-bitset} expands as follows.
 
\paragraph{Reduction to reachable sums.}
$F_g(Q)$ depends on $Q$ only through $\Delta_s(Q) = \sum_{i \in Q} s_i$. Hence, for each cardinality $k \leq R_{\max}$, it suffices to enumerate
\[
\mathcal{R}_k \;:=\; \{ \Delta_s(Q) : Q \subseteq R_{\mathrm{wait}}(k),\, |Q| = k \}.
\]
We encode $\mathcal{R}_k$ as a bitmask $\mathrm{dp}[k]$ whose bit $b$ is set iff $b \in \mathcal{R}_k$. The standard 0/1-knapsack recurrence becomes a single shift-OR per item, $\mathrm{dp}[j] \mathrel{|}{=} (\mathrm{dp}[j-1] \ll s_i)$, scanning $j$ from $k$ down to $1$. Storing $\mathrm{dp}$ snapshots at each item allows the chosen subset to be recovered by $O(n)$ backtracking after the optimum is identified.
 
For BR-0, the score has a single kink and the two-probe rule is exact. For BR-H, the score has up to $H+$ 1 kinks. In our implementation with $R_{\max }=4$, we use exhaustive enumeration over the candidate window; this is only $2^{R_{\max }} \leq 16$ subsets per worker and is negligible in the measured dispatch overhead. A bitset implementation can also be made exact by evaluating $F$ over all reachable sums, or by probing around every distinct horizon margin.
 
\paragraph{Cost.}
The dominant cost is $O(R_{\max} \cdot S/w)$ word-level operations per worker, where $w$ is the machine word size and $S$ is an upper bound on per-request prompt length used to size the bitmask. This is exponentially smaller than the $O(2^{R_{\max}})$ cost of naive enumeration in $R_{\max}$, while remaining an exact solver under the structure of $F_g$.
 
\subsection{Engine-side instrumentation for evaluation}
\label{app:engineering-instrumentation}
 
For closed-loop validation against the ground-truth straggler dynamics, we apply an optional, lightweight monkey patch to the engine-step loop in vLLM that timestamps every step and records $(\texttt{step\_cost\_ms}, \texttt{running\_reqs}, \texttt{total\_tokens}, \texttt{waiting\_reqs}, \texttt{kv\_cache\_usage})$ to a per-DP-rank file. The DP rank is recovered from process metadata, with a fallback via launch-argument and PID. The patch is the source of the empirical step traces used in Section~\ref{sec:experiments}; it is purely diagnostic and disabled in production runs.
 
\subsection{Pool-bypass deployment path for BR-0}
\label{app:engineering-h0}
 
For BR-0 deployments where end-to-end latency is the operator's priority, the proxy can be configured to bypass the prompt pool. The motivation is structural: in the pool-based path, dispatch waits for a decode slot to free before sending, after which the cross-tier KV transfer is serialized with engine scheduling on the receiving worker. The path described here trades some routing flexibility for an earlier start to that transfer.
 
\paragraph{Mechanism.}
On prefill completion, the proxy immediately evaluates the BR-0 single-step score $F_g(s) = \alpha s - \beta(s - m_g)_+$ over all decode workers, using a \emph{virtual load} that counts both running and already-dispatched-but-not-yet-running requests on each worker, and forwards the request directly. The target worker enters the receive-pending state immediately, allowing the KV transfer to overlap with the residual compute of currently-running requests rather than starting after slot release. To keep the connector's send buffers bounded, the dispatcher caps the per-worker inflight to a small fixed margin above the batch limit $B$; in our deployment a tight cap was sufficient to keep slots utilized without saturating the buffers at the traffic levels we tested.
 
\paragraph{Scope.}
This path is available only for BR-0 ($H{=}0$): with the pool removed, the dispatcher sees one waiting request at a time, so Stage-2 subset selection collapses to Stage-1 by construction. For BR-H ($H{>}0$) the pool is required, and the late-binding cost noted in Section~\ref{sec:system} applies. A controlled latency benchmark of pool-based BR-0 against the pool-bypass variant, and an extension that preserves Stage-2 leverage while overlapping KV transfer for BR-H, are directions we leave to future work.
\subsection{Dispatch overhead}
\label{subsec:exp_overhead}
\begin{figure}[t]
\centering
\includegraphics[width=\linewidth]{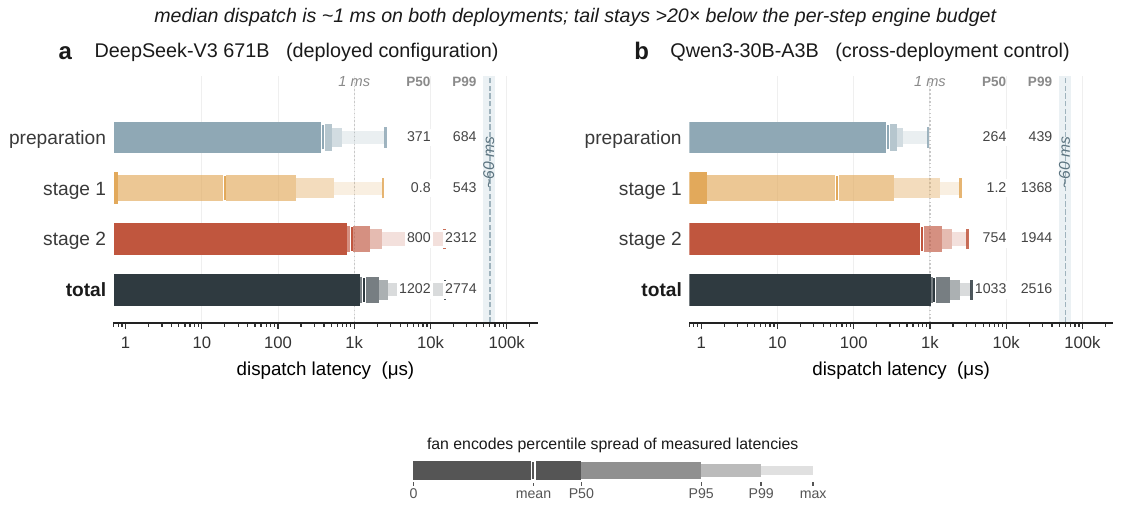}
\caption{\textbf{Per-tick dispatch overhead, two deployments.} $G{=}8$, $R_{\max}{=}4$. \textbf{(a)} DeepSeek-V3 671B. \textbf{(b)} Qwen3-30B-A3B (cross-deployment control). Each row is a percentile fan ($0\to P_{50}$ solid, then to $P_{95}$, $P_{99}$, max); the white tick marks the mean. Shaded band ($\sim\!60$\,ms) marks the per-step engine latency on DeepSeek-V3.}
\label{fig:overhead}
\end{figure}

Figure~\ref{fig:overhead} measures the per-tick cost of the BR-H dispatcher itself. We instrument the proxy as described in Appendix~\ref{app:engineering-instrumentation} and record per-stage latency on two deployments: the DeepSeek-V3 671B configuration of Section~\ref{subsec:exp_main} (Figure~\ref{fig:overhead}a) and a Qwen3-30B-A3B deployment held to identical $G{=}8$ and $R_{\max}{=}4$ (Figure~\ref{fig:overhead}b).

Total per-loop dispatch cost has $P_{50}{=}1.20$\,ms (mean $1.34$\,ms, $P_{99}{=}2.77$\,ms) on DeepSeek-V3 and $P_{50}{=}1.03$\,ms (mean $1.15$\,ms, $P_{99}{=}2.52$\,ms) on Qwen3. The two distributions agree to within $17\%$ at $P_{50}$ and $10\%$ at $P_{99}$, consistent with dispatcher cost being determined by control-plane work in the proxy rather than engine-side compute. Against the $\sim\!60$\,ms per-step engine budget on the deployed configuration (shaded band), this places the dispatcher roughly $50\times$ below the engine step at the median and $\sim\!22\times$ below at $P_{99}$, with a small number of Stage-2 outliers above $10$\,ms that are still within one engine step.

The per-stage decomposition shows Stage 2 (the bit-set DP) accounting for $\sim\!69\%$ of total dispatch time on both deployments (mean $924\,\mu$s on DeepSeek, $794\,\mu$s on Qwen3). Stage 1 is sub-microsecond at the median ($P_{50}{=}0.8$--$1.2\,\mu$s) but has a longer tail ($P_{99}{=}0.54$\,ms on DeepSeek, $1.37$\,ms on Qwen3) when the abundant-capacity loop iterates many times before crossing the $S_{\mathrm{greedy}}$ threshold. Preparation has the tightest distribution ($P_{99}/P_{50} \approx 1.8$), consistent with its $O(GH)$ structure once $G$ and $H$ are fixed.

\section{Supplementary trace-level analysis: Survival predictor on Proprietary Data}
\label{app:survival-30002}
 
This appendix supplies the trace-level analysis of BR-H deployed with the empirical-survival predictor on Proprietary Data, the comparison instantiation reported in Table~\ref{tab:main_combined}. The motivation is to verify that the secondary deployment behaves as the offline characterization in Section~\ref{subsec:exp_predictor} predicts: on this workload empirical-survival has weaker offline signal than the per-prompt memorization variant, so the corresponding deployed router should do worse end-to-end---but still substantially better than the baselines and better than BR-0, because the Stage-1 confidence gate at $\widehat{p}_{\mathrm{fin}} \geq 0.5$ closes cleanly when the predictor lacks information rather than injecting noise (Section~\ref{app:predictor-refresh}).
 
\begin{figure}[htbp]
\centering
\includegraphics[width=0.5\linewidth]{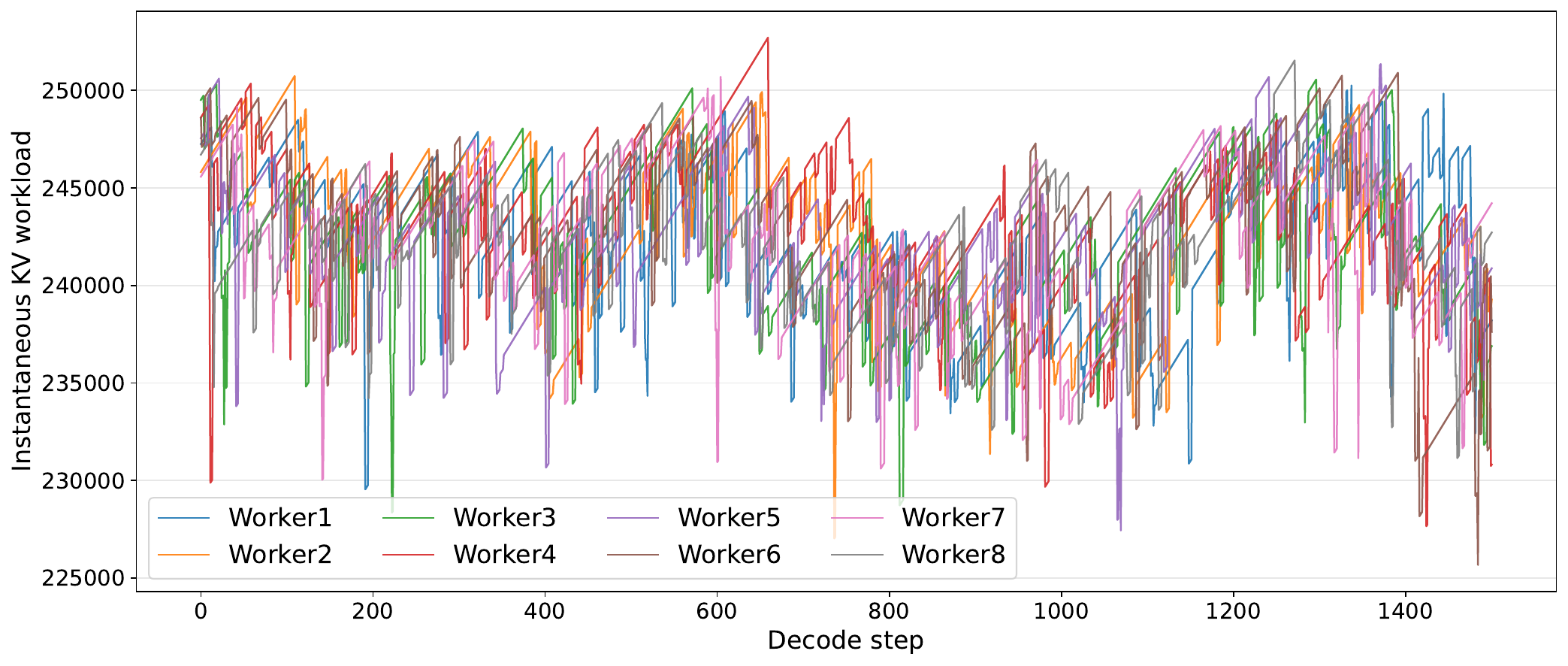}
\caption{\textbf{Per-worker instantaneous KV-cache workload on Proprietary Data under BR-H deployed with the empirical-survival predictor}, $H{=}80$, $G{=}8$ decode workers, same $1{,}500$-step trace segment as Figure~\ref{fig:kv-workload-grid}. Trace-mean imbalance is $40{,}757$ tokens. The band remains visibly tighter than BR-0 ($51{,}927$) and far tighter than the strongest baseline JSQ ($215{,}110$); it is somewhat wider than the ExactMatch deployment ($36{,}277$). The Stage-1 confidence gate (Section~\ref{app:predictor-refresh}) closes for the long-tail majority of requests on this workload, so $\widehat{c}_i(k)$ falls back to the conservative anchor $H$ rather than injecting a noisy point estimate; the gain over BR-0 comes from the requests for which the predictor is confident.}
\label{fig:kv-workload-30002-survival}
\end{figure}

\paragraph{Trace-level worker spread.}
Figure~\ref{fig:kv-workload-30002-survival} plots the per-worker instantaneous KV-cache footprint over the same $1{,}500$-step segment used in Figure~\ref{fig:kv-workload-grid}, with the BR-H router driven by the empirical-survival predictor. The trace-mean imbalance is $40{,}757$ tokens, against $36{,}277$ for ExactMatch (Figure~\ref{fig:kv-workload-grid}, bottom-right), $51{,}927$ for BR-0, and $215{,}110$ for the strongest baseline JSQ. Visually, the band of eight worker curves remains tightly clustered throughout the trace; the spread is $\sim 1.1\times$ that of ExactMatch and $\sim 1.3\times$ tighter than BR-0, consistent with the predictor's role of adding short-horizon resolution to BR-0's prediction-free baseline even when its Stage-1 separation on this workload is weak.
 
\paragraph{Headline numbers in context.}
On every metric reported in Table~\ref{tab:main_combined}, BR-H with empirical-survival improves over BR-0: imbalance reduces by $21.5\%$ ($51{,}927 \to 40{,}757$), TPOT P95 reduces by $0.7\%$ ($79.32 \to 78.75$\,ms), and throughput rises by $3.9\%$ ($942.5 \to 979.2$\,tok/s). Compared to JSQ, the same router achieves a $5.3\times$ imbalance reduction and a $16.2\%$ throughput improvement. Compared to its better-aligned counterpart ExactMatch, the empirical-survival deployment is $\sim 12\%$ behind on imbalance ($40{,}757$ vs.\ $36{,}277$) and essentially tied on throughput.
 
\paragraph{Reading this in context.}
The Survival deployment on Proprietary Data is the failure-mode case for the predictor: the workload offers no auxiliary signal beyond the marginal CDF, and the marginal CDF itself is heavy-tailed, so the predictor's output is conservative. The empirical observation is that the BR-H rule degrades gracefully under this regime---the gate closes, $\widehat{c}_i(k)$ collapses to $H$, and the router behaves like a horizon-anchored variant of BR-0 rather than an oracle approximation. This is the design property targeted by the Stage-1 confidence gate: the predictor abstains cleanly on requests it cannot resolve, so a worse predictor yields a degraded but still-useful router rather than a router that would perform worse than the prediction-free baseline.

 
\section{Per-worker KV-workload panels}
\label{app:full-kv-grids}
 
This appendix provides the per-worker KV-cache workload panels that were omitted from Figures~\ref{fig:kv-workload-grid} for space and the panels for Azure-2024. Layout matches the corresponding main-text figure: each panel plots all eight workers on the same $1{,}500$-step decode segment, with the trace-mean imbalance annotated under each panel.
 
\begin{figure}[htbp]
\centering
\begin{subfigure}[t]{0.48\linewidth}
    \centering
    \includegraphics[width=\linewidth]{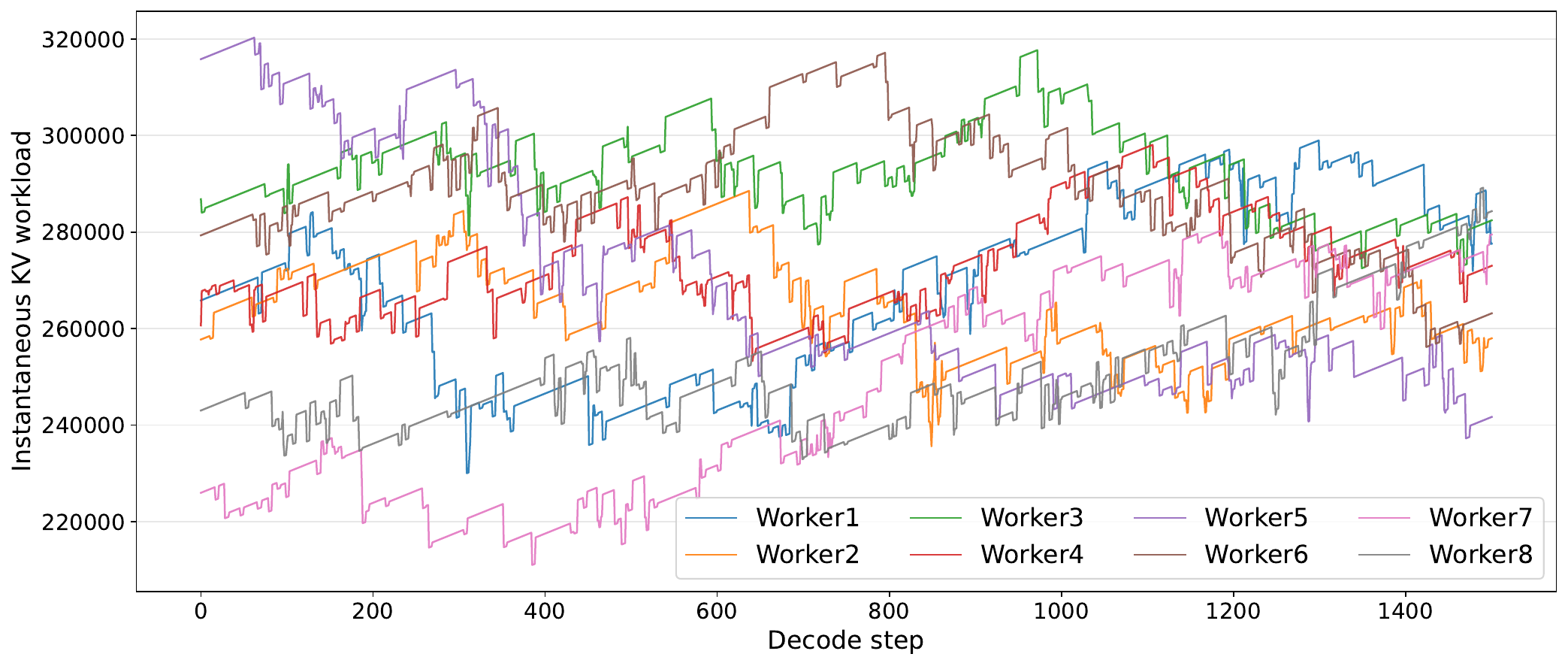}
    \caption*{\centering\footnotesize Round-Robin \\ \textit{imbal.\ 255k}}
\end{subfigure}\hfill
\begin{subfigure}[t]{0.48\linewidth}
    \centering
    \includegraphics[width=\linewidth]{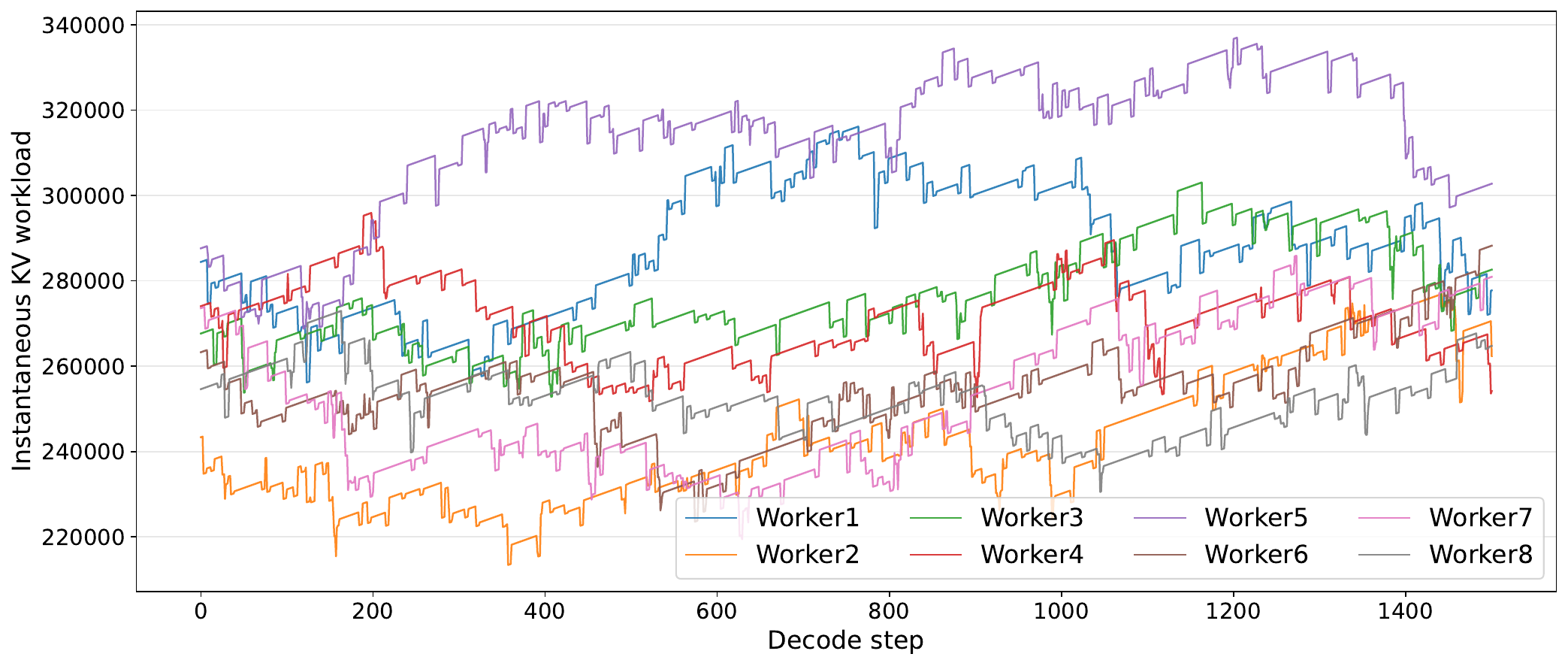}
    \caption*{\centering\footnotesize P2C \\ \textit{imbal.\ 263k}}
\end{subfigure}
 
\vspace{0.7em}
\begin{subfigure}[t]{0.48\linewidth}
    \centering
    \includegraphics[width=\linewidth]{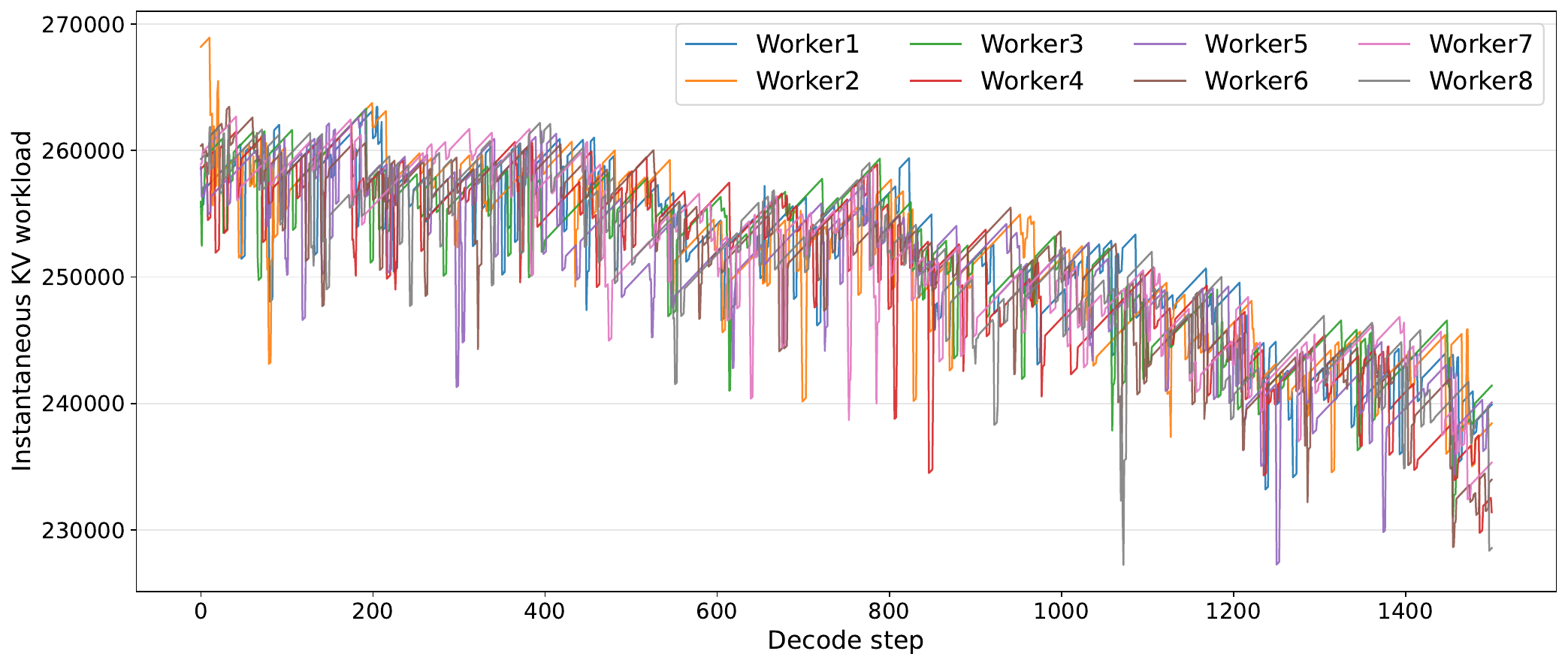}
    \caption*{\centering\footnotesize \textbf{BR-H} oracle, $(\beta,\gamma){=}(43, 0.86)$ \\ \textit{imbal.\ 24.0k}}
\end{subfigure}\hfill
\begin{subfigure}[t]{0.48\linewidth}
    \centering
    \includegraphics[width=\linewidth]{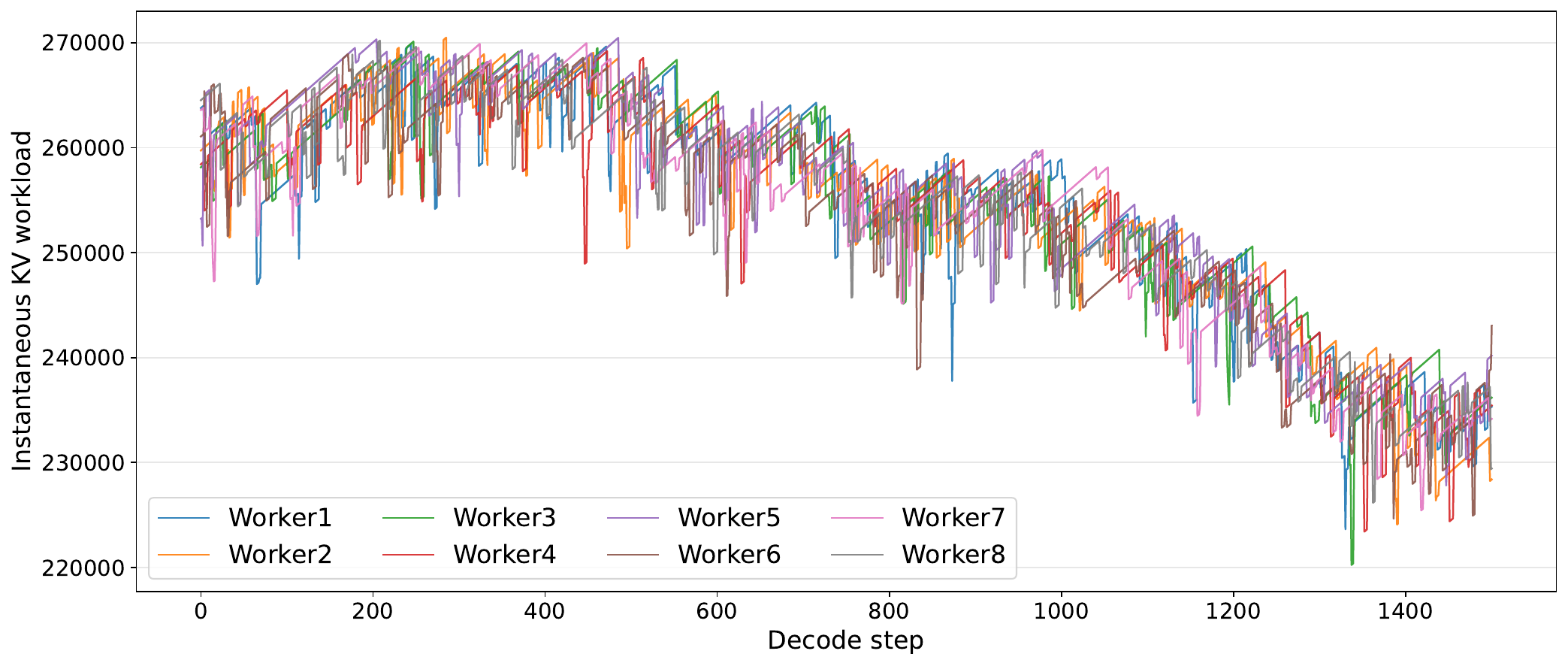}
    \caption*{\centering\footnotesize \textbf{BR-H} oracle, $(\beta,\gamma){=}(14.67, 0.64)$ \\ \textit{imbal.\ 23.6k}}
\end{subfigure}
 
\caption{\textbf{Remaining per-worker panels for Proprietary Data, $G{=}8$ (companion to Figure~\ref{fig:kv-workload-grid}).} Top: RR and P2C baselines. Bottom: two BR-H oracle operating points.}
\label{fig:kv-workload-grid-app}
\end{figure}

\begin{figure}[htbp]
\centering
\begin{subfigure}[t]{0.48\linewidth}
    \centering
    \includegraphics[width=\linewidth]{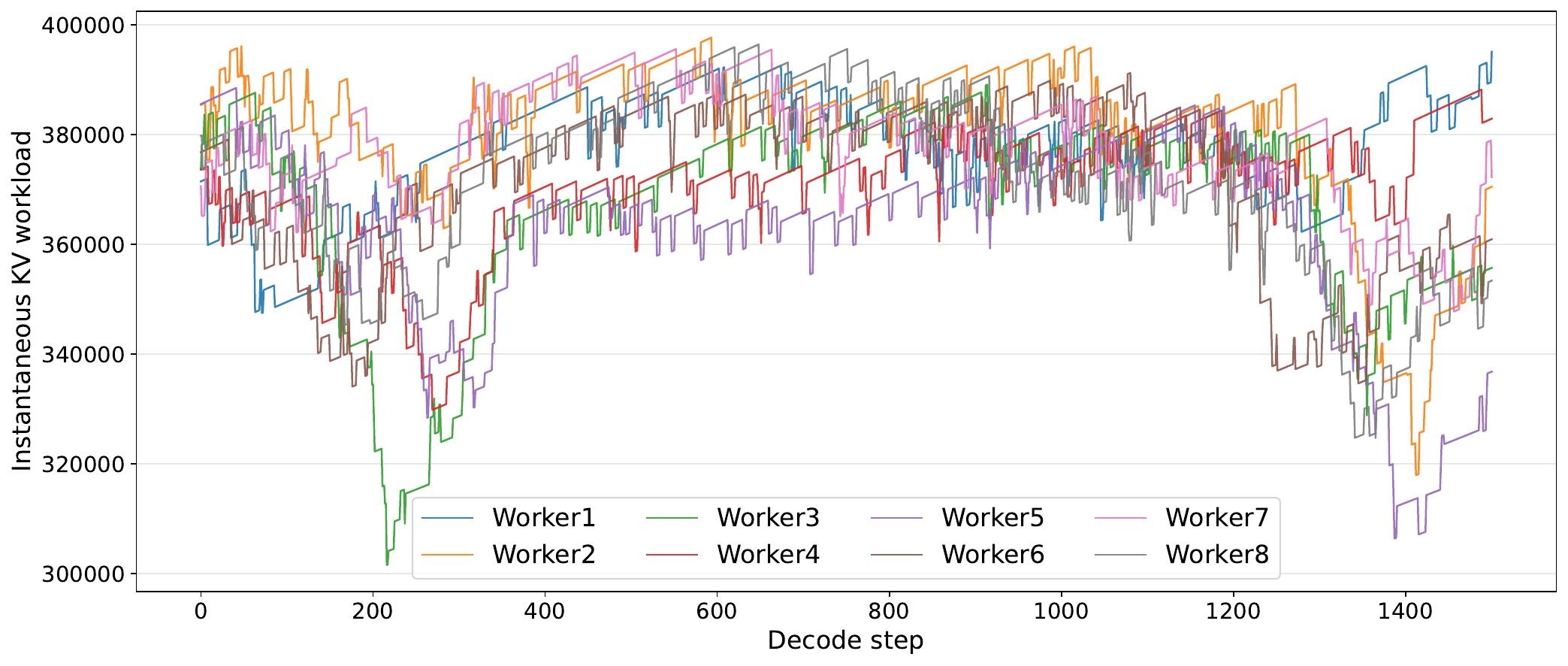}
    \caption*{\centering\footnotesize Random \\ \textit{imbal.\ 147k}}
\end{subfigure}\hfill
\begin{subfigure}[t]{0.48\linewidth}
    \centering
    \includegraphics[width=\linewidth]{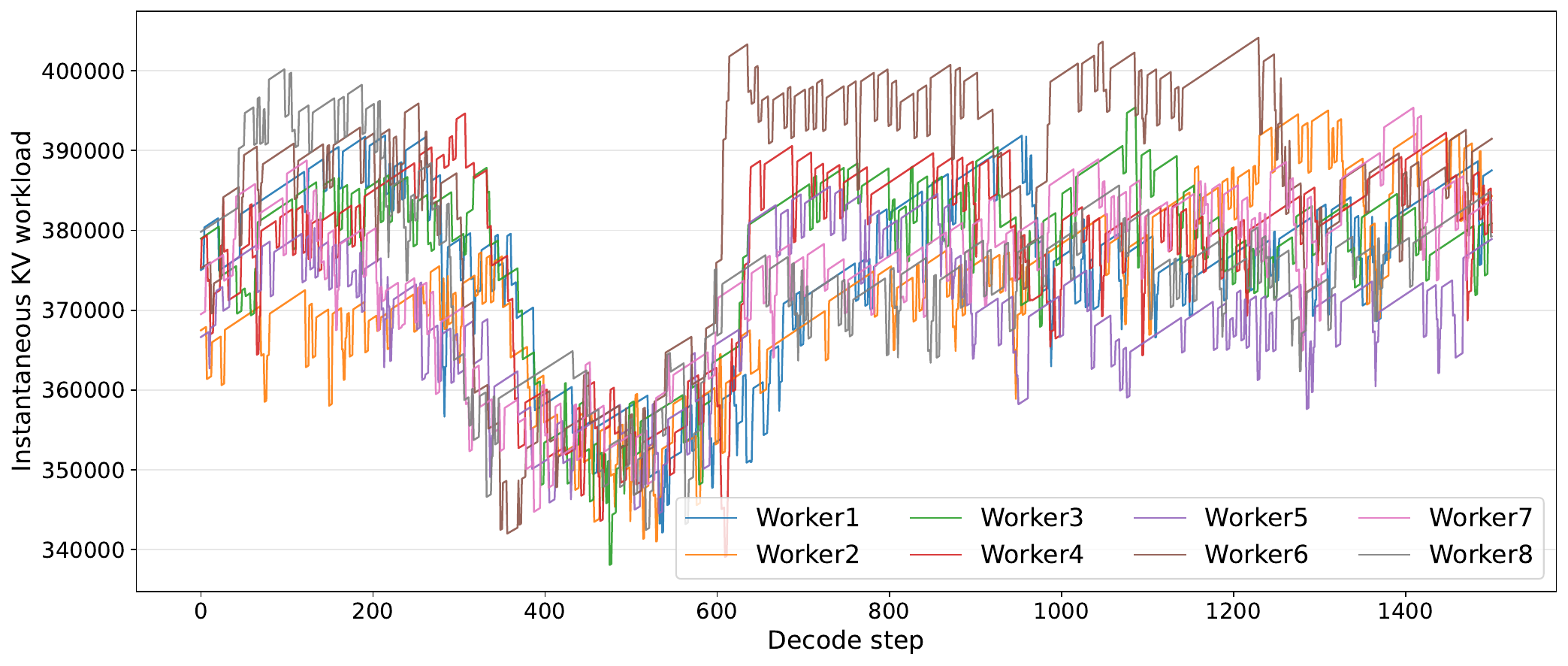}
    \caption*{\centering\footnotesize JSQ \\ \textit{imbal.\ 105k}}
\end{subfigure}

\vspace{0.7em}
\begin{subfigure}[t]{0.48\linewidth}
    \centering
    \includegraphics[width=\linewidth]{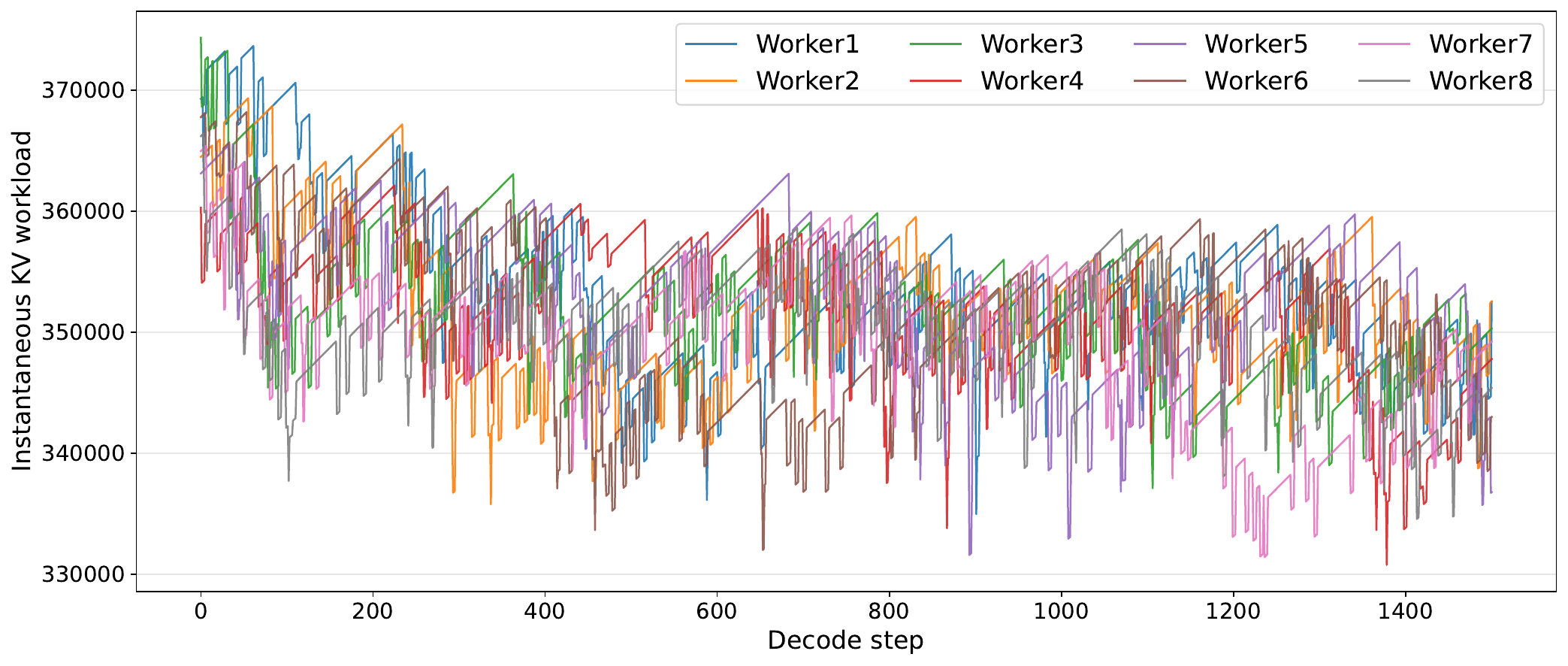}
    \caption*{\centering\footnotesize \textbf{BR-0} (ours) \\ \textit{imbal.\ 54k}}
\end{subfigure}\hfill
\begin{subfigure}[t]{0.48\linewidth}
    \centering
    \includegraphics[width=\linewidth]{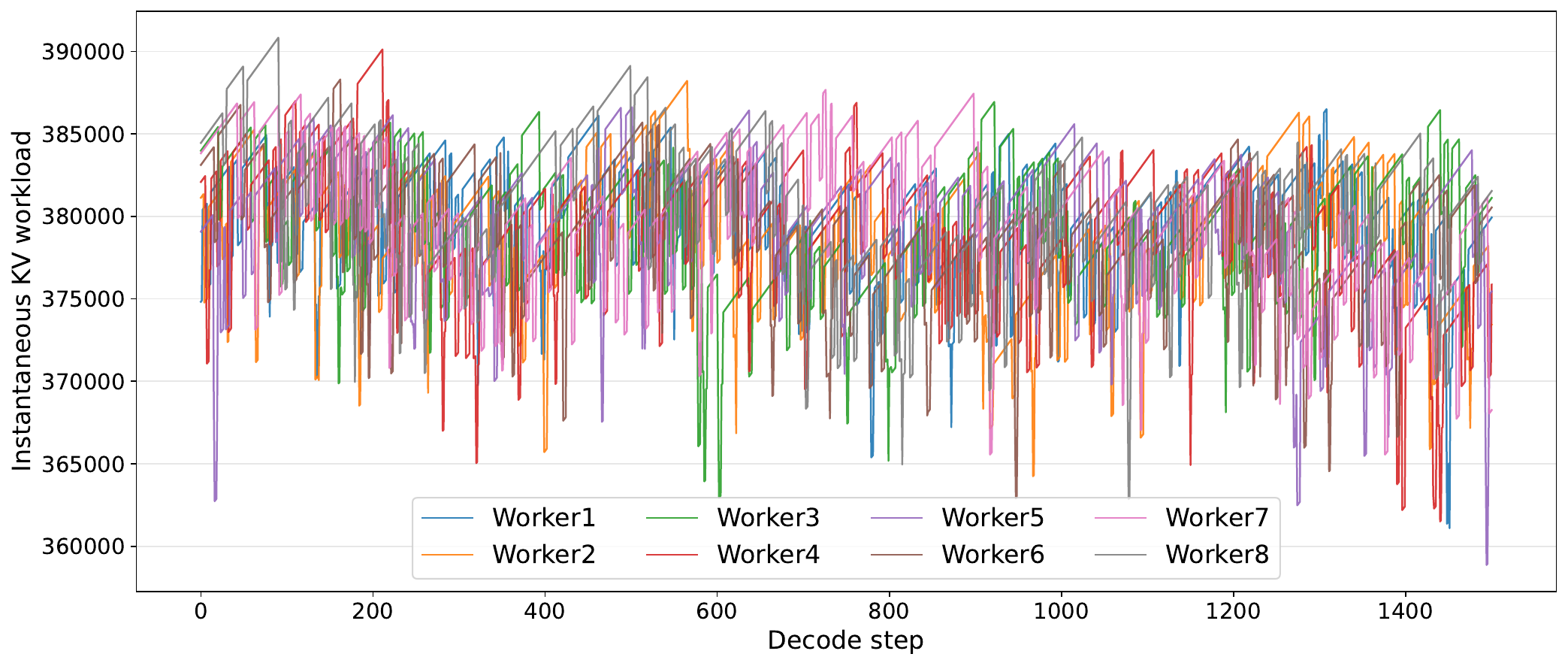}
    \caption*{\centering\footnotesize \textbf{BR-H} predicted, ExactMatch \\ \textit{imbal.\ 38.5k}}
\end{subfigure}

\caption{\textbf{Per-worker KV-cache workload on Azure-2024, $G{=}8$.} Same layout as Figure~\ref{fig:kv-workload-grid}. }
\label{fig:kv-workload-azure-grid-paper}
\end{figure}

\begin{figure}[htbp]
\centering
\begin{subfigure}[t]{0.48\linewidth}
    \centering
    \includegraphics[width=\linewidth]{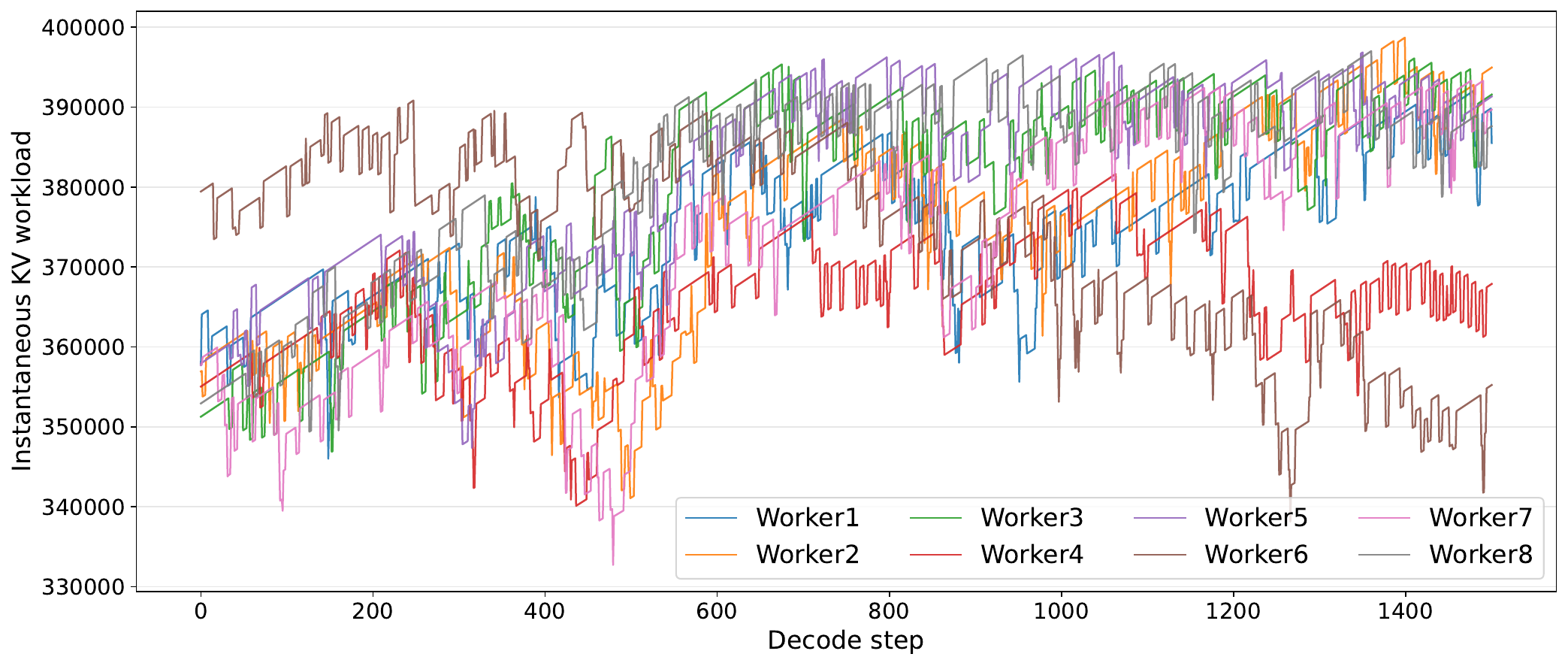}
    \caption*{\centering\footnotesize Round-Robin \\ \textit{imbal.\ 110k}}
\end{subfigure}\hfill
\begin{subfigure}[t]{0.48\linewidth}
    \centering
    \includegraphics[width=\linewidth]{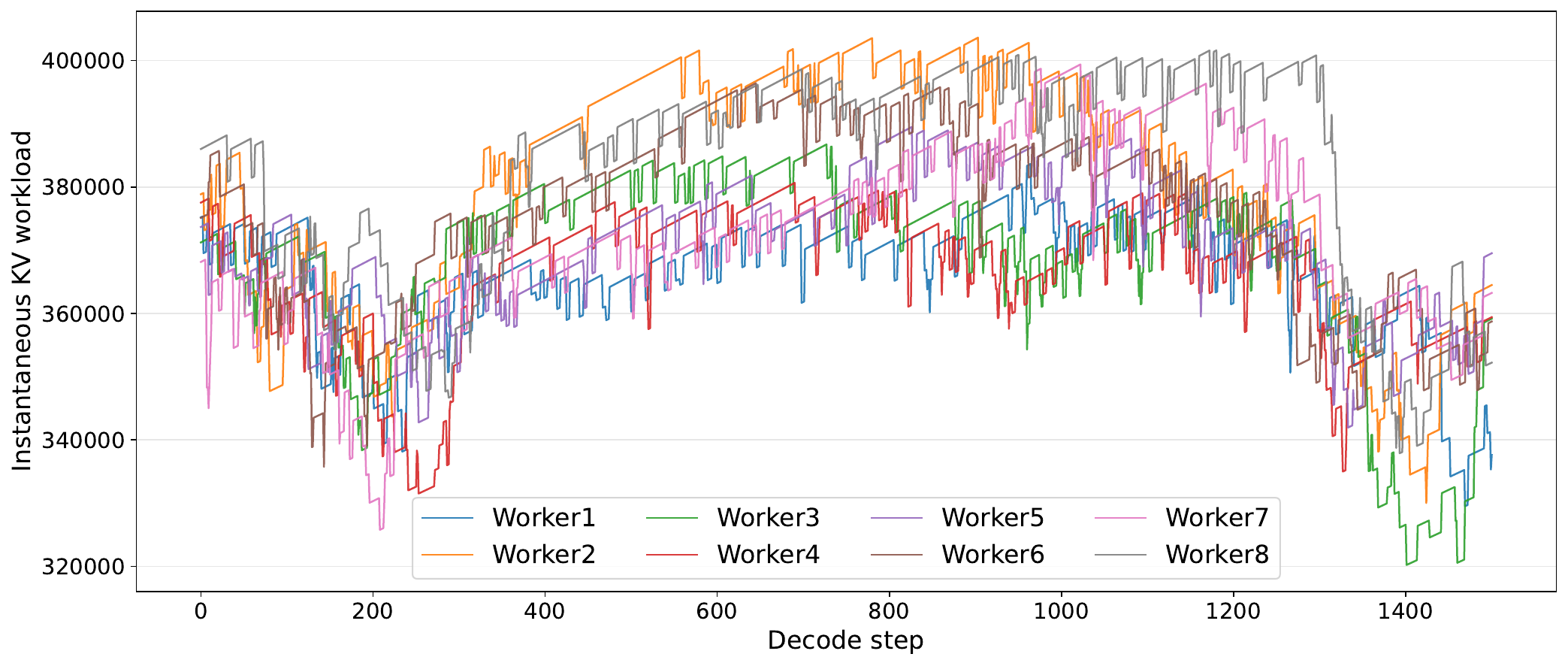}
    \caption*{\centering\footnotesize P2C \\ \textit{imbal.\ 109k}}
\end{subfigure}
 
\vspace{0.7em}
\begin{subfigure}[t]{0.48\linewidth}
    \centering
    \includegraphics[width=\linewidth]{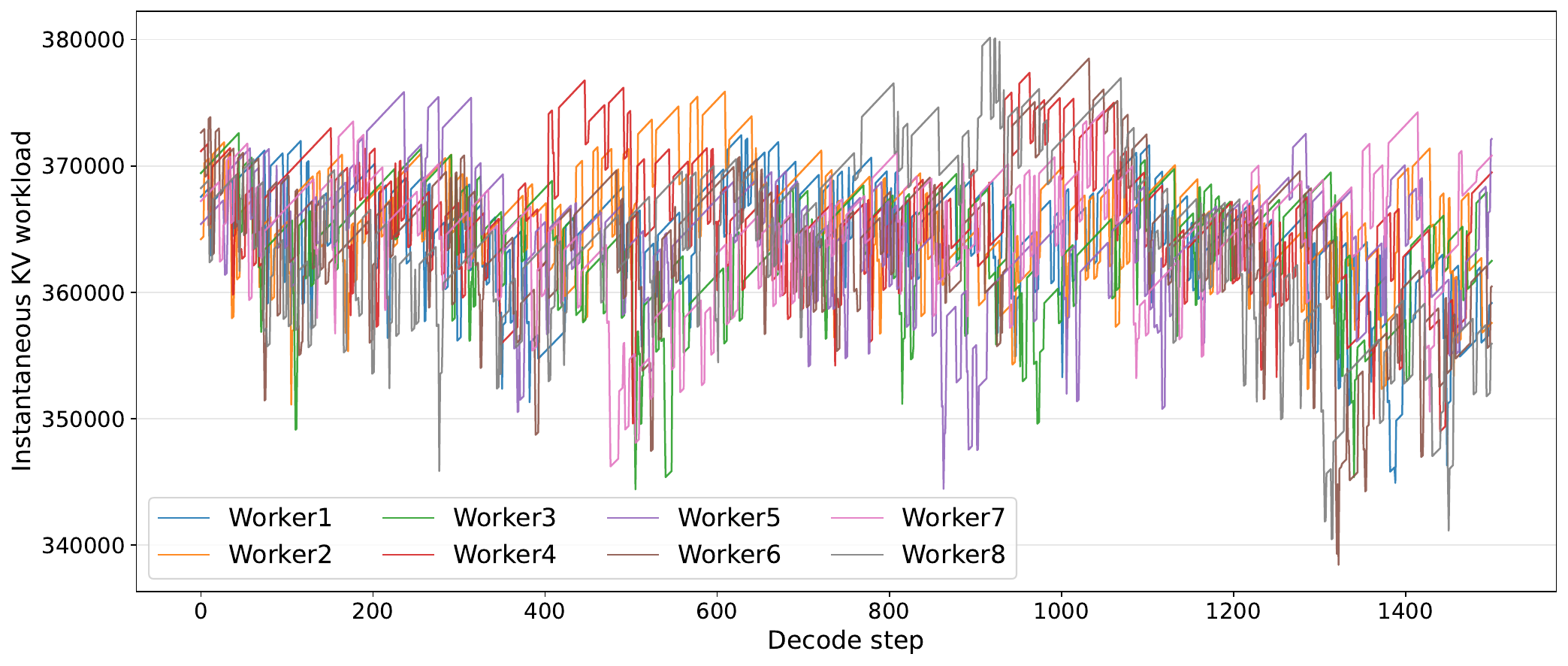}
    \caption*{\centering\footnotesize \textbf{BR-H} oracle, $(\beta,\gamma){=}(43, 0.86)$ \\ \textit{imbal.\ 37.8k}}
\end{subfigure}\hfill
\begin{subfigure}[t]{0.48\linewidth}
    \centering
    \includegraphics[width=\linewidth]{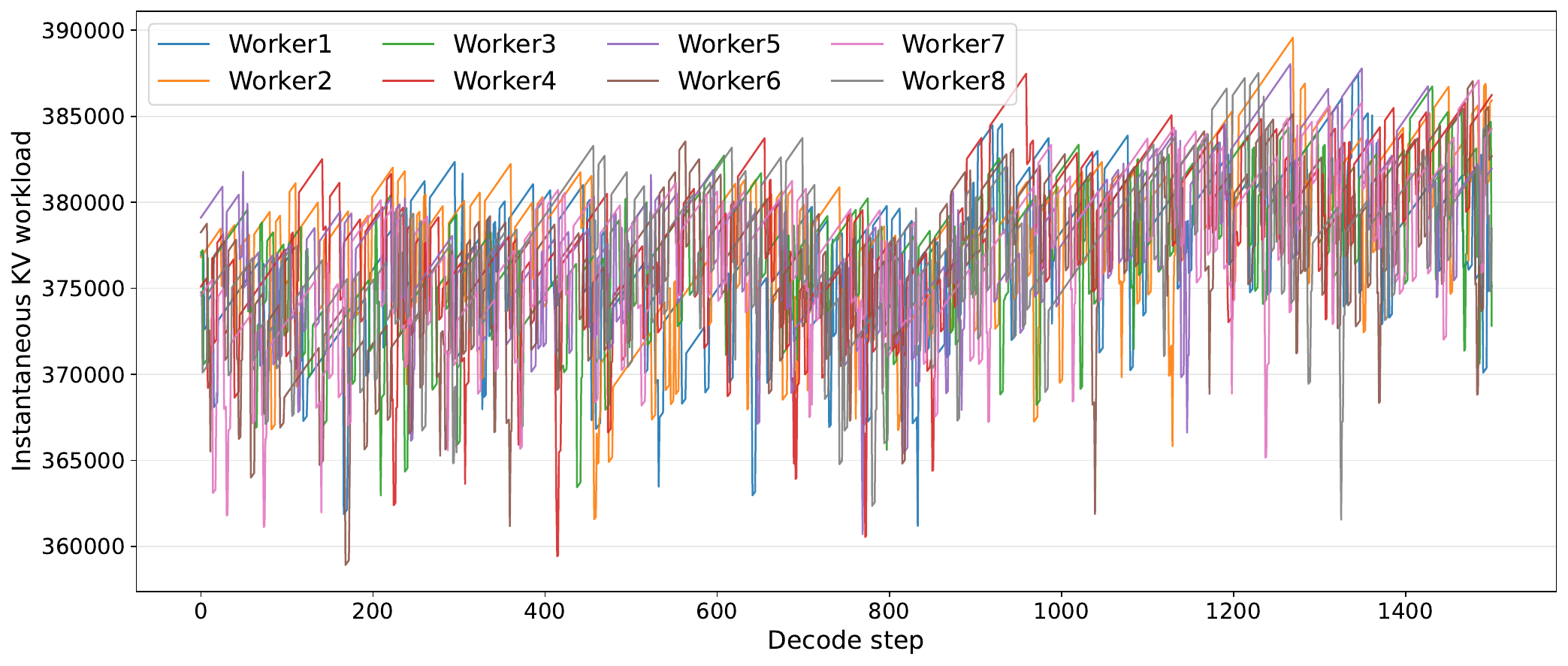}
    \caption*{\centering\footnotesize \textbf{BR-H} oracle, $(\beta,\gamma){=}(14.67, 0.64)$ \\ \textit{imbal.\ 35.3k}}
\end{subfigure}
 
\caption{\textbf{Remaining per-worker panels for Azure-2024, $G{=}8$ (companion to Figure~\ref{fig:kv-workload-azure-grid-paper}).} Top: RR and P2C baselines. Bottom: two BR-H oracle operating points.}
\label{fig:kv-workload-azure-grid-app}
\end{figure}

\section{Trace-level evidence for the scaling sweep at \texorpdfstring{$G \in \{4, 16\}$}{G in {4, 16}}}
\label{app:scaling}
 
\begin{figure}[htbp]
\centering
\begin{subfigure}[t]{0.48\linewidth}
    \centering
    \includegraphics[width=\linewidth]{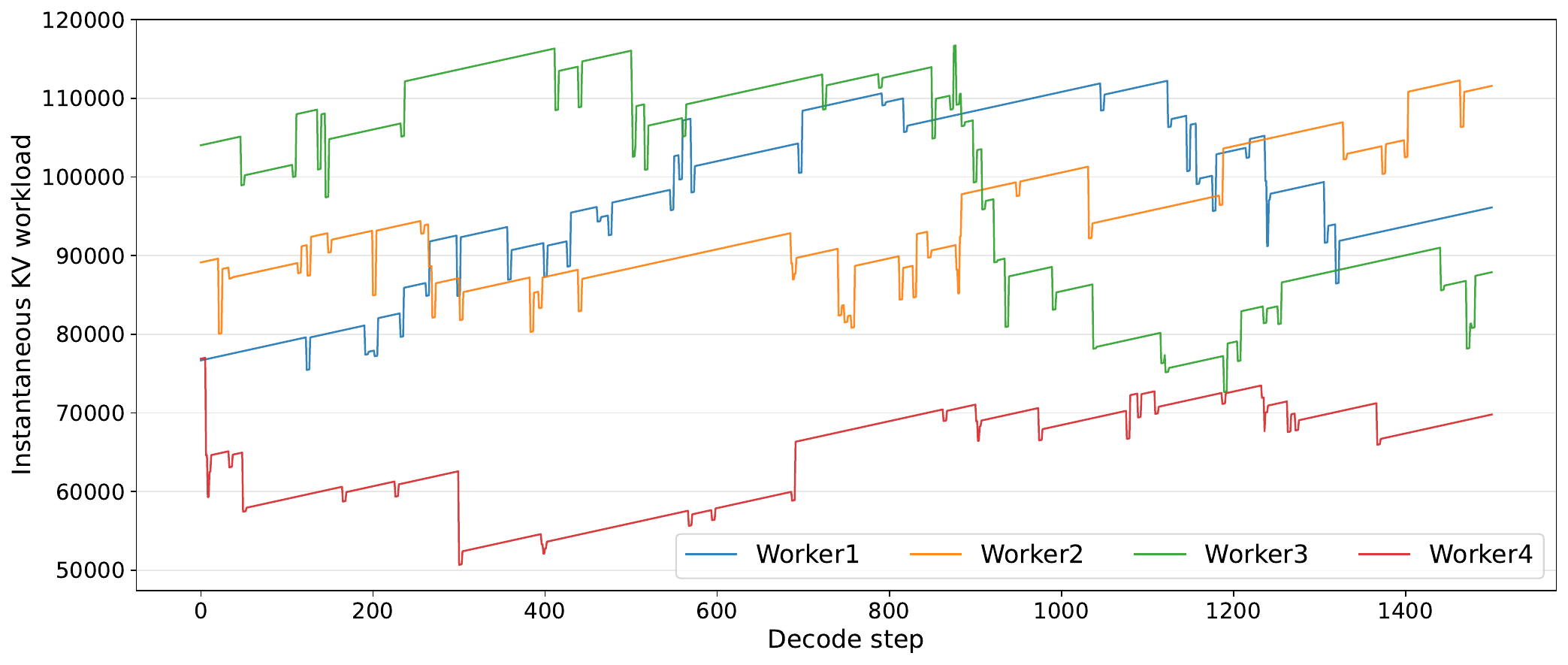}
    \caption*{\centering\footnotesize Random \\ \textit{imbal.\ 80k}}
\end{subfigure}\hfill
\begin{subfigure}[t]{0.48\linewidth}
    \centering
    \includegraphics[width=\linewidth]{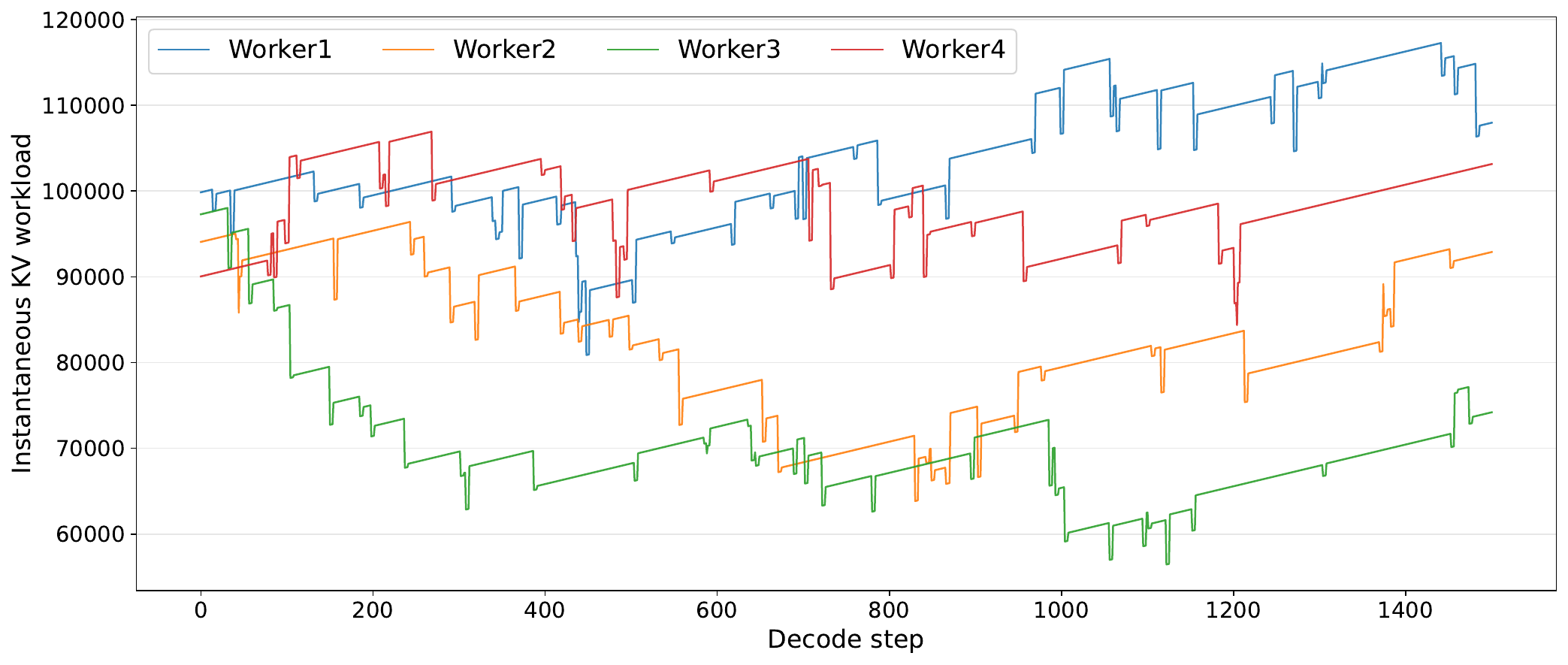}
    \caption*{\centering\footnotesize Round-Robin \\ \textit{imbal.\ 68k}}
\end{subfigure}\hfill
\begin{subfigure}[t]{0.48\linewidth}
    \centering
    \includegraphics[width=\linewidth]{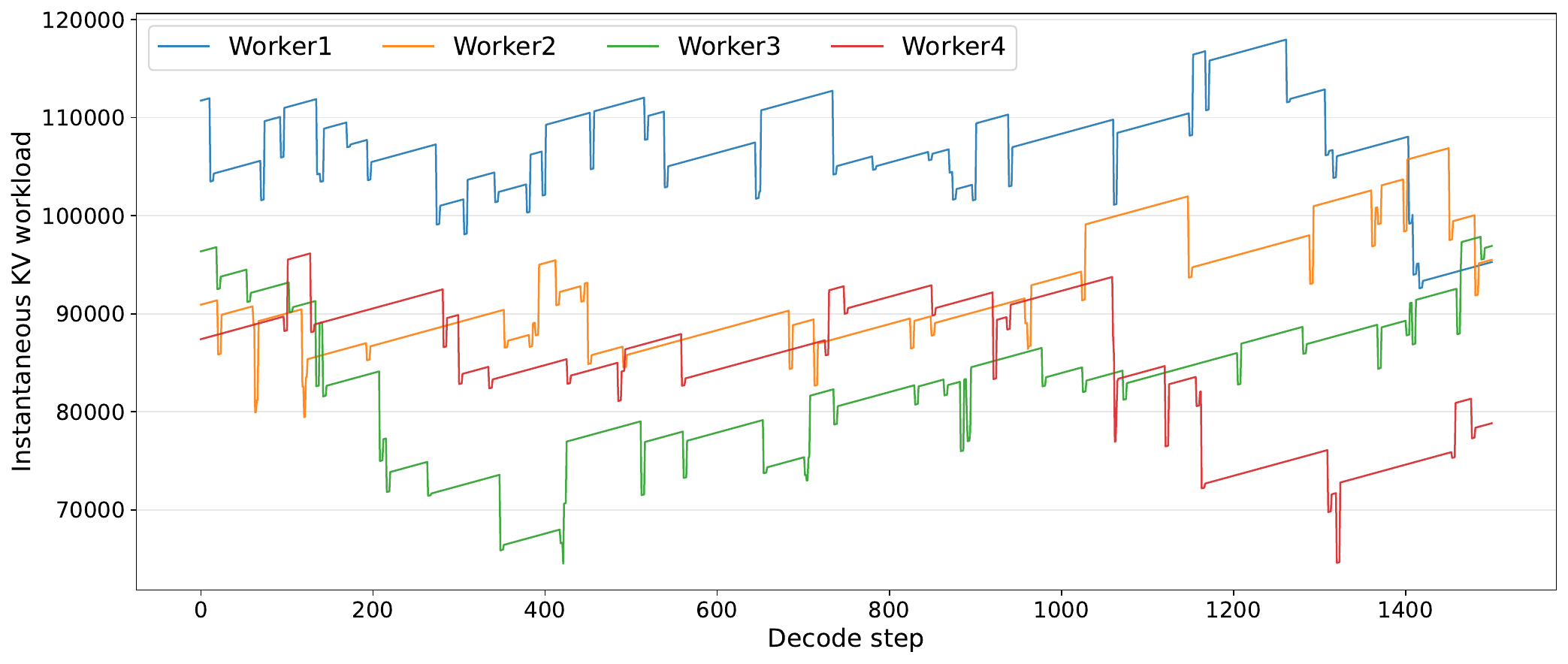}
    \caption*{\centering\footnotesize P2C \\ \textit{imbal.\ 64k}}
\end{subfigure}\hfill
\begin{subfigure}[t]{0.48\linewidth}
    \centering
    \includegraphics[width=\linewidth]{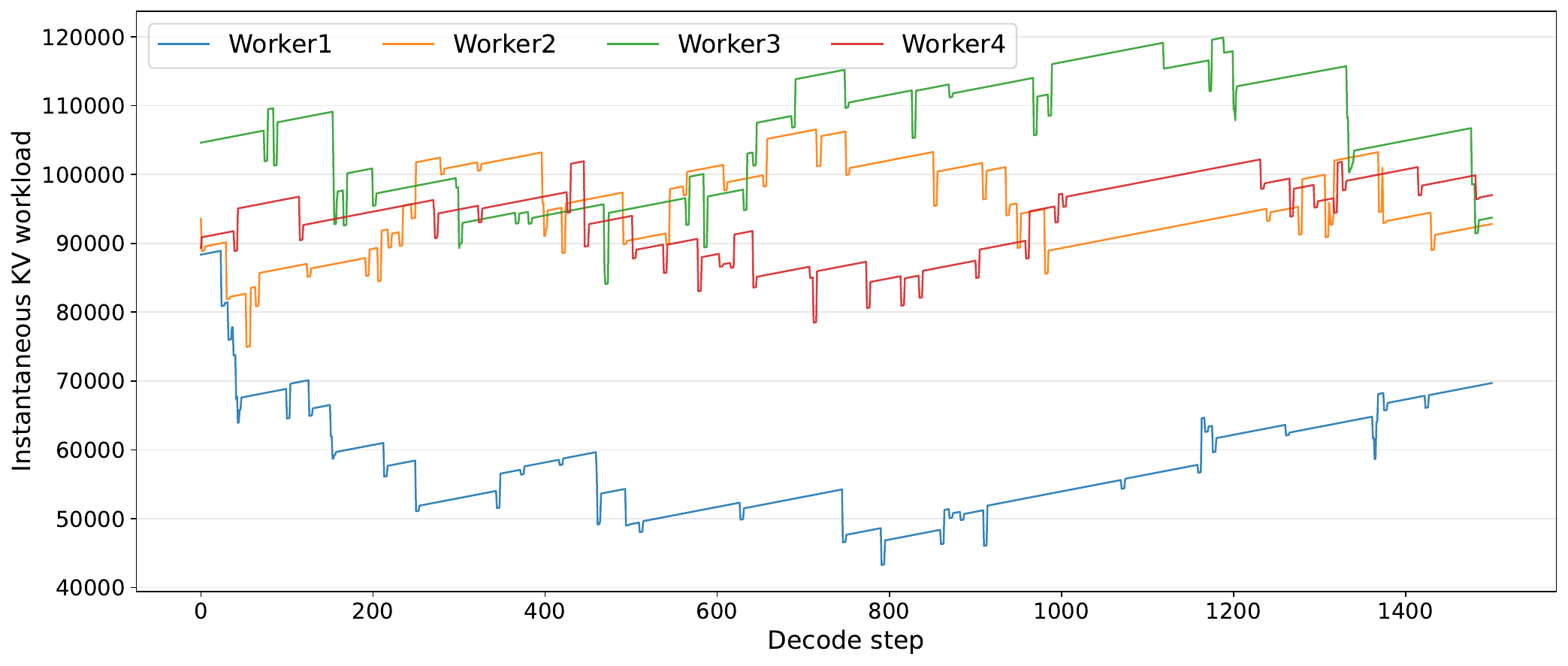}
    \caption*{\centering\footnotesize JSQ \\ \textit{imbal.\ 75k}}
\end{subfigure}
 
\vspace{0.7em}
 
\begin{subfigure}[t]{0.48\linewidth}
    \centering
    \includegraphics[width=\linewidth]{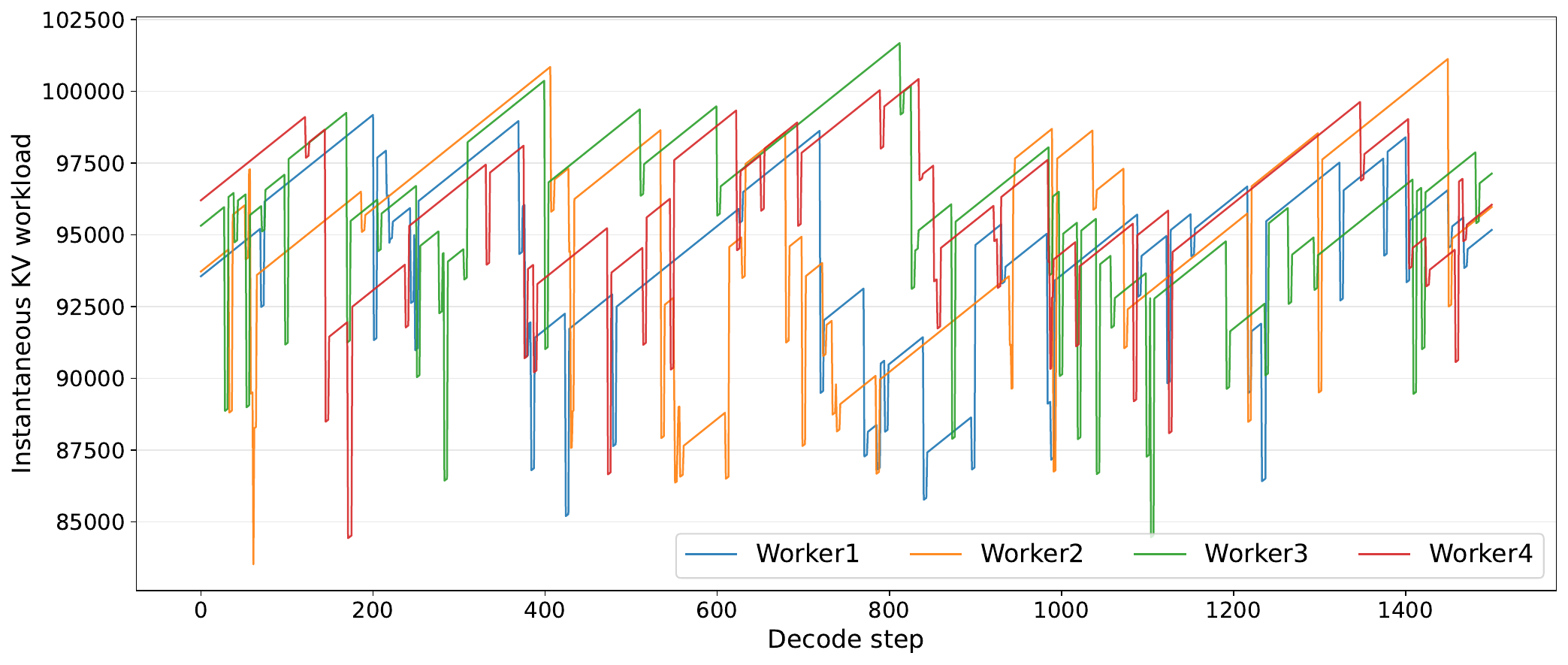}
    \caption*{\centering\footnotesize \textbf{BR-0} (ours) \\ \textit{imbal.\ 9.7k}}
\end{subfigure}\hfill
\begin{subfigure}[t]{0.48\linewidth}
    \centering
    \includegraphics[width=\linewidth]{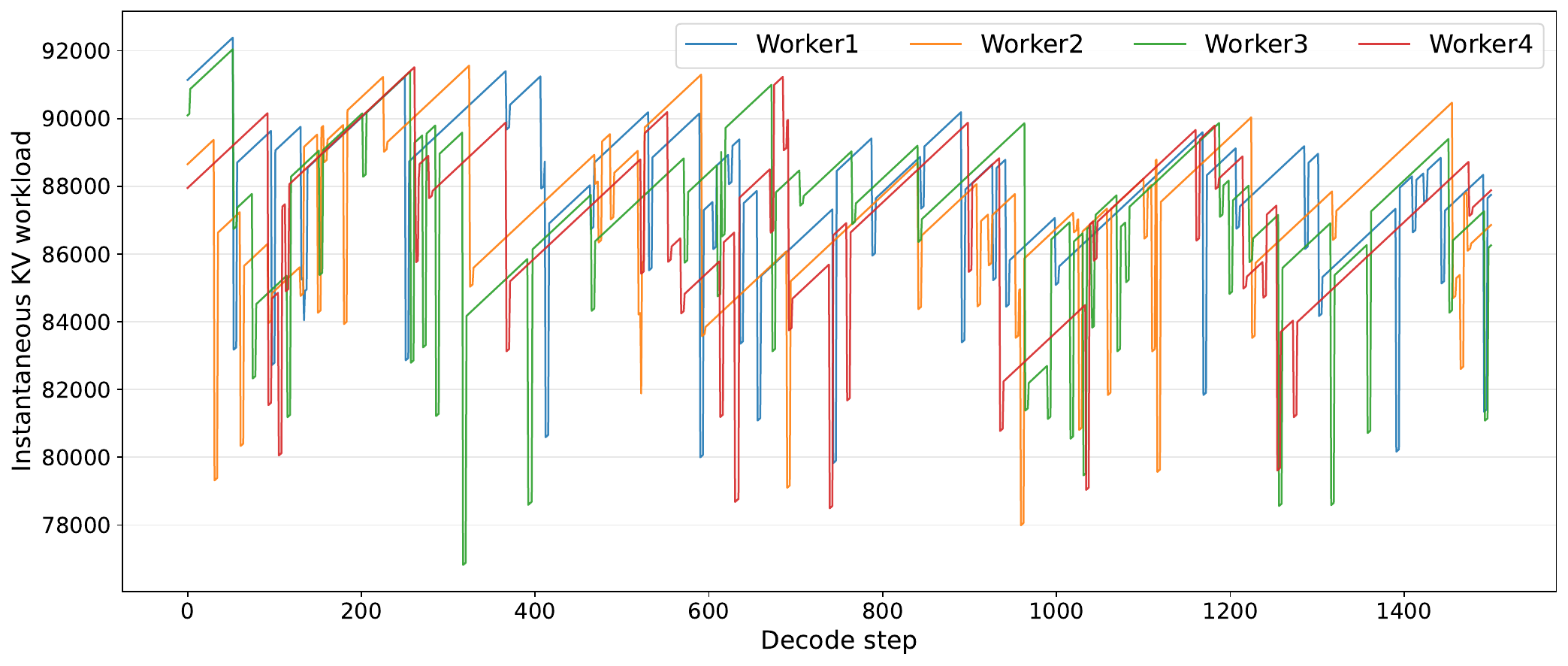}
    \caption*{\centering\footnotesize \textbf{BR-H} (oracle, $H{=}80$) \\ \textit{imbal.\ 6.6k}}
\end{subfigure}
\hspace{0.48\linewidth}
\hspace{0.48\linewidth}
 
\caption{\textbf{Per-worker KV-cache traces on Proprietary Data at $G{=}4$ (2P1D, 48 NPUs).} The four DP workers' KV-cache footprints over a $1{,}500$-step decode segment under each method; trace-mean imbalance under each panel matches the $G{=}4$ column of Table~\ref{tab:scaling_summary}.}
\label{fig:scaling-traces-g4}
\end{figure}

\begin{figure}[htbp]
\centering
\begin{subfigure}[t]{0.48\linewidth}
    \centering
    \includegraphics[width=\linewidth]{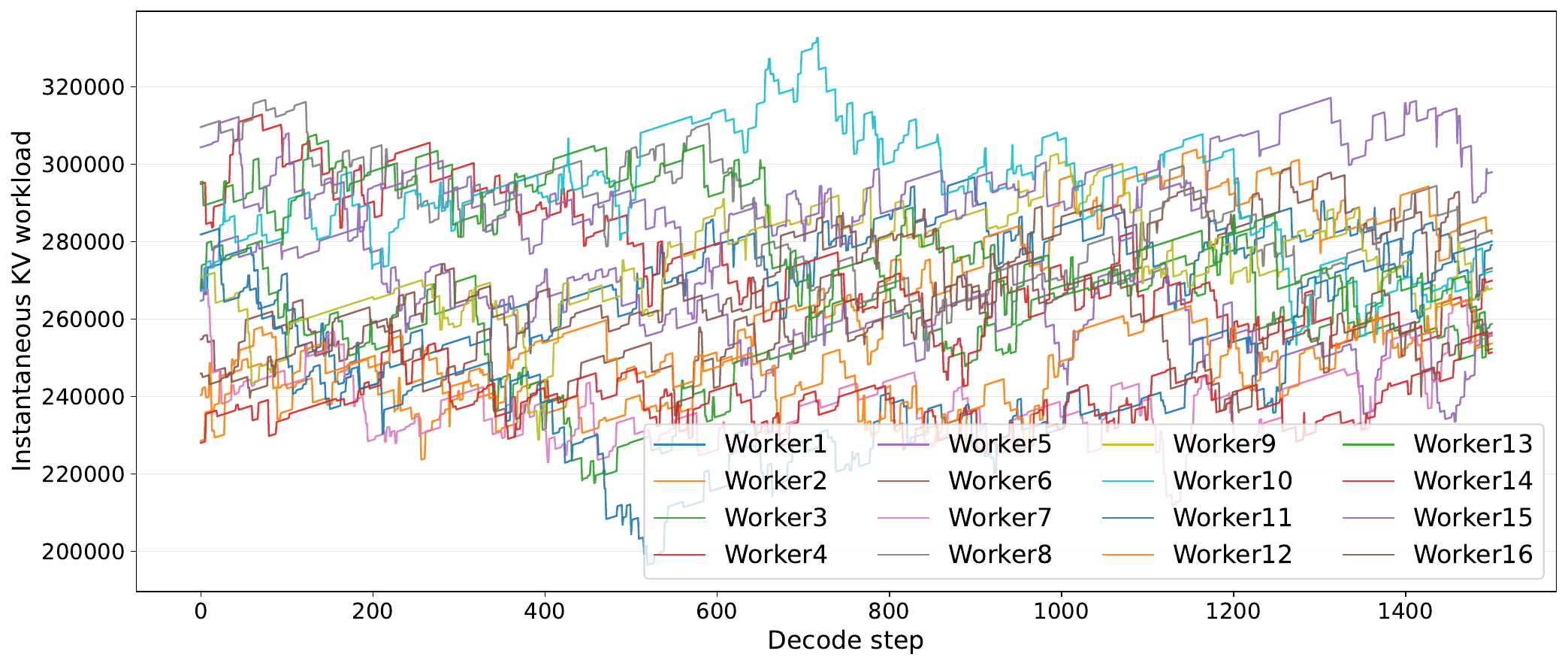}
    \caption*{\centering\footnotesize Random \\ \textit{imbal.\ 638k}}
\end{subfigure}\hfill
\begin{subfigure}[t]{0.48\linewidth}
    \centering
    \includegraphics[width=\linewidth]{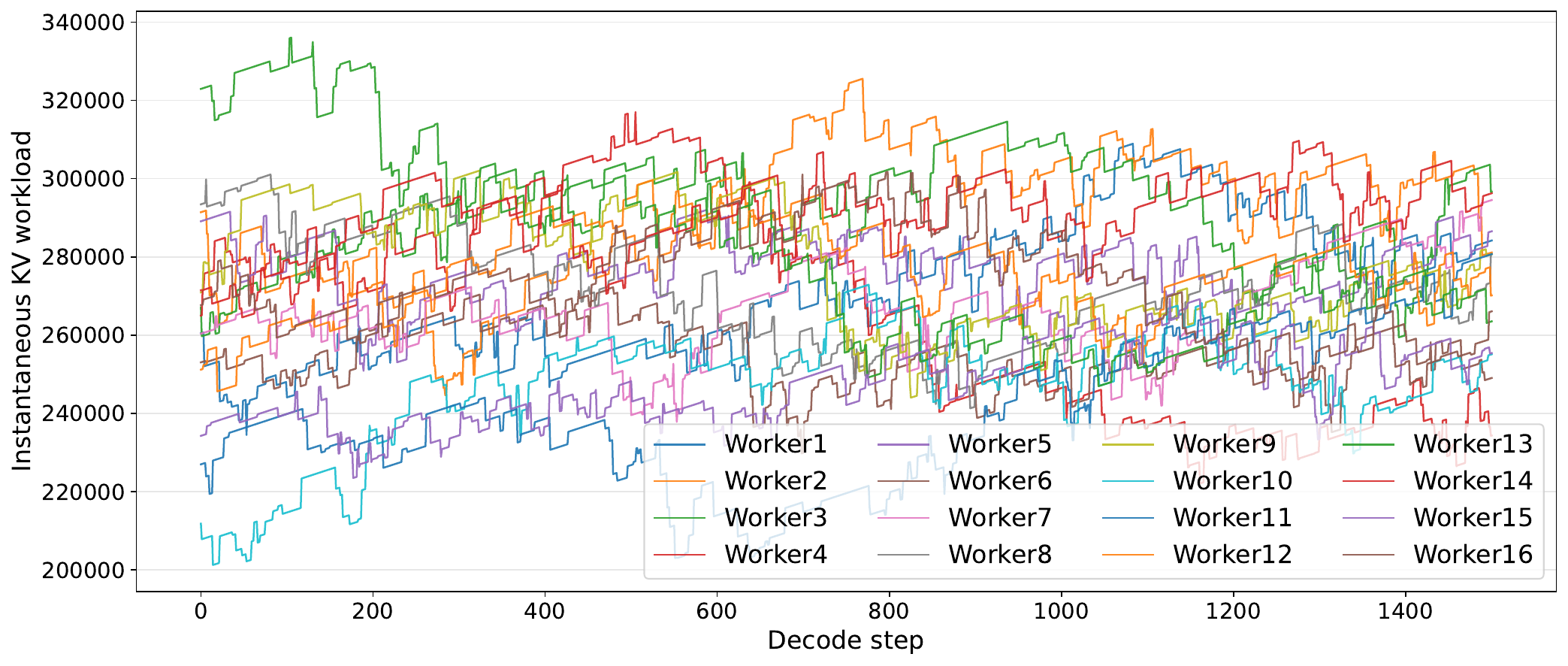}
    \caption*{\centering\footnotesize Round-Robin \\ \textit{imbal.\ 616k}}
\end{subfigure}\hfill
\begin{subfigure}[t]{0.48\linewidth}
    \centering
    \includegraphics[width=\linewidth]{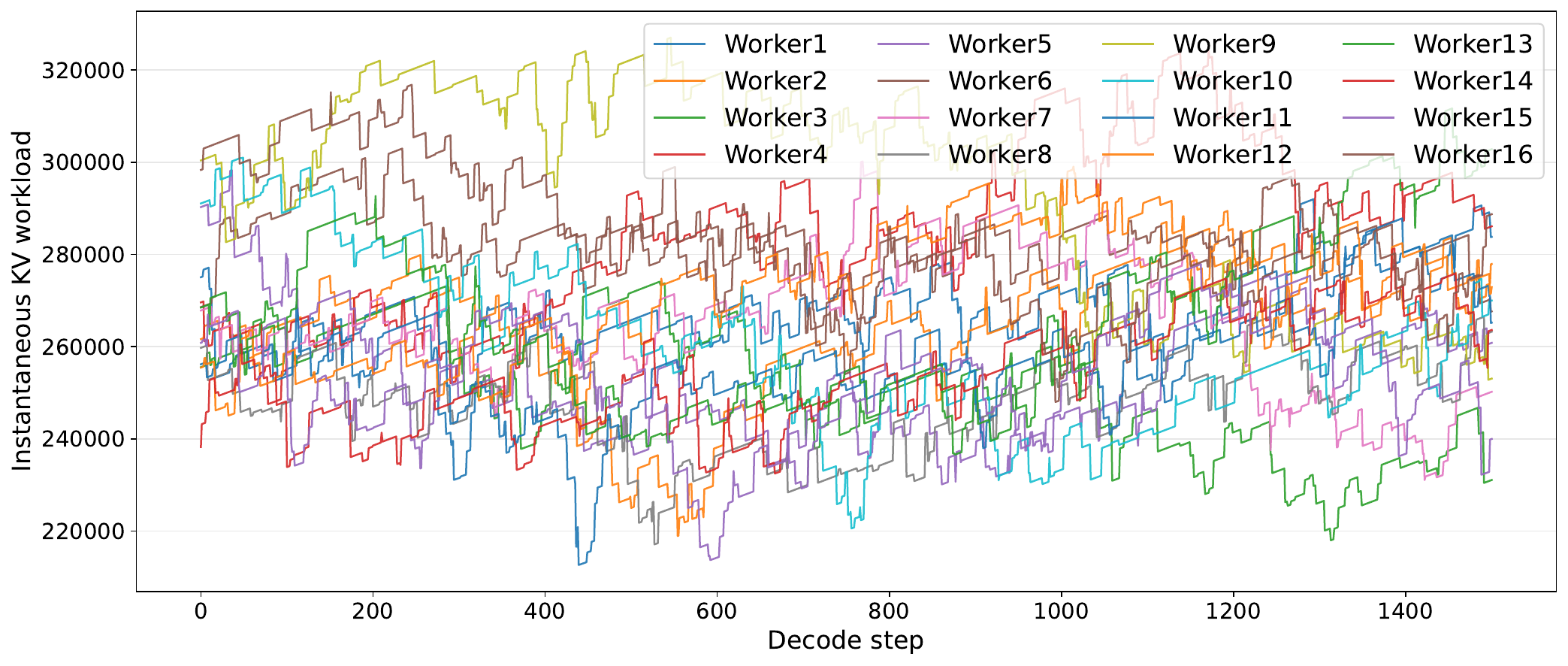}
    \caption*{\centering\footnotesize P2C \\ \textit{imbal.\ 688k}}
\end{subfigure}\hfill
\begin{subfigure}[t]{0.48\linewidth}
    \centering
    \includegraphics[width=\linewidth]{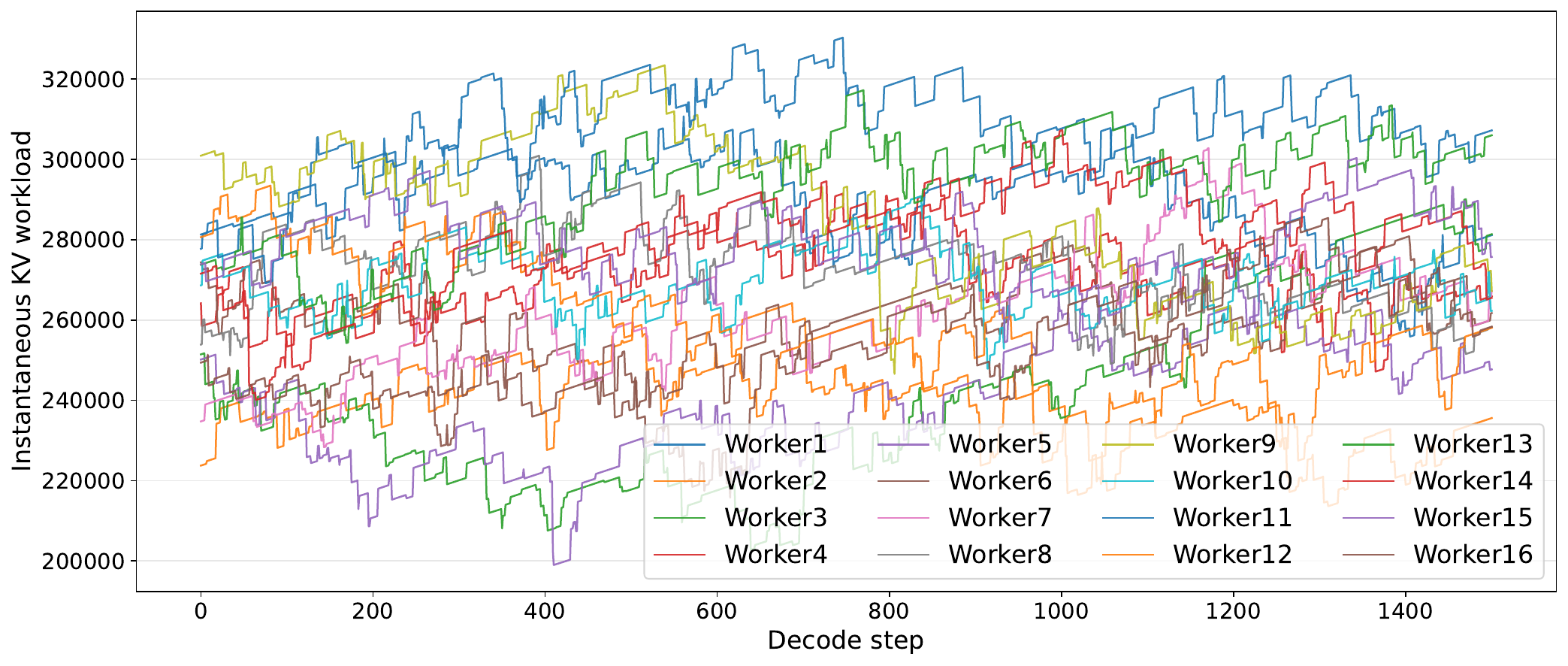}
    \caption*{\centering\footnotesize JSQ \\ \textit{imbal.\ 676k}}
\end{subfigure}
 
\vspace{0.7em}
 
\begin{subfigure}[t]{0.48\linewidth}
    \centering
    \includegraphics[width=\linewidth]{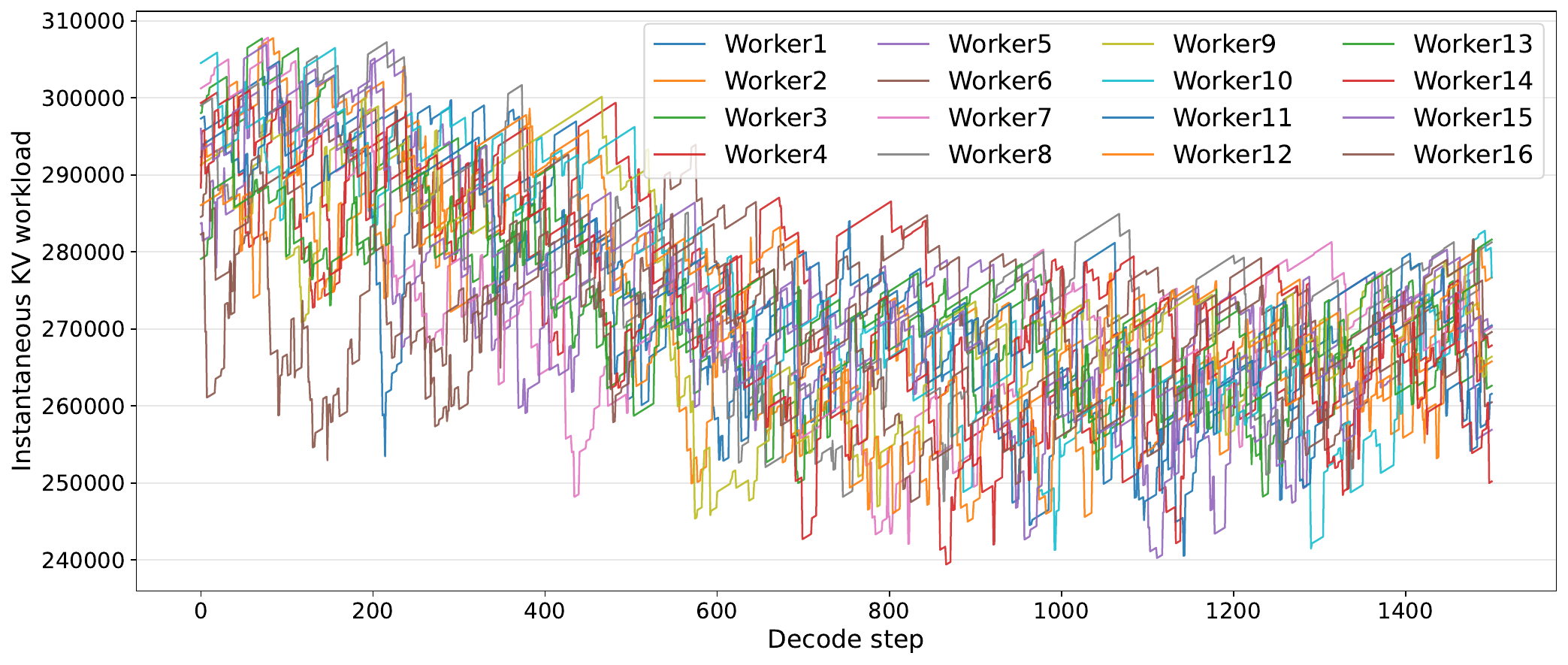}
    \caption*{\centering\footnotesize \textbf{BR-0} (ours) \\ \textit{imbal.\ 215k}}
\end{subfigure}\hfill
\begin{subfigure}[t]{0.48\linewidth}
    \centering
    \includegraphics[width=\linewidth]{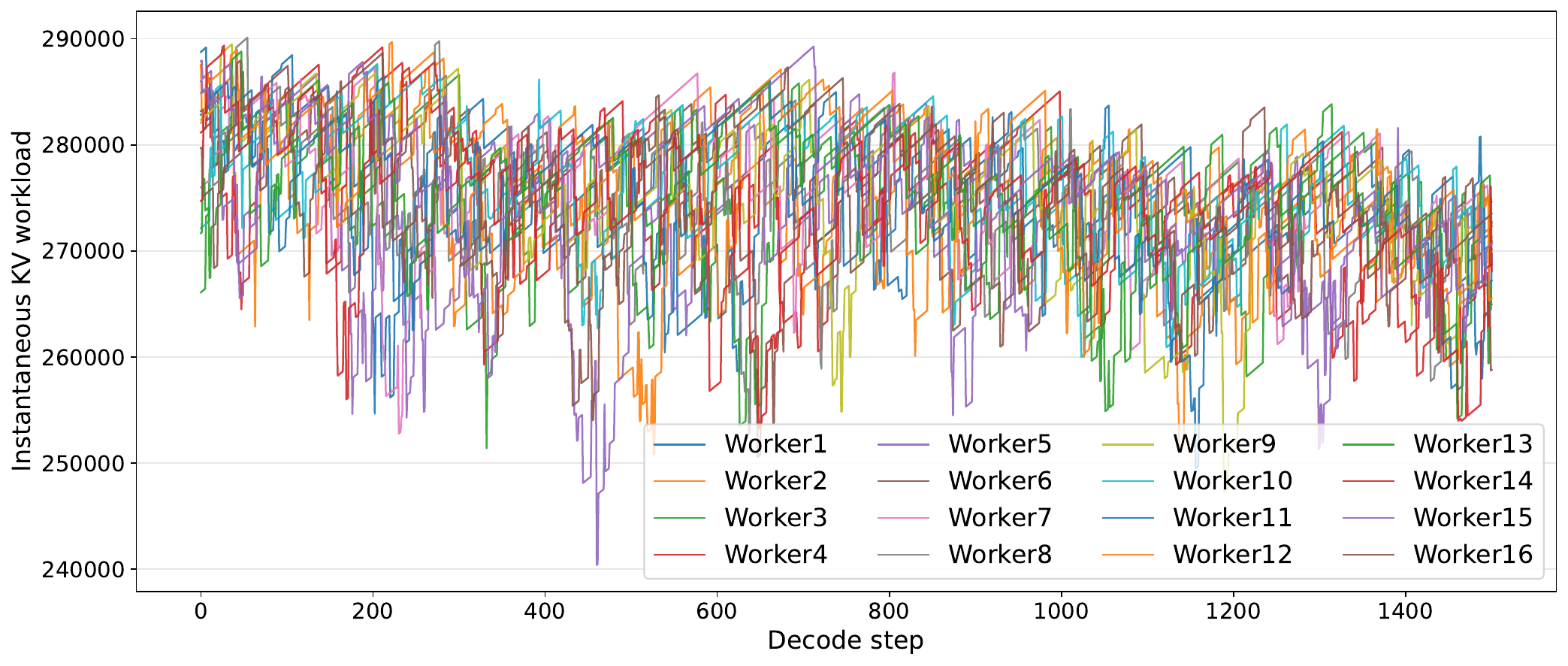}
    \caption*{\centering\footnotesize \textbf{BR-H} oracle \\ $\beta{=}43, \gamma{=}0.86$ \\ \textit{imbal.\ 124k}}
\end{subfigure}\hfill
\begin{subfigure}[t]{0.48\linewidth}
    \centering
    \includegraphics[width=\linewidth]{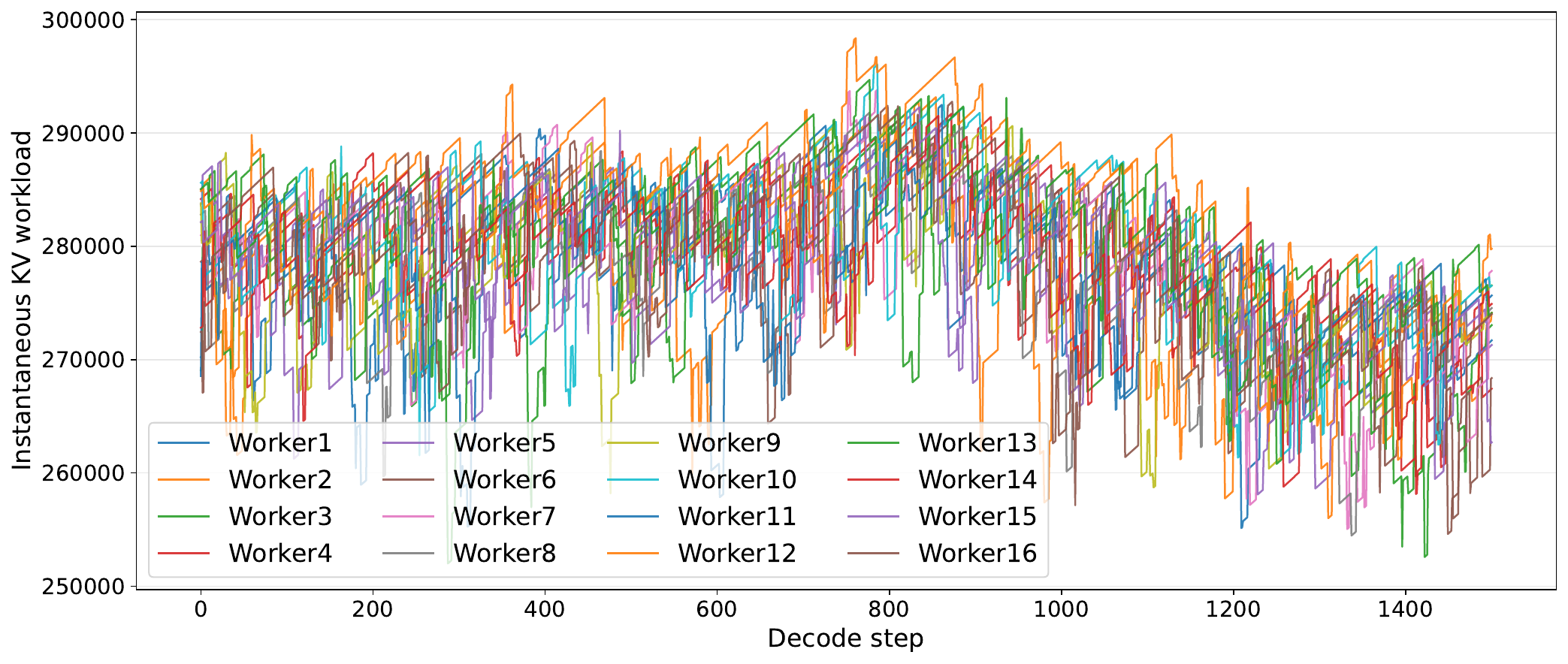}
    \caption*{\centering\footnotesize \textbf{BR-H} oracle \emph{(main)} \\ $\beta{=}14.67, \gamma{=}0.64$ \\ \textit{imbal.\ 117k}}
\end{subfigure}\hfill
\hspace{0.48\linewidth}
 
\caption{\textbf{Per-worker KV-cache traces on Proprietary Data at $G{=}16$ (5P1D, 144 NPUs).} Layout matches Figure~\ref{fig:scaling-traces-g4}; trace-mean imbalance under each panel matches the $G{=}16$ column of Table~\ref{tab:scaling_summary}. Both oracle BR-H operating points are shown for direct comparison; $(\beta, \gamma){=}(14.67, 0.64)$ is the configuration plotted in the main-text Figure~\ref{fig:scaling}.}
\label{fig:scaling-traces-g16-main}
\section{Trace-level evidence for the sensitivity sweep at \texorpdfstring{$G \in \{8, 16\}$}{G in {8, 16}}}
\label{app:sensitivity-traces}
\end{figure}

This appendix complements Section~\ref{subsec:exp_scaling} with per-cluster-size headline numbers (Tables~\ref{tab:scaling_g4_app} and~\ref{tab:scaling_g16_app}) and per-worker KV-workload trace galleries (Figures~\ref{fig:scaling-traces-g4} and~\ref{fig:scaling-traces-g16-main}) for the two cluster sizes outside the $G{=}8$ main configuration: $G{=}4$ on 2P1D and $G{=}16$ on 5P1D. The trace counterpart at $G{=}8$ is Figure~\ref{fig:kv-workload-grid} in the main text. Each $G$'s table and figure are co-located here so that the appendix is self-contained; the same numbers also appear in compressed form in Table~\ref{tab:scaling_summary} of the main text.
 
\paragraph{On the choice of oracle prediction.}
Section~\ref{subsec:exp_scaling} reports BR-H with oracle prediction at every $G$ in order to isolate the router's scaling behaviour from any confounding effect of predictor quality. The deployed (predictor-driven) BR-H is characterised against its oracle counterpart at $G{=}8$ in Section~\ref{subsec:exp_main} on the main Proprietary Data configuration; the per-worker traces at $G{=}4$ and $G{=}16$ shown below are the corresponding oracle traces only, since we did not run the deployed predictor at the other two cluster sizes. The widening of the BR-H advantage with $G$ reported in Section~\ref{subsec:exp_scaling} is a structural consequence of the order-statistics gap (Section~\ref{subsec:dp_decode}) and is captured by the router itself; quantifying how predictor noise interacts with $G$ is left to future work.
 
\paragraph{Headline numbers and traces at $G{=}4$.}
Table~\ref{tab:scaling_g4_app} reports the average imbalance, TPOT P95, and throughput of every method at the smallest cluster size in the sweep ($G{=}4$ on 2P1D, 48 NPUs); Figure~\ref{fig:scaling-traces-g4} shows the corresponding per-worker KV-cache traces. The four standard baselines fall in the $64$--$80$k imbalance range and around $365$\,tok/s throughput; BR-0 reaches $9.7$k imbalance and $407$\,tok/s; BR-H with oracle prediction at $H{=}80$ reaches $6.6$k imbalance and $415$\,tok/s, the lowest TPOT P95 in the table.
 
\begin{table}[h]
\centering
\caption{\textbf{All methods on Proprietary Data at $G{=}4$ (2P1D, 48 NPUs).} Reproduces the $G{=}4$ column of Table~\ref{tab:scaling_summary} as a self-contained companion to Figure~\ref{fig:scaling-traces-g4}. Best in each column in \textbf{bold}.}
\label{tab:scaling_g4_app}
\begin{tabular}{lccc}
\toprule
Method & Avg.\ imbalance $\downarrow$ & TPOT P95 (ms) $\downarrow$ & Throughput (tok/s) $\uparrow$ \\
\midrule
Random       &  80{,}118 & 66.30 & 364.7 \\
Round-Robin  &  68{,}108 & 65.27 & 365.7 \\
P2C          &  64{,}012 & 65.40 & 366.4 \\
JSQ          &  75{,}326 & 66.45 & 365.4 \\
\midrule
BR-0         &   9{,}711 & 63.19 & 407.1 \\
BR-H \emph{oracle}, $H{=}80$ & \textbf{6{,}576} & \textbf{62.52} & \textbf{415.3} \\
\bottomrule
\end{tabular}
\end{table}
 
\paragraph{Headline numbers and traces at $G{=}16$.}
Table~\ref{tab:scaling_g16_app} reports the analogous numbers at the largest cluster size in the sweep ($G{=}16$ on 5P1D, 144 NPUs), including both oracle BR-H operating points used in the main paper; Figure~\ref{fig:scaling-traces-g16-main} shows the corresponding traces. The four baselines fall in the $616$--$688$k imbalance range and around $926$\,tok/s throughput; BR-0 reaches $215$k imbalance and $1{,}118$\,tok/s; the two BR-H oracle operating points reach $117$--$124$k imbalance and $1{,}247$--$1{,}251$\,tok/s, with $(\beta, \gamma){=}(14.67, 0.64)$ achieving the lowest imbalance and lowest TPOT P95, and $(43, 0.86)$ marginally higher throughput by $\sim 4$\,tok/s. We use $(14.67, 0.64)$ as the representative BR-H point in the main-text Figure~\ref{fig:scaling}.

\begin{table}[htbp]
\centering
\caption{\textbf{All methods on Proprietary Data at $G{=}16$ (5P1D, 144 NPUs).} Reproduces the $G{=}16$ column of Table~\ref{tab:scaling_summary} as a self-contained companion to Figure~\ref{fig:scaling-traces-g16-main}, with both oracle BR-H operating points reported. Best in each column in \textbf{bold}.}
\label{tab:scaling_g16_app}
\begin{tabular}{lccc}
\toprule
Method & Avg.\ imbalance $\downarrow$ & TPOT P95 (ms) $\downarrow$ & Throughput (tok/s) $\uparrow$ \\
\midrule
Random       & 638{,}062 & 77.80 &  923.6 \\
Round-Robin  & 616{,}089 & 77.91 &  925.3 \\
P2C          & 688{,}283 & 77.46 &  927.3 \\
JSQ          & 675{,}607 & 77.40 &  926.6 \\
\midrule
BR-0                                                            & 215{,}170 & 74.92 & 1{,}117.8 \\
BR-H \emph{oracle}, $(43, 0.86)$              & 123{,}700 & 72.15 & \textbf{1{,}251.1} \\
BR-H \emph{oracle}, $(14.67, 0.64)$ & \textbf{117{,}067} & \textbf{72.04} & 1{,}247.4 \\
\bottomrule
\end{tabular}
\end{table}
 
This appendix complements Section~\ref{subsec:exp_robust_overhead} with per-worker KV-workload traces under each swept $(\beta, \gamma)$ configuration on Proprietary Data at the two cluster sizes for which we ran a sweep: $G{=}8$ (4P1D, the main configuration) and $G{=}16$ (5P1D, the largest configuration). Each sub-panel of the trace galleries plots all $G$ DP workers' instantaneous KV-cache footprints simultaneously over the same $1{,}500$-step decode segment used in Section~\ref{subsec:exp_main}; the vertical spread of the curves is the instantaneous imbalance.
 
\paragraph{Sensitivity at $G{=}8$.}
Table~\ref{tab:sensitivity_sweep_g8_app} provides the cross-shaped sweep around $(\beta{=}48, \gamma{=}0.9)$; we display it here for reference next to its trace gallery. The sweep covers $\beta \in \{1, 24, 48, 96\}$ at $\gamma{=}0.9$ and $\gamma \in \{0.5, 0.7, 0.9, 1.0\}$ at $\beta{=}48$. Trace-mean imbalance ranges from $24.7$k to $32.3$k tokens across the swept region; throughput from $945$ to $1{,}042$ tok/s; TPOT P95 from $77.3$ to $79.5$\,ms. By comparison, every standard baseline at $G{=}8$ on Proprietary Data (Table~\ref{tab:main_combined}) sits at $215$--$438$k imbalance and $810$--$847$ tok/s throughput---an order-of-magnitude imbalance gap and a $> 100$\,tok/s throughput gap. Every swept configuration sits well inside the high-throughput, low-imbalance regime.
 
\begin{table}[htbp]
\centering
\caption{\textbf{$(\beta, \gamma)$ sensitivity sweep at $G{=}8$ (4P1D, 96 NPUs) on Proprietary Data, BR-H oracle, $H{=}80$.} Cross-shaped sweep around $(\beta{=}48, \gamma{=}0.9)$.}
\label{tab:sensitivity_sweep_g8_app}
\begin{tabular}{lccccc}
\toprule
Sweep & $\beta$ & $\gamma$ & Avg.\ imbalance $\downarrow$ & TPOT P95 (ms) $\downarrow$ & Throughput (tok/s) $\uparrow$ \\
\midrule
\emph{$\beta$ at $\gamma{=}0.9$} &  1 & 0.9 & 32{,}304 & 78.120 &  986.41 \\
                                 & 24 & 0.9 & 24{,}706 & 77.933 & 1{,}022.67 \\
                                 & 48 & 0.9 & 25{,}608 & 77.517 & 1{,}042.61 \\
                                 & 96 & 0.9 & 31{,}862 & 79.460 &  945.41 \\
\midrule
\emph{$\gamma$ at $\beta{=}48$}  & 48 & 0.5 & 28{,}327 & 78.040 & 1{,}024.04 \\
                                 & 48 & 0.7 & 26{,}599 & 77.611 & 1{,}020.78 \\
                                 & 48 & 0.9 & 25{,}608 & 77.517 & 1{,}042.61 \\
                                 & 48 & 1.0 & 29{,}318 & 77.310 & 1{,}026.77 \\
\bottomrule
\end{tabular}
\end{table}
 
\begin{figure}[htbp]
\centering
\begin{subfigure}[t]{0.48\linewidth}
    \centering
    \includegraphics[width=\linewidth]{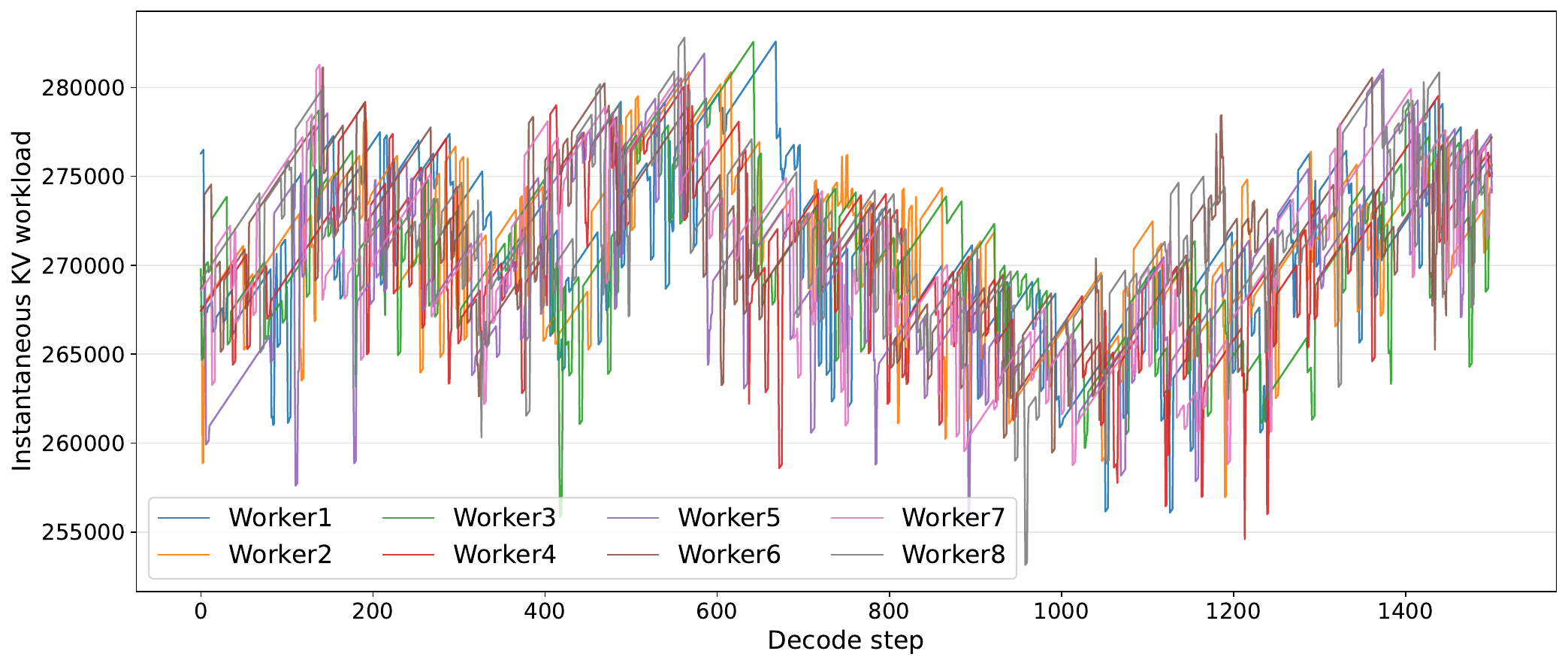}
    \caption*{\centering\footnotesize $\beta{=}1,\,\gamma{=}0.9$ \\ \textit{imbal.\ 32.3k}}
\end{subfigure}\hfill
\begin{subfigure}[t]{0.48\linewidth}
    \centering
    \includegraphics[width=\linewidth]{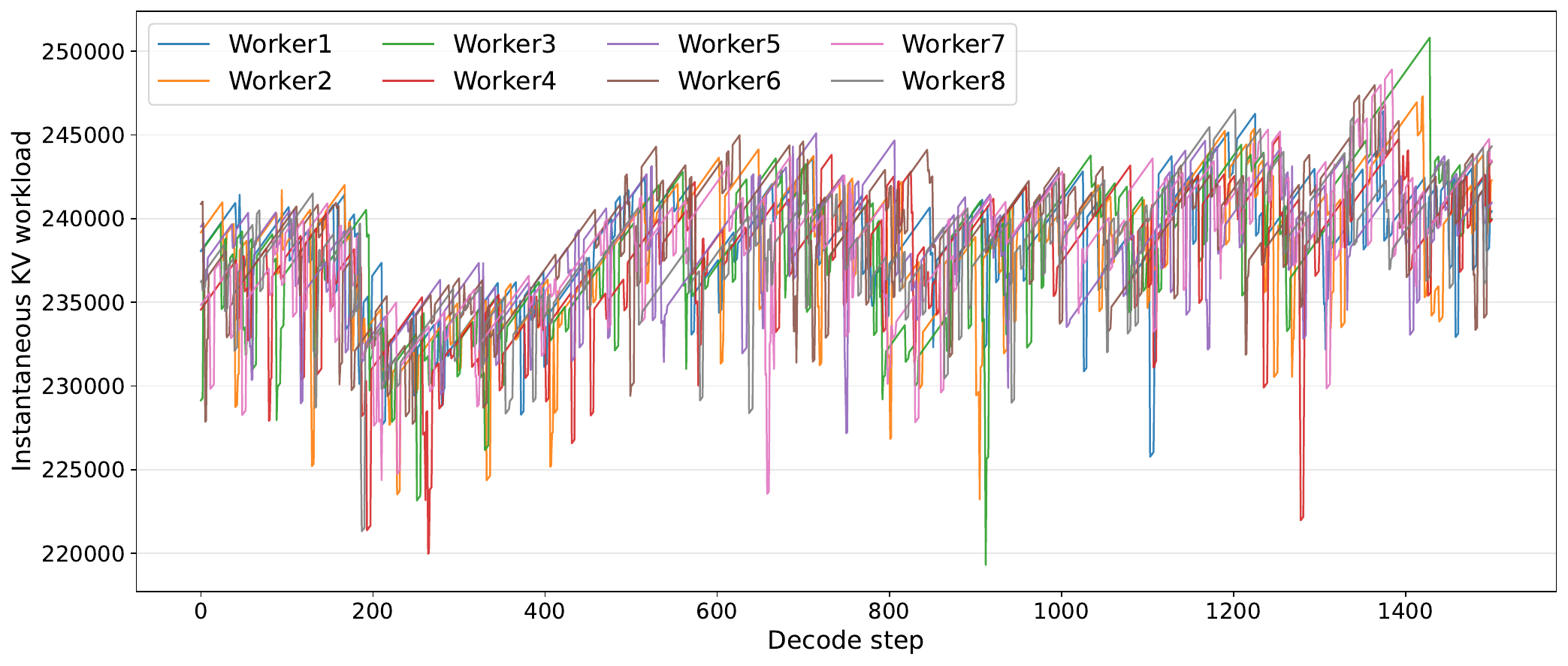}
    \caption*{\centering\footnotesize $\beta{=}24,\,\gamma{=}0.9$ \\ \textit{imbal.\ 24.7k}}
\end{subfigure}\hfill
\begin{subfigure}[t]{0.48\linewidth}
    \centering
    \includegraphics[width=\linewidth]{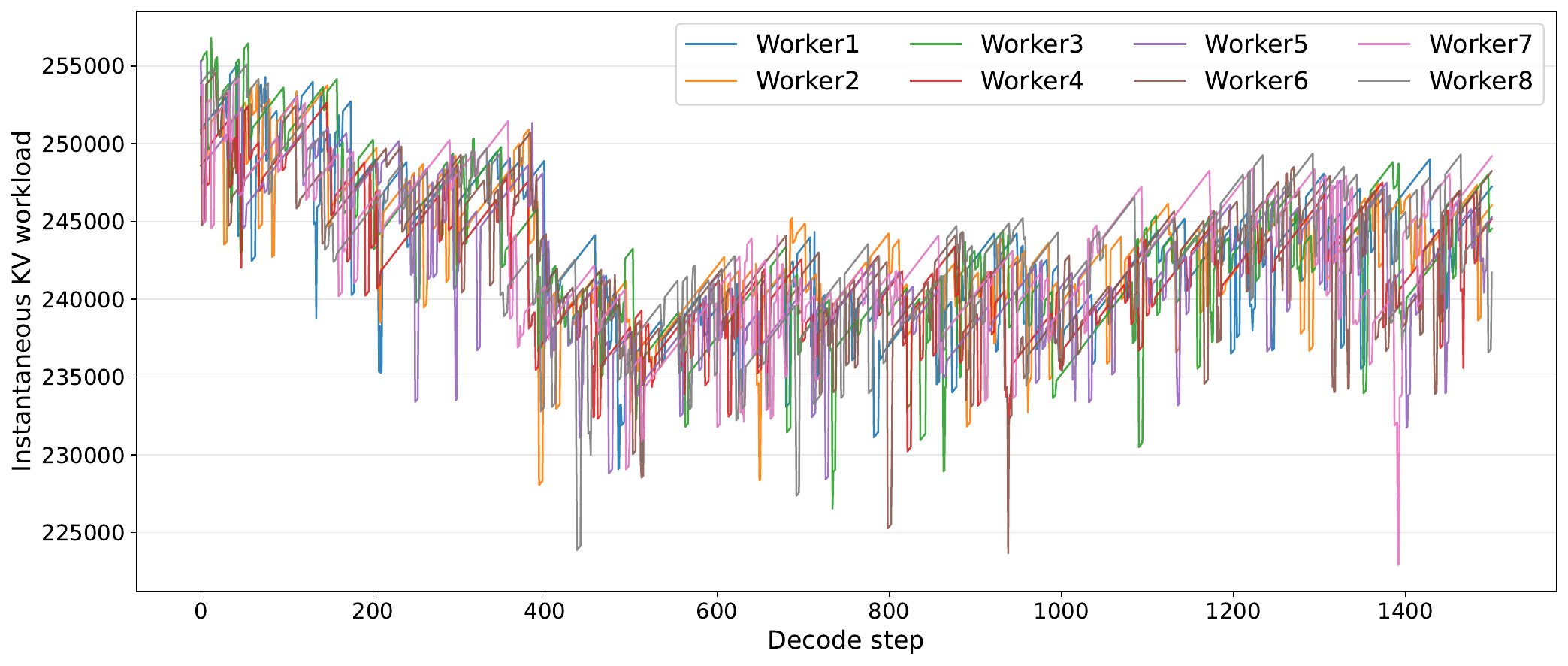}
    \caption*{\centering\footnotesize $\beta{=}48,\,\gamma{=}0.9$ \\ \textit{imbal.\ 25.6k}}
\end{subfigure}\hfill
\begin{subfigure}[t]{0.48\linewidth}
    \centering
    \includegraphics[width=\linewidth]{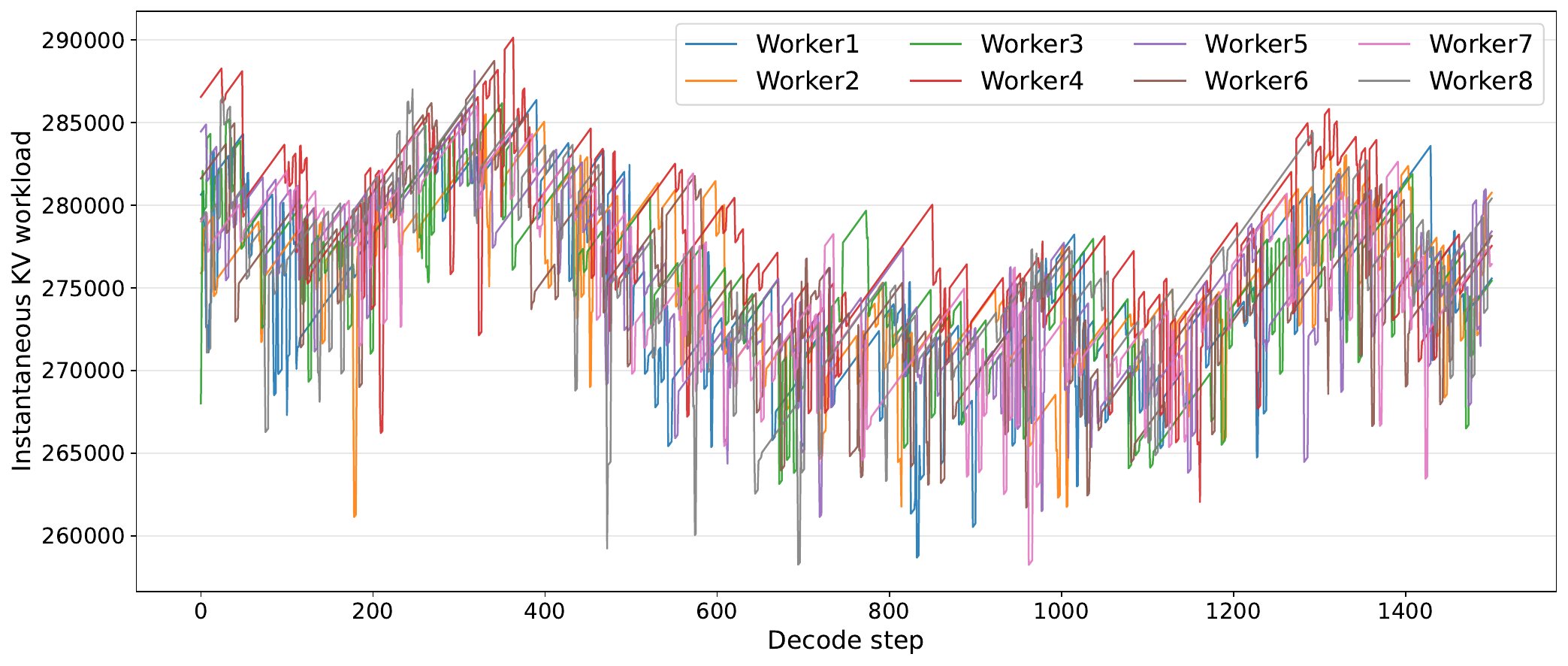}
    \caption*{\centering\footnotesize $\beta{=}96,\,\gamma{=}0.9$ \\ \textit{imbal.\ 31.9k}}
\end{subfigure}
 
\vspace{0.7em}
 
\begin{subfigure}[t]{0.48\linewidth}
    \centering
    \includegraphics[width=\linewidth]{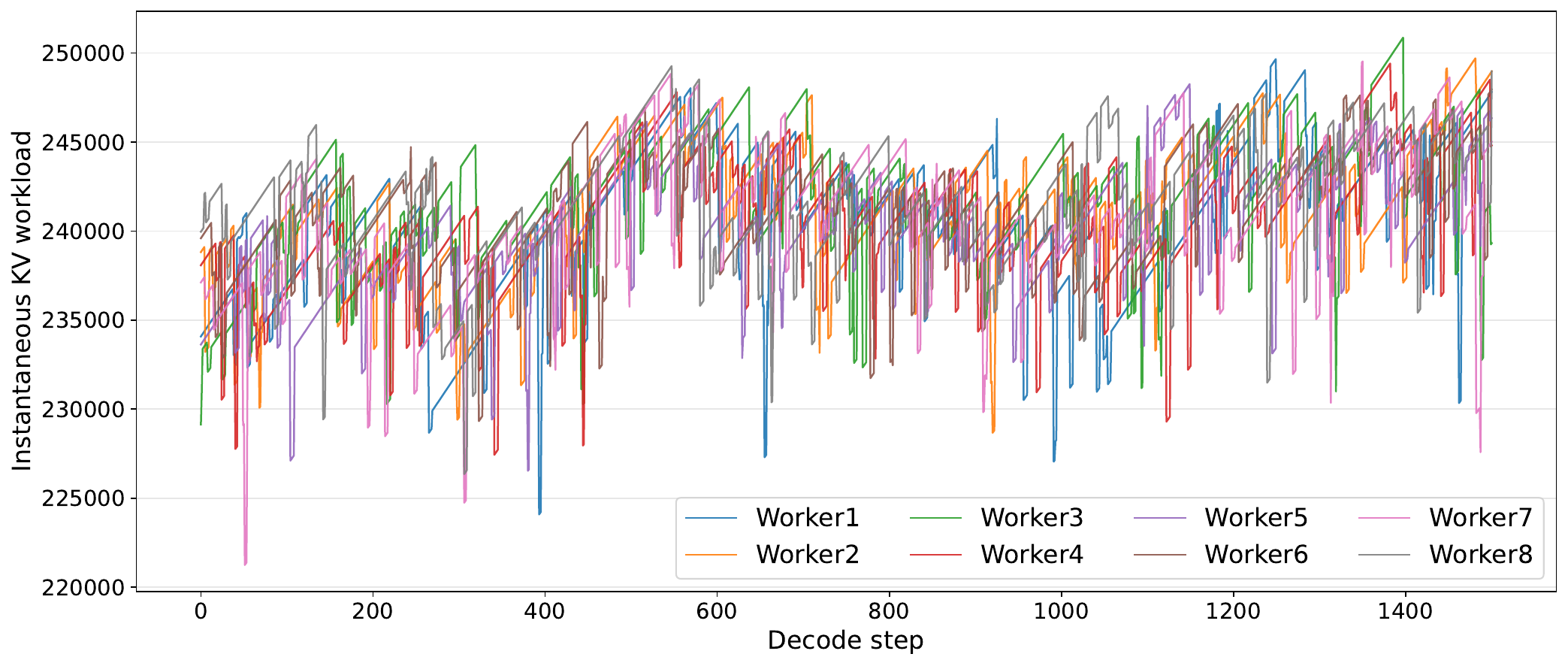}
    \caption*{\centering\footnotesize $\beta{=}48,\,\gamma{=}0.5$ \\ \textit{imbal.\ 28.3k}}
\end{subfigure}\hfill
\begin{subfigure}[t]{0.48\linewidth}
    \centering
    \includegraphics[width=\linewidth]{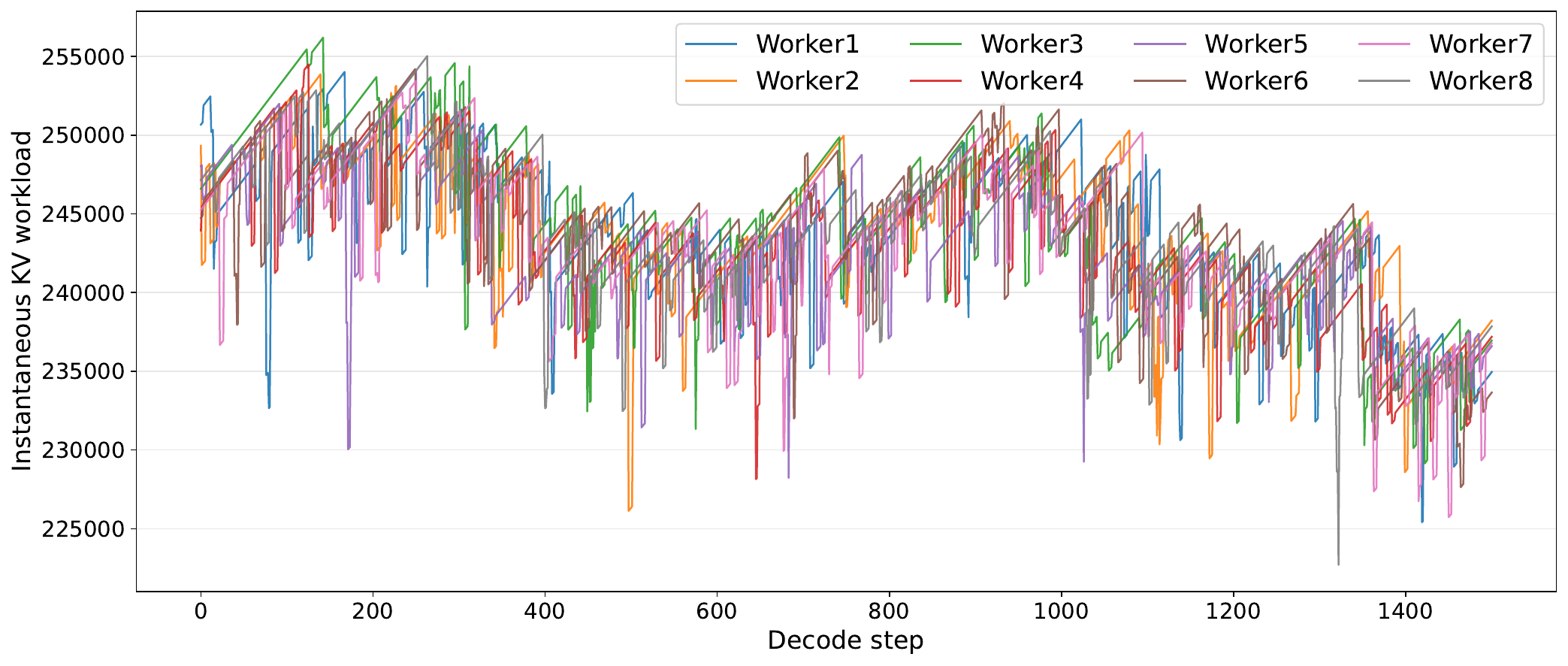}
    \caption*{\centering\footnotesize $\beta{=}48,\,\gamma{=}0.7$ \\ \textit{imbal.\ 26.6k}}
\end{subfigure}\hfill
\begin{subfigure}[t]{0.48\linewidth}
    \centering
    \includegraphics[width=\linewidth]{plots/Appendix/sensitivity/kv_4p1d_b48_gamma0.9.pdf}
    \caption*{\centering\footnotesize $\beta{=}48,\,\gamma{=}0.9$ \\ \textit{imbal.\ 25.6k}}
\end{subfigure}\hfill
\begin{subfigure}[t]{0.48\linewidth}
    \centering
    \includegraphics[width=\linewidth]{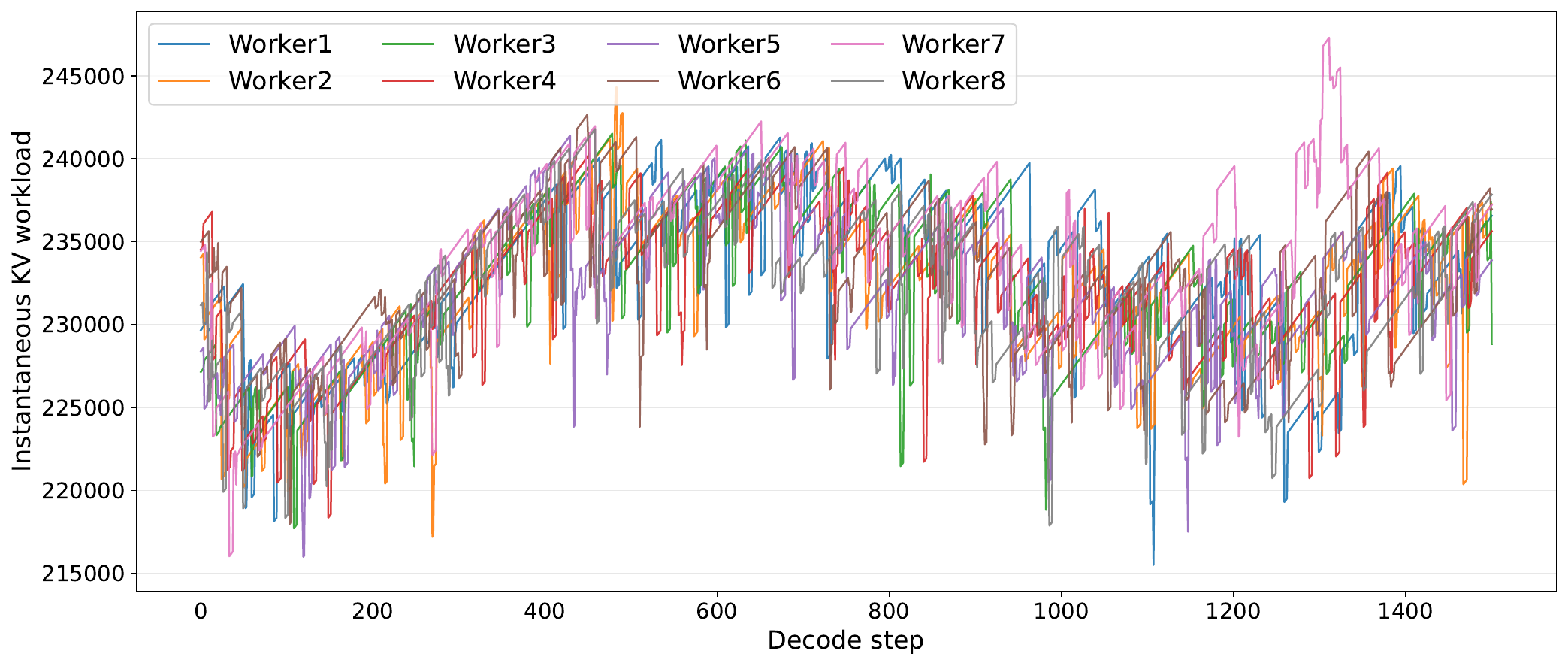}
    \caption*{\centering\footnotesize $\beta{=}48,\,\gamma{=}1.0$ \\ \textit{imbal.\ 29.3k}}
\end{subfigure}
 
\caption{\textbf{Per-worker KV-cache traces under each swept $(\beta, \gamma)$ on Proprietary Data at $G{=}8$, BR-H oracle, $H{=}80$.} Each panel is annotated with its $(\beta, \gamma)$ and trace-mean imbalance. The configuration $(\beta{=}48, \gamma{=}0.9)$ is shared between the $\beta$- and $\gamma$-sweep blocks and therefore appears in both. Per-panel $y$-axes are independently autoscaled.}
\label{fig:sensitivity-traces-g8}
\end{figure}

\begin{figure}[htbp]
\centering
\includegraphics[width=0.92\linewidth]{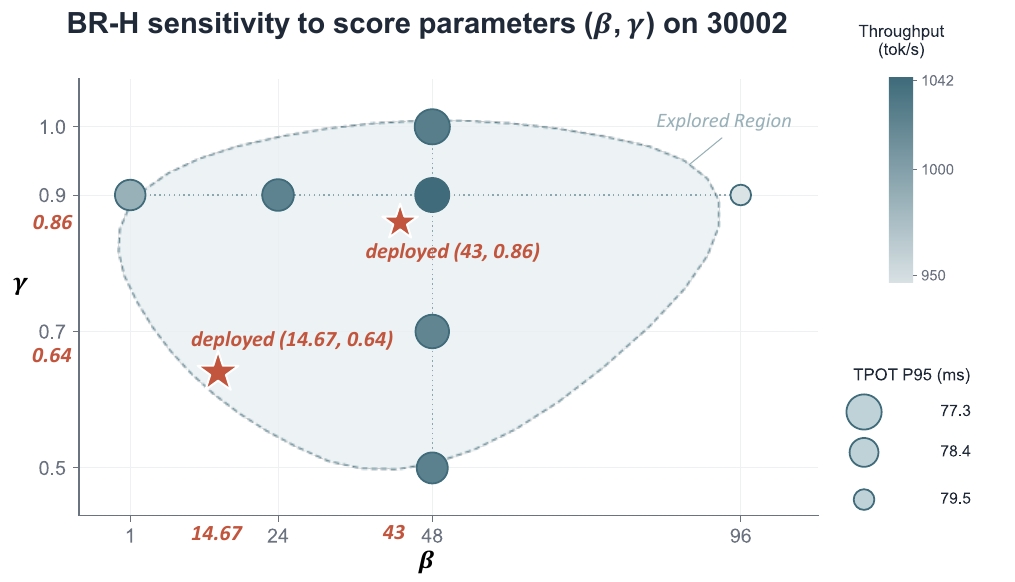}
\caption{\textbf{BR-H sensitivity to $(\beta, \gamma)$ on Proprietary Data, oracle, $H{=}80$.} Marker color: throughput (darker $=$ higher). Marker size: $1/\textsc{TPOT P95}$ (larger $=$ faster). Circles: 7 swept configurations. Stars: 2 deployed configurations.}
\label{fig:betagamma}
\end{figure}

Figure~\ref{fig:betagamma} reports BR-H sensitivity to $(\beta, \gamma)$ in~\eqref{eq:brh-Fscore} on Proprietary Data under oracle prediction at $H{=}80$. We sweep the parameter plane along two axes intersecting at $(\beta{=}48, \gamma{=}0.9)$: $\beta \in \{1, 24, 48, 96\}$ at fixed $\gamma{=}0.9$, and $\gamma \in \{0.5, 0.7, 0.9, 1.0\}$ at fixed $\beta{=}48$, giving a cross-shaped sweep of seven configurations.

\paragraph{Sensitivity at $G{=}16$.}
Table~\ref{tab:sensitivity_sweep_g16_app} reports the analogous cross-shaped sweep at $G{=}16$ (5P1D), again under BR-H oracle prediction with $H{=}80$. The same seven $(\beta, \gamma)$ operating points are evaluated. Trace-mean imbalance ranges from $104$k to $127$k tokens across the swept region; throughput from $1{,}243$ to $1{,}257$ tok/s; TPOT P95 from $71.1$ to $73.2$\,ms. The relative spread of imbalance (factor $\sim 1.2\times$ across the sweep) and throughput (factor $\sim 1.01\times$ across the sweep) is even tighter than at $G{=}8$, indicating that the robustness pattern observed at $G{=}8$ persists at $G{=}16$. By comparison, every standard baseline at $G{=}16$ on Proprietary Data sits at $616$--$688$k imbalance and $924$--$927$ tok/s throughput (Table~\ref{tab:scaling_summary}); the BR-H sweep is $5$--$6\times$ tighter on imbalance and $\sim 34\%$ higher on throughput than every baseline at this scale.
 
\begin{table}[htbp]
\centering
\caption{\textbf{$(\beta, \gamma)$ sensitivity sweep at $G{=}16$ (5P1D, 144 NPUs) on Proprietary Data, BR-H oracle, $H{=}80$.} Same cross-shaped sweep as Table~\ref{tab:sensitivity_sweep_g8_app}, evaluated at the largest cluster size.}
\label{tab:sensitivity_sweep_g16_app}
\begin{tabular}{lccccc}
\toprule
Sweep & $\beta$ & $\gamma$ & Avg.\ imbalance $\downarrow$ & TPOT P95 (ms) $\downarrow$ & Throughput (tok/s) $\uparrow$ \\
\midrule
\emph{$\beta$ at $\gamma{=}0.9$} &  1 & 0.9 & 104{,}229 & 71.583 & 1{,}256.30 \\
                                 & 24 & 0.9 & 117{,}879 & 72.198 & 1{,}248.51 \\
                                 & 48 & 0.9 & 118{,}396 & 72.610 & 1{,}255.77 \\
                                 & 96 & 0.9 & 112{,}771 & 72.260 & 1{,}256.17 \\
\midrule
\emph{$\gamma$ at $\beta{=}48$}  & 48 & 0.5 & 116{,}074 & 72.227 & 1{,}247.32 \\
                                 & 48 & 0.7 & 109{,}763 & 71.114 & 1{,}256.53 \\
                                 & 48 & 0.9 & 118{,}396 & 72.610 & 1{,}255.77 \\
                                 & 48 & 1.0 & 127{,}207 & 73.167 & 1{,}243.79 \\
\bottomrule
\end{tabular}
\end{table}
 
\begin{figure}[htbp]
\centering
\begin{subfigure}[t]{0.48\linewidth}
    \centering
    \includegraphics[width=\linewidth]{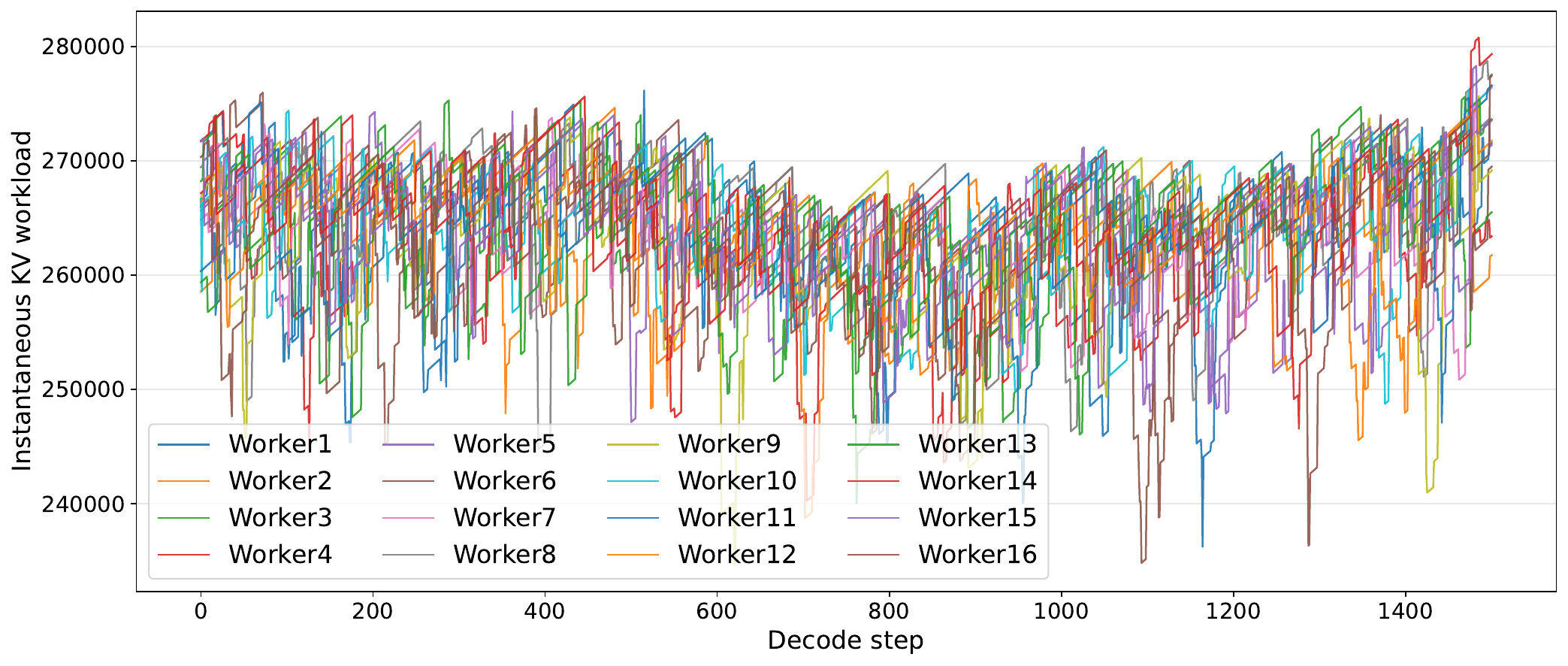}
    \caption*{\centering\footnotesize $\beta{=}1,\,\gamma{=}0.9$ \\ \textit{imbal.\ 104k}}
\end{subfigure}\hfill
\begin{subfigure}[t]{0.48\linewidth}
    \centering
    \includegraphics[width=\linewidth]{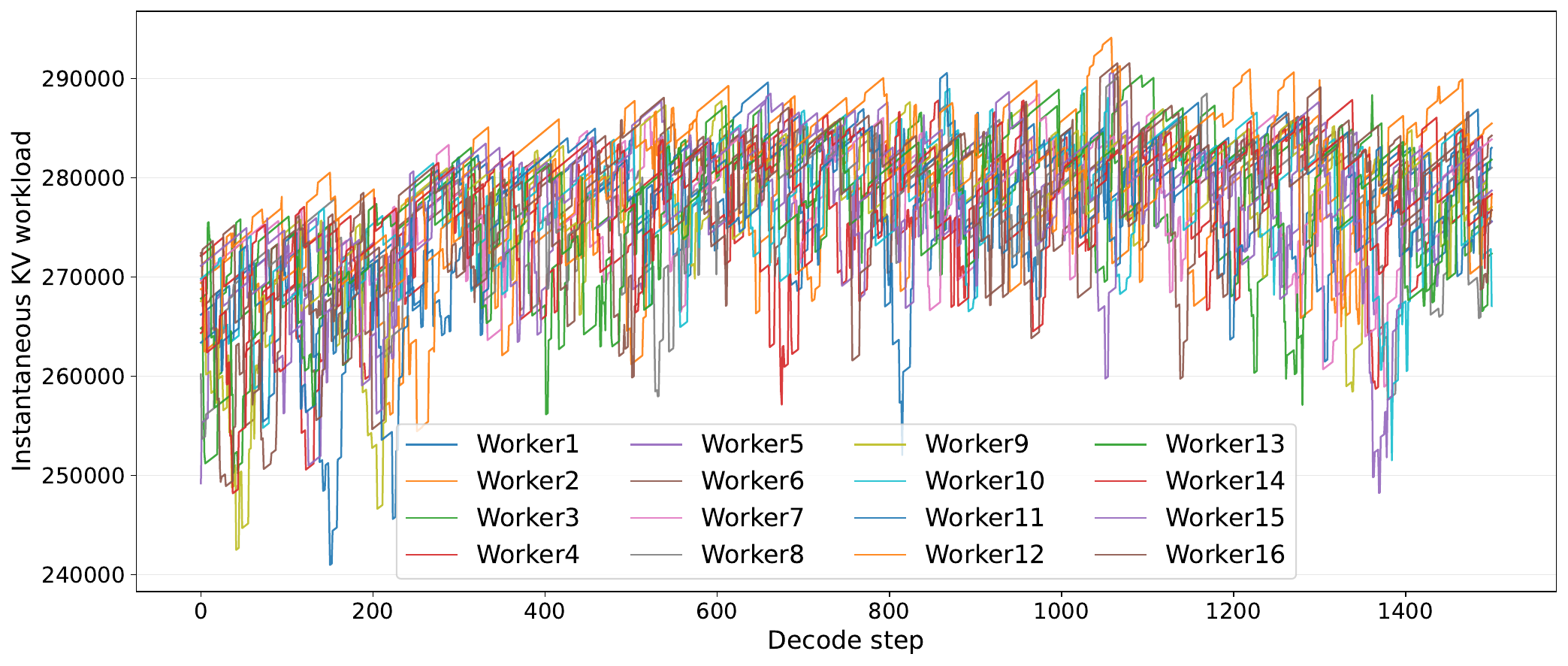}
    \caption*{\centering\footnotesize $\beta{=}24,\,\gamma{=}0.9$ \\ \textit{imbal.\ 118k}}
\end{subfigure}\hfill
\begin{subfigure}[t]{0.48\linewidth}
    \centering
    \includegraphics[width=\linewidth]{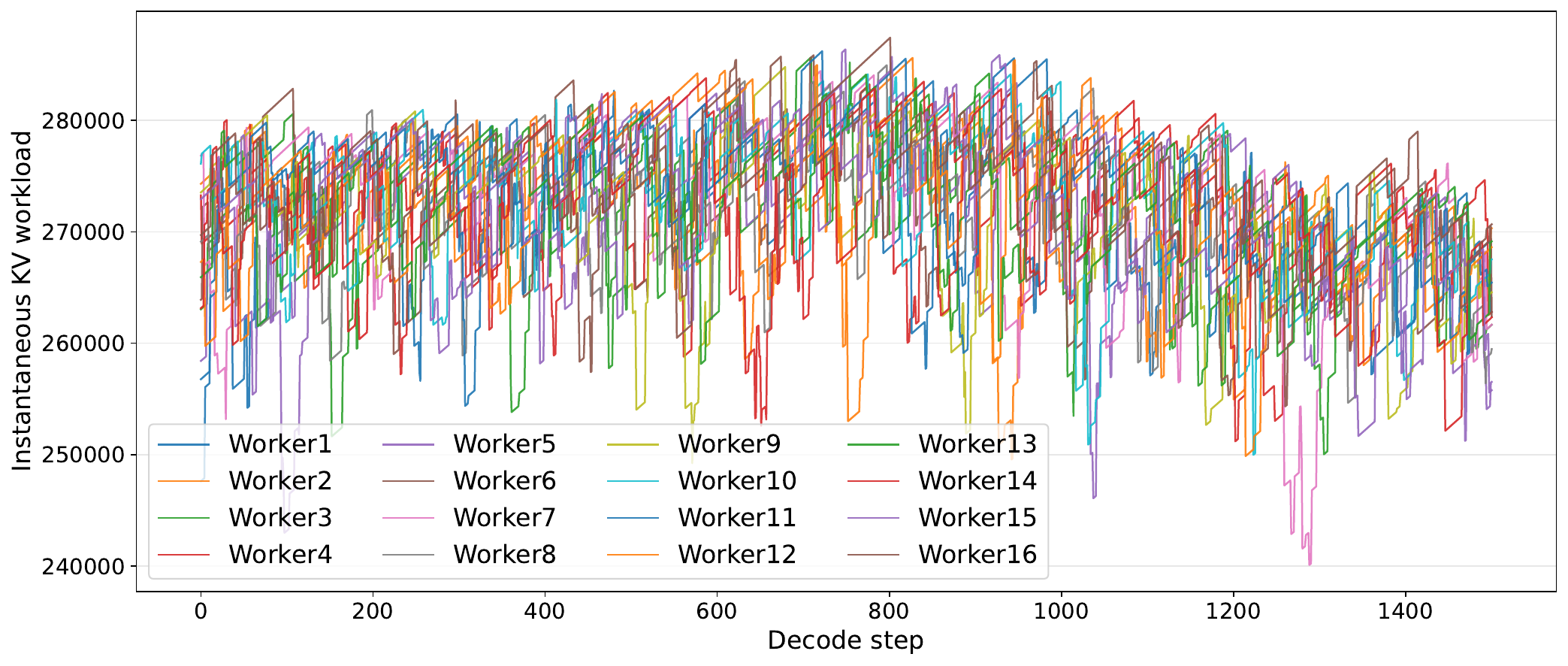}
    \caption*{\centering\footnotesize $\beta{=}48,\,\gamma{=}0.9$ \\ \textit{imbal.\ 118k}}
\end{subfigure}\hfill
\begin{subfigure}[t]{0.48\linewidth}
    \centering
    \includegraphics[width=\linewidth]{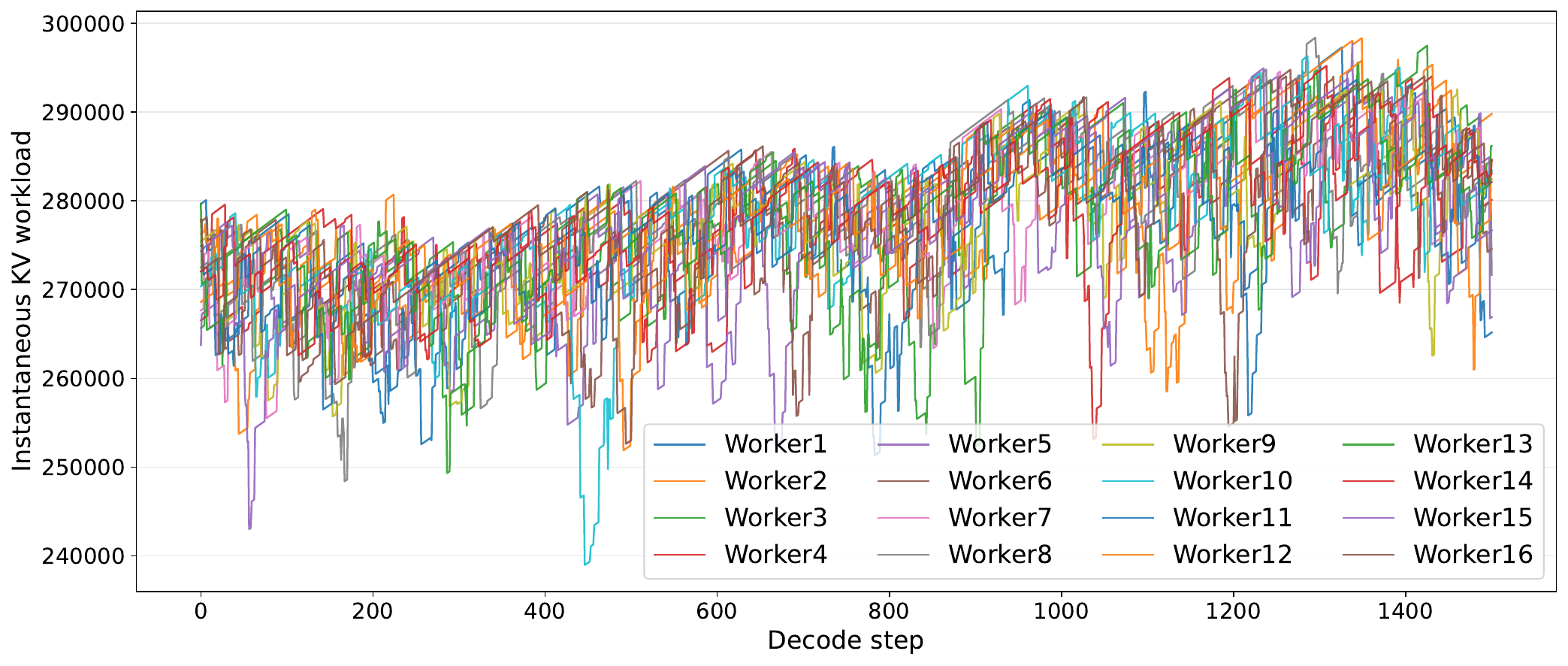}
    \caption*{\centering\footnotesize $\beta{=}96,\,\gamma{=}0.9$ \\ \textit{imbal.\ 113k}}
\end{subfigure}
 
\vspace{0.7em}
 
\begin{subfigure}[t]{0.48\linewidth}
    \centering
    \includegraphics[width=\linewidth]{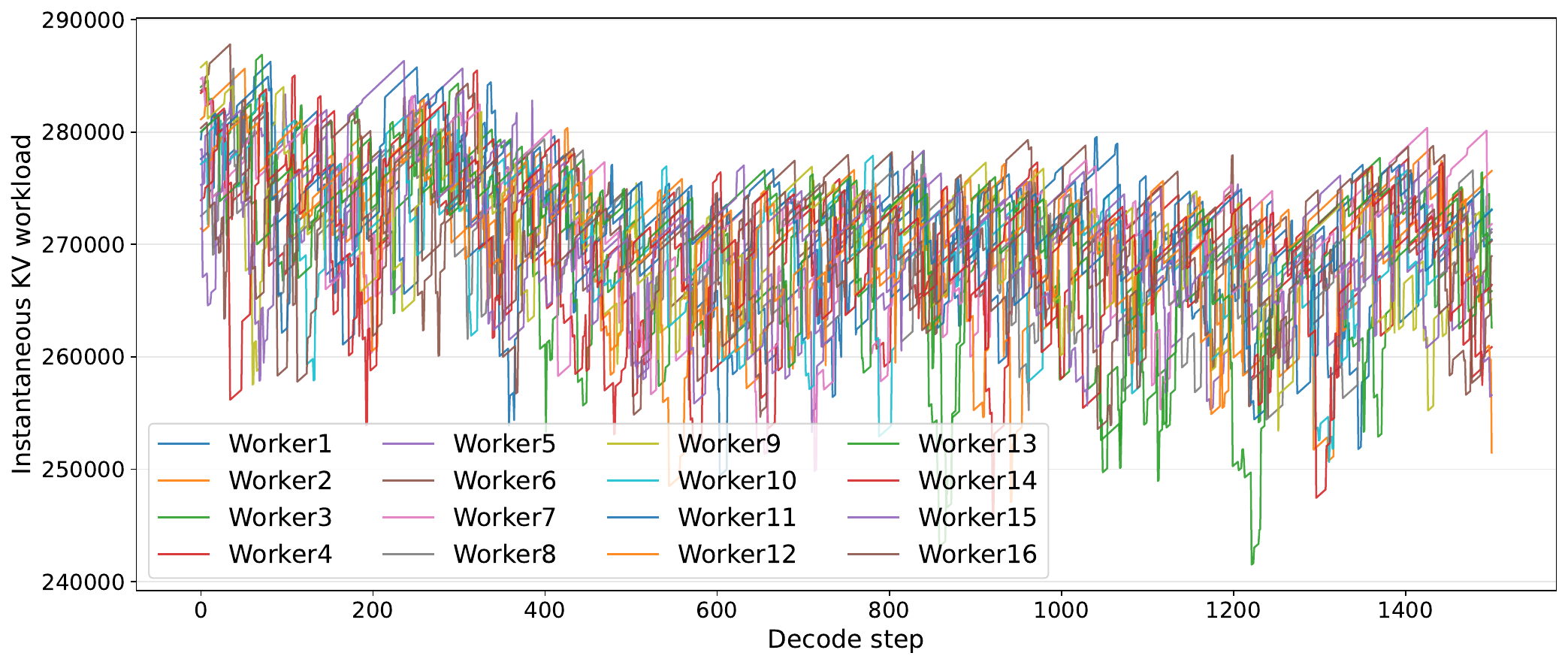}
    \caption*{\centering\footnotesize $\beta{=}48,\,\gamma{=}0.5$ \\ \textit{imbal.\ 116k}}
\end{subfigure}\hfill
\begin{subfigure}[t]{0.48\linewidth}
    \centering
    \includegraphics[width=\linewidth]{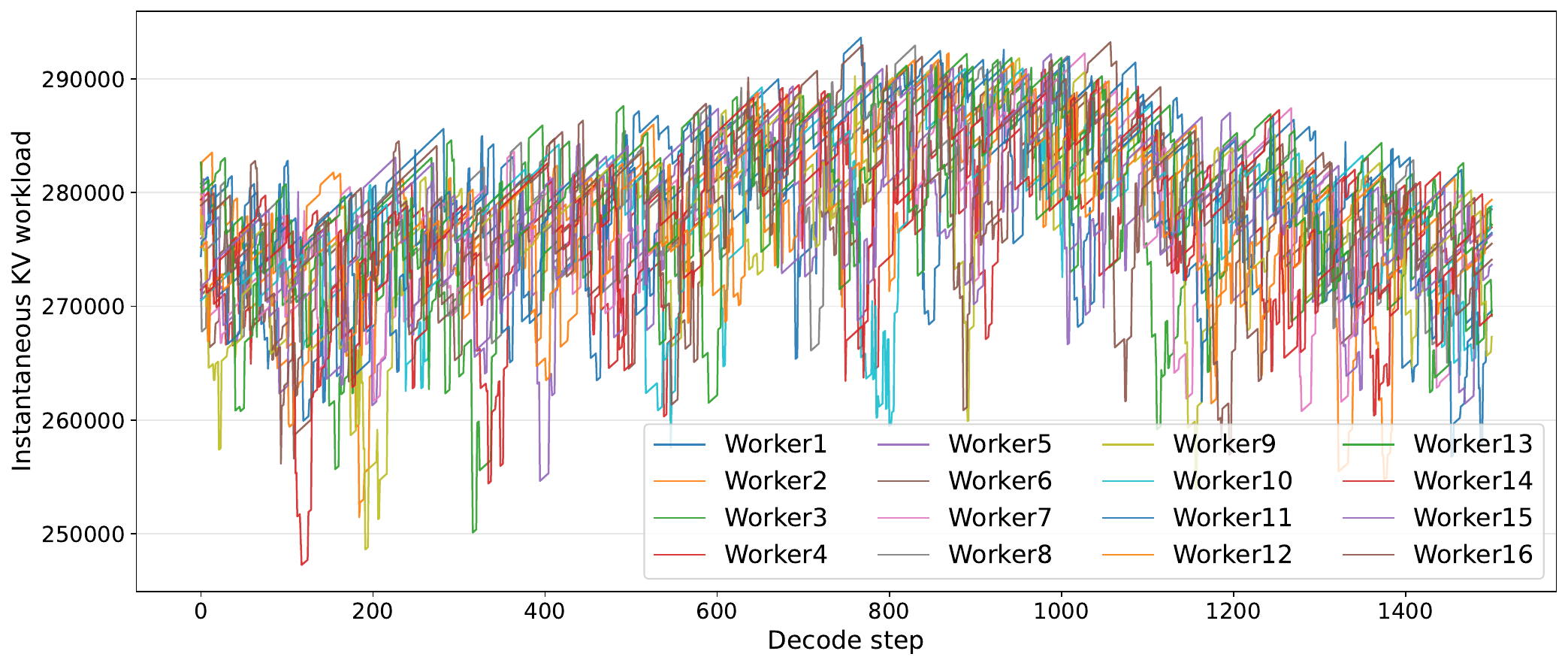}
    \caption*{\centering\footnotesize $\beta{=}48,\,\gamma{=}0.7$ \\ \textit{imbal.\ 110k}}
\end{subfigure}\hfill
\begin{subfigure}[t]{0.48\linewidth}
    \centering
    \includegraphics[width=\linewidth]{plots/Appendix/sensitivity/kv_16g_b48_gamma0.9.pdf}
    \caption*{\centering\footnotesize $\beta{=}48,\,\gamma{=}0.9$ \\ \textit{imbal.\ 118k}}
\end{subfigure}\hfill
\begin{subfigure}[t]{0.48\linewidth}
    \centering
\includegraphics[width=\linewidth]{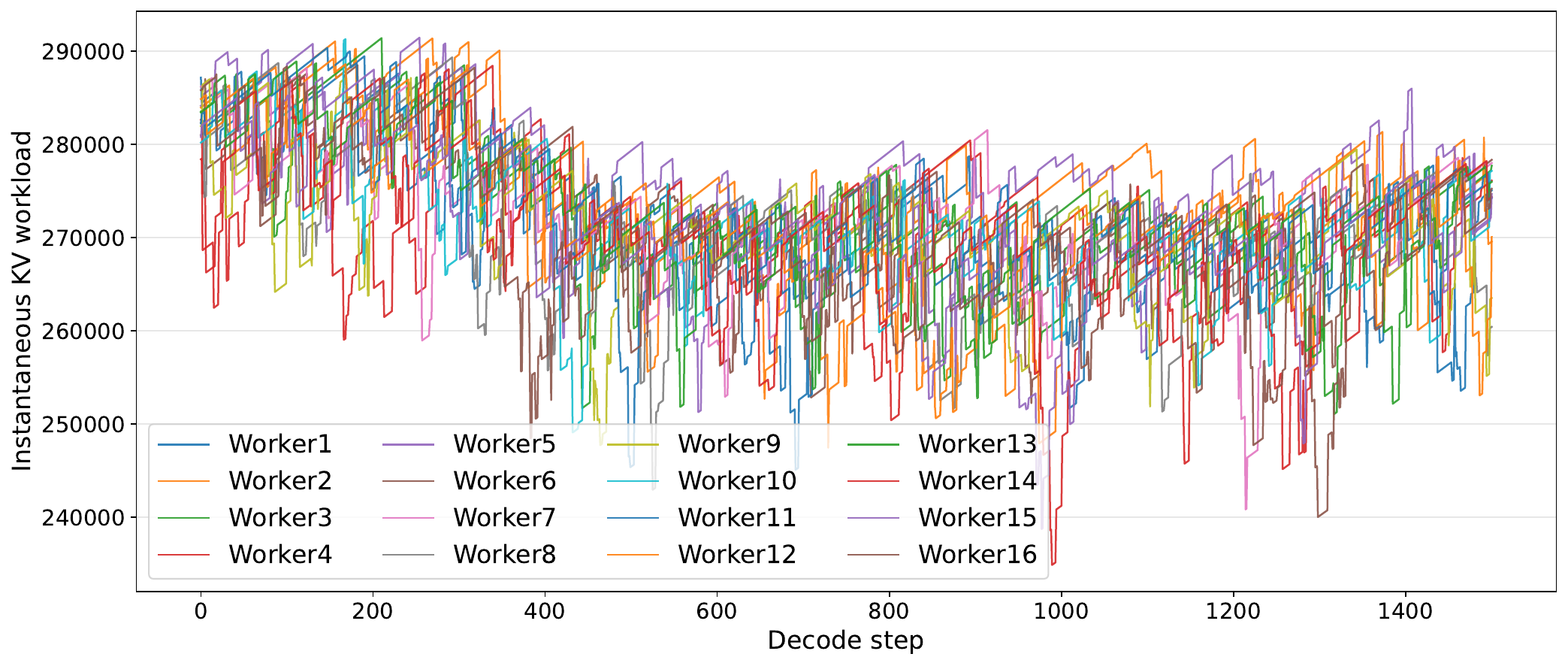}
    \caption*{\centering\footnotesize $\beta{=}48,\,\gamma{=}1.0$ \\ \textit{imbal.\ 127k}}
\end{subfigure}
 
\caption{\textbf{Per-worker KV-cache traces under each swept $(\beta, \gamma)$ on Proprietary Data at $G{=}16$, BR-H oracle, $H{=}80$.} Same convention as Figure~\ref{fig:sensitivity-traces-g8}; each panel is labelled by $(\beta, \gamma)$ and trace-mean imbalance.}
\label{fig:sensitivity-traces-g16}
\end{figure}
 
\paragraph{Reading the $G{=}16$ gallery.}
The qualitative pattern matches the $G{=}8$ gallery: every panel shows a tightly clustered $16$-worker band with no visibly diverging worker, and the band shape is similar across the $(\beta, \gamma)$ region. Visually, the bands at $G{=}16$ are wider in absolute terms than at $G{=}8$---consistent with the order-statistics scaling discussed in Section~\ref{subsec:exp_scaling}---but the spread \emph{within} the swept region remains a small fraction of the gap to any baseline panel of Figure~\ref{fig:scaling-traces-g16-main}, the trace counterpart at the same scale. The robustness conclusion of Section~\ref{subsec:exp_robust_overhead} therefore extends to the largest cluster size we tested: BR-H is not a knife-edge configuration even at $G{=}16$.
 